\documentclass[aps,prb,twocolumn,nofootinbib,longbibliography]{revtex4-2}

\usepackage{microtype} 
\usepackage[]{newtxtext} 
\usepackage[]{newtxmath}

\usepackage{amsmath}
\usepackage{amsfonts}
\usepackage{graphicx}
\usepackage{booktabs}
\usepackage{braket}

\usepackage{lipsum}
\usepackage{mwe}
\usepackage{caption}
\usepackage{subcaption}
\usepackage{ntheorem}

\newcommand{\ii}{\mathrm{i}}

\newcommand{\kv}{\boldsymbol{k}}

\usepackage{multirow}
\usepackage{xcolor}
\usepackage{hyperref}

\hypersetup{colorlinks,
	linkcolor={blue!75!black!80!yellow},
	citecolor={blue!75!black!80!yellow},
	urlcolor={blue!75!black!80!yellow}
}

\begin{document}
\begin{abstract}

The extension of topological band theory to non-Hermitian Hamiltonians with line energy gaps remains largely unexplored, despite early indications of rich underlying physics. In these systems, Wilson loops, the objects characterizing polarization, become nonunitary. Yet, the physical consequences of this nonunitarity have remained unclear. Using biorthonormal quantum mechanics, we introduce the concept of complex Wannier centers, defined from the gauge-invariant eigenvalues of nonunitary Wilson loops. Complex Wannier centers acquire physical meaning through the breaking of reciprocity in their associated Wannier functions. When the centroid of a Wannier function shifts into the complex plane, it acquires an effective momentum offset that produces directional drift over time. We analyze how symmetries constrain complex Wannier centers and identify symmetry-protected Wannier configurations in pseudo-Hermitian Hamiltonians, where the centers are either real or form complex-conjugate pairs, as determined by conserved ``Krein signatures'' of the projected metric operator of pseudo-Hermiticity. We further show that the Krein structure of the Wilson loop can establish a bulk–boundary correspondence: in a system with anticommuting pseudo-Hermitian metric and (pseudo) inversion symmetries, the behavior of complex Wannier centers predicts the existence of a filling anomaly in the occupied bands and whether the resulting edge modes experience gain or loss. Finally, we propose an implementation of this system that enables experimental tests of our predictions.

\end{abstract}

\preprint{APS/123-QED}
\author{Pedro Fittipaldi de Castro}
\author{Yifan Wang}
\author{Wladimir A. Benalcazar}
\email{benalcazar@emory.edu}
\affiliation{
Department of Physics, Emory University, Atlanta, GA 30322, USA
}
\title{Complex Wannier centers and drifting Wannier functions in non-Hermitian Hamiltonians}
\maketitle

\section{Introduction}

The past few years have witnessed the rapid development of non-Hermitian (NH) topological band theory—an effective single-particle description of open quantum systems that undergo non-unitary time evolution. Two hallmarks of NH Hamiltonians are exceptional points~\cite{Heiss_2012,Ding_Fang_Ma_2022,Bergholtz_2021,Kawabata_2017_Exceptional,konig_braid-protected_2023} and the non-Hermitian skin effect (NHSE)~\cite{shen_topological_2018,yao_edge_2018,Xiong_2018,Song_2019,PhysRevLett.124.056802,okuma_topological_2020,Zhang_Yang_Fang_2022,Zhang_2020,Yang_2020,Yokomizi_2019,Jiang_2024} in Hamiltonians with point gaps. Such gaps give rise to intrinsically non-Hermitian topological phases~\cite{Okugawa_2021,Vecsei_2021,schindler_hermitian_2023,Shiozaki_2021,wang2024classifyingordertwospatialsymmetries,wang2024higherordertopologicalknotsnonreciprocal,tanaka2024exceptionalsecondordertopologicalinsulators} and enrich the ten-fold AZ classification~\cite{altland1997nonstandard} of topological insulators and superconductors to 38 symmetry classes~\cite{kawabata_symmetry_2019}. However, systems with point gaps generally exhibit unstable dynamics, manifested in the damping or amplification of wave functions.

By contrast, line-gapped NH Hamiltonians can exhibit real spectra and stable dynamics in the presence of pseudo-Hermiticity~\cite{Ali,mostafazadeh2002pseudo,mostafazadeh2001pseudo,mostafazadeh2002pseudo3} or parity-time symmetry~\cite{PhysRevLett.80.5243,Bender_2015}. Because line-gapped NH Hamiltonians can be smoothly deformed into Hermitian or anti-Hermitian ones without closing the gap, their band topology is often assumed to reduce to the Hermitian case. Nevertheless, important questions about the band geometry of line-gapped NH systems remain open.

Notoriously, the splitting of energy eigenstates into right and left eigenkets~\cite{Curtright_2007,Pati_2009,Kunst,Brody_2013,Ashida_2020,edvardsson2023biorthogonalrenormalization} introduces an amplitude gauge freedom in the definition of state vectors: one may rescale pairs of right and left Bloch eigenstates by arbitrary reciprocal factors without violating biorthonormality. In contrast to the unitary gauge transformations of Hermitian Hamiltonians, this enlarged freedom renders the amplitudes of Wilson-line elements locally gauge dependent under parallel transport in momentum space. However, because admissible gauge transformations must be periodic over the Brillouin zone, any local rescaling necessarily cancels along a closed non-contractible path. Consequently, deviations from unitarity in the Wilson loops of non-Hermitian energy bands cannot be gauged away and define a global, gauge-invariant property. For the Abelian case, the nonunitarity of Wilson loops has already been recognized in early formulations of a complex-valued Berry phase in~\cite{GARRISON1988177,PhysRevA.98.053833,PhysRevA.87.012118,PhysRevB.107.085122,lane2025complex,PhysRevResearch.5.L032026}. The more detailed structure of the non-Abelian nonunitary Wilson loops demands a reconsideration of the geometric objects that characterize the Wannier center configurations and the bulk-boundary correspondence (BBC)~\cite{Benalcazar,PhysRevB.100.195135,Fidkowski,Hwang,vanderbilt_2018,Yu,Alexandradinata}.

In this work, we identify the geometric origin and physical meaning of Wilson-loop nonunitarity in line-gapped systems. In Hermitian Hamiltonians, the eigenvalues of unitary Wilson loops are gauge-invariant $U(1)$ numbers $\exp(- 2\pi \ii\nu)$ that encode the Wannier centers (WCs) $\nu$. The main contribution of our paper is two-fold. First, we show that, as NH Hamiltonians are characterized by biorthogonal Wilson loops that break unitarity, the Wilson loop eigenvalues $\lambda$ ---while still gauge invariant--- are generally not unimodular, $|\lambda| \neq1$. Correspondingly, the Wilson loops eigenvalues can be thought of as encoding \emph{complex Wannier centers},  $\nu+\mathrm{i}\kappa=\frac{\mathrm{i}}{2\pi}\log\lambda$. Second, we demonstrate that these complex Wannier centers correspond to Wannier functions (WFs) with nonreciprocal dynamics. Heuristically, the imaginary components $\kappa$ of the WCs shift the centroids of the associated WFs into the complex plane, thereby imposing a linear phase gradient on an otherwise transform-limited wave packet and acting as its effective momentum. This nonreciprocal behavior is manifested in the dynamics of the WFs, which we verify numerically by time evolving the eigenstates of the position operator projected onto the bands of interest.

We also investigate how symmetries constrain the spectrum and eigenvectors of the Wilson loop $\mathcal{W}$ and the associated Wannier Hamiltonian $H_{W}=\frac{\mathrm{i}}{2\pi}\log\mathcal{W}$, giving rise to multiple symmetry classes and symmetry-protected Wannier configurations. The corresponding Wannier transitions, which occur when Wannier centers become degenerate in the complex plane, need not coincide with bandgap closings in the energy spectrum.

We first show that pseudo-Hermitian (pH) Bloch Hamiltonians lead to pseudo-unitary Wilson loops, and that the projected pH metric acts as an effective symmetry operator for the Wannier Hamiltonian. This property endows the Wannier spectrum with a Krein signature structure~\cite{geyer2025stability,chernyavsky2018krein,schindler_hermitian_2023}: Wannier functions are either associated with real Wannier centers carrying well-defined Krein signatures or occur in complex-conjugate pairs related by the projected metric. Transitions between these regimes can only occur through collisions of real Wannier centers with opposite Krein signatures.

In addition, we show that the intrinsically non-Hermitian phenomenon of symmetry ramification~\cite{kawabata_symmetry_2019} further enriches the classification of Wannier configurations. As a concrete example, we consider inversion and pseudo-inversion symmetries, which ramify into distinct symmetry branches in non-Hermitian systems. When both branches are simultaneously present, they enforce pseudo-Hermiticity of the Wilson loop. If the corresponding symmetry operators mutually anticommute, a BBC follows: the real parts of the Wannier centers track the presence of a filling anomaly in the occupied bands and the associated edge-localized modes, while the imaginary parts determine dynamical stability, predicting whether these modes undergo amplification or attenuation.

The theoretical foundation of the paper is laid out in Sec.~II, where we introduce biorthogonal and nonunitary Wilson loops and complex Wannier centers, together with a physical interpretation of the imaginary parts of Wannier centers. We corroborate these theoretical findings through dynamical simulations of Wannier functions. In Sec.~III, we focus on pseudo-Hermitian Hamiltonians and the Wannier configurations protected by the total Krein signature carried by the Wannier functions. Using a one-dimensional model, we verify that Krein signatures determine whether degenerate Wannier centers remain real or split into complex-conjugate pairs. In Sec.~IV, we analyze the ramification of inversion symmetry induced by non-Hermiticity and derive the constraints that its symmetry branches impose on the Wilson loop. We investigate both commuting and anticommuting realizations of inversion and pseudo-inversion symmetries. While both cases constrain the Wannier spectrum, the commuting case is more restrictive and forbids nontrivial Krein collisions, whereas the anticommuting case allows such collisions and thereby gives rise to a bulk–boundary correspondence, as demonstrated by an explicit example that can be implemented in multimode photonic systems. 

\section{Complex Wannier centers and drifting Wannier functions from nonunitary Wilson loops}{\label{TheoryLayout}}

Let us consider a system whose non-Hermitian Hamiltonian $H$ acts on a Hilbert space $\mathcal{H}$ of dimension $D$. In general, there is no orthonormal eigenbasis for $H$, but we can almost always construct a biorthonormal set of right and left eigenvectors, $\ket{\psi^{R}_{j}}$ and $\ket{\psi^{L}_{j}}$, such that $\langle\psi^{L}_{i}|\psi^{R}_{j}\rangle=\delta_{ij}
$ and
\begin{align}
    H\ket{\psi^{R}_{j}}&=E_j\ket{\psi^{R}_{j}}, \nonumber \\
    H^{\dagger}\ket{\psi^{L}_{j}} &= E_{j}^{*} \ket{\psi^{L}_{j}}, \label{BiOrthonormalHamiltonian}
\end{align}
as we show in Appendix \ref{AppendixBiorthonormal}. From now on, we will always assume that we are working with a biorthonormal basis, unless stated otherwise.

Let us also restrict ourselves to non-Hermitian Hamiltonians with a real \emph{line gap} in their spectrum, so we can define the right and left \emph{occupied subspaces}, or more generally the \emph{subspaces of interest}~\footnote{The notion of a Fermi level and occupied bands does not apply to photonic platforms, where one can in principle excite any band of interest. For brevity, we adopt the term ``occupied'' throughout the text, but our results apply to arbitrary gapped subspaces.}, as the subsets of $\mathcal{H}$ spanned by the right and left eigenvectors whose real part of $E_{j}$ is less than a reference value $E_0 \in \mathbb{R}$:
\begin{align}
    \mathcal{R}_{\mathrm{occ}} &=\mathrm{span}(\ket{\psi^{R}_{1}},\dots,\ket{\psi^{R}_{D_{\mathrm{occ}}}}), \nonumber \\
    \mathcal{L}_{\mathrm{occ}} &=\mathrm{span}(\ket{\psi^{L}_{1}},\dots,\ket{\psi^{L}_{D_{\mathrm{occ}}}}), \label{RightandLeftsubspace}
\end{align}
where $D_{\mathrm{occ}} \leq D$. Throughout the paper, we work with operators restricted to the occupied subspaces.
It is convenient to introduce the auxiliary space $\mathbb{C}^{D_{\mathrm{occ}}}$
and the maps $\mathbb{C}^{D_{\mathrm{occ}}} \to \mathcal{H}$
\begin{align}
    R &= \sum_{j=1}^{D_{\mathrm{occ}}} \ket{\psi_{j}^{R}}\bra{j}, \quad
    L = \sum_{j=1}^{D_{\mathrm{occ}}} \ket{\psi_{j}^{L}}\bra{j},
    \label{OccupiedSubspaces}
\end{align}
where $\ket{j}$ is the canonical basis of $\mathbb{C}^{D_{\mathrm{occ}}}$ and
$L^\dagger R = \mathbb{I}_{D_{\mathrm{occ}}}$. Any operator
$X:\mathcal{H} \to \mathcal{H}$ admits the occupied-subspace representation
\begin{equation}
    X_{\mathrm{occ}} \equiv L^\dagger X R.
    \label{Projection}
\end{equation}

Alternatively, one may work with the projected operator
\begin{equation}
    X_P = P X P,
    \label{Xp}
\end{equation}
where the biorthogonal projector onto the occupied subspace is
\begin{equation}
    P = R L^\dagger
      = \sum_{j=1}^{D_{\mathrm{occ}}} \ket{\psi^{R}_j}\bra{\psi^{L}_j}.
    \label{Pdef}
\end{equation}
While $X_{\mathrm{occ}}$ is a full-rank $D_{\mathrm{occ}} \times D_{\mathrm{occ}}$ matrix
acting on $\mathbb{C}^{D_{\mathrm{occ}}}$, $X_P$ acts on $\mathcal{H}$ and vanishes on
the unoccupied sector. We use either representation according to convenience, noting that $X_p=RX_{\mathrm{occ}}L^{\dagger}$.

While this projection formalism is general, we now focus on crystalline systems with discrete translation invariance, where the biorthogonal set of occupied energy eigenstates naturally leads to Wilson loops and Wannier centers, the central objects of our analysis.

\subsection{Non-unitary Wilson loops and complex Wannier centers}

Let us define biorthogonal Wilson loops in a way that is consistent with the formulations of Refs.~\cite{PhysRevLett.123.073601,PhysRevA.87.012118,PhysRevLett.123.073601,masuda2022relationship,Kunst,Edvardsson,Hu,OrtegaTaberner,Chen,PhysRevA.87.012118}.
Discrete translation invariance in a crystal with $\ell$ unit cells allows us to express the $D$-dimensional Hamiltonian in block diagonal form $H = \oplus_k h_k$, where each Bloch Hamiltonian $h_k$ acts on a sector of dimension $N=D/\ell$ corresponding to the internal (orbital) degrees of freedom at fixed crystal momentum $k=m \Delta$, where $m \in \{0,\dots,\ell-1\}$ and $\Delta = 2\pi/\ell$, which implies our unit of length is the unit cell. As a result, the eigenstates of the full Hamiltonian $H$, as defined in \eqref{BiOrthonormalHamiltonian}, can be written in Bloch form
\begin{align}
    \ket{\psi^{R,L}_{k,n}} &= \ket{k} \otimes \ket{u^{R,L}_{k,n}}, \label{BlochState}
\end{align}
where $n=1,\dots,N$ is the band index. Here, $\ket{u^{R,L}_{k,n}}$ are the right/left eigenvectors of the Bloch Hamiltonian,
\begin{align}
    h_k\ket{u^{R}_{k,n}}&=E_{k,n}\ket{u^{R}_{k,n}}, \nonumber \\
    h_k^{\dagger}\ket{u^{L}_{k,n}} &= E_{k,n}^{*} \ket{u^{L}_{k,n}} \label{BiOrthonormalBloch},
\end{align}
satisfying the biorthonormality condition $\langle u^{L}_{k,m}|u^{R}_{k,n}\rangle=\delta_{mn}$. Now, let $N_{\mathrm{occ}} = D_{\mathrm{occ}}/\ell$ be the number of occupied energy bands and, in the spirit of \eqref{OccupiedSubspaces}, let us collect the right and left occupied eigenvectors of each momentum sector as
\begin{align}
    R_k &= \sum_{n=1}^{N_{\mathrm{occ}}}\ket{u^{R}_{n,k}}\bra{n}, \nonumber \\
    L_k &= \sum_{n=1}^{N_{\mathrm{occ}}}\ket{u^{L}_{n,k}}\bra{n}, \label{RandL}
\end{align}
where $\ket{n}$ are the canonical basis vectors of the auxiliary space $\mathbb{C}^{N_{\mathrm{occ}}}$ and $L_k^{\dagger}R_{k}=\mathbb{I}_{N_{\mathrm{occ}}}$.

The biorthogonal \emph{Wilson line element} connecting two neighboring points of the BZ is given by the $N_{\mathrm{occ}} \times N_{\mathrm{occ}}$ matrix
\begin{equation}
    G_{k} = L_{k+\Delta}^{\dagger}R_{k}. \label{WilsonLineElement}
\end{equation}
In the thermodynamic limit $\ell \to \infty$ ($\Delta \to 0$), we may expand the Wilson line element up to first order in $\Delta$ as
\begin{align}
    G_{k} &= L_k^{\dagger}R_{k} + \Delta(\partial_kL^{\dagger}_k)R_k + \mathcal{O}(\Delta^2) \nonumber \\
    &= \mathbb{I}_{N_{\mathrm{occ}}} - \mathrm{i}\Delta A_k+\mathcal{O}(\Delta^2),
\end{align}
where
\begin{equation}
    A_k\equiv \mathrm{i}(\partial_kL^{\dagger}_k)R_k = -\mathrm{i}L_{k}^{\dagger}\partial_kR_k \label{BiorthogonalBerry}
\end{equation}
is the biorthogonal Berry connection~\cite{PhysRevLett.123.073601,PhysRevA.87.012118}. Hence, the Wilson line from point $k_i$ to $k_f$ can be written in terms of the Berry connection as
\begin{align}
    W_{k_f \leftarrow k_i}&=\lim_{\Delta \to 0}\prod_{k=k_i}^{k_f-\Delta}G_{k} \nonumber \\
    &= \mathcal{P}\exp\left(-\mathrm{i}\int_{k_i}^{k_f}A(k)dk\right), \label{WilsonLine}
\end{align}
where $\mathcal{P}$ is the path-ordering operator~\footnote{We define $\mathcal{P}$ such that, if $k_2>k_1$, then $\mathcal{P}\{O(k_1)O(k_2)\}=\mathcal{P}\{O(k_2)O(k_1)\}=O(k_2)O(k_1)$, where $O$ is some operator.} in momentum space. The biorthogonal \emph{Wilson loop} at base point $k$ is a Wilson line over a closed path $C$ in the BZ~\footnote{ In $D>1$ spatial dimensions, Wilson loops exist for any closed path in the BZ. However, only Wilson loops that cross the entire BZ are physically relevant, as they are not contractible to a point and their eigenvalues are gauge-invariant quantities that indicate the locations of electrons in crystalline insulators. In this paper, ``Wilson loops'' refer exclusively to those taken over non-contractible loops of the BZ.},
\begin{equation}
    \mathcal{W}_k = \mathcal{P}\exp\left(-\mathrm{i}\oint_{C}A(k')dk'\right). \label{BiorthW}
\end{equation}
where $C$ has endpoints $k$ and $k+2\pi$.

Because the biorthogonal Bloch basis \eqref{RandL} admits complex gauge transformations [see Appendix \eqref{Appendix:GaugeFreedom}], the Wilson line elements \eqref{WilsonLineElement} are not constrained to be unitary. Correspondingly, the Berry connection \eqref{BiorthogonalBerry} is generally non-Hermitian~\cite{PhysRevLett.123.073601,PhysRevA.87.012118} and gauge dependent under this larger gauge freedom. However, the nonunitarity of the Wilson loops signals a global amplifications and attenuations that cannot be removed by any gauge choice. These nonunitary Wilson loops \eqref{BiorthW} will therefore have nonunimodular eigenvalues.

Writing the Wilson loop eigenvalues as $\lambda=e^{-2\pi \mathrm{i}z}$, we can define complex Wannier centers $z=\nu+\mathrm{i}\kappa$. As in the case of Hermitian Hamiltonians, the real part $\nu=-\frac{1}{2\pi}\mathrm{Arg}\lambda \in [-\frac{1}{2},\frac{1}{2})$ specifies the position of a Wannier function within the unit cell of length $1$, whereas the imaginary part $\kappa = \frac{1}{2\pi}\log|\lambda|$ quantifies the deviation of the Wilson loop from unitarity and captures the accumulated amplification/attenuation of the biorthogonal parallel transport of Bloch states over the BZ. Refs.~\cite{PhysRevA.87.012118,PhysRevB.107.085122,lane2025complex,PhysRevResearch.5.L032026} have described such dissipative adiabatic transport in the Abelian case in terms of complex Berry phases.

In Appendix \ref{AppendixNonUnitaryWL}, we show that a nonzero anti-Hermitian component of the Berry connection,
\begin{equation}
    A^{(a)}(k) = \frac{A(k)-A^{\dagger}(k)}{2} \neq 0, \label{NecessaryCondition}
\end{equation}
is a \emph{necessary} condition for breaking the unitarity of the Wilson loop. If $A(k)$ is Hermitian for all $k$, the operator inside the brackets in expression \eqref{BiorthW} is anti-Hermitian at every $k$ point, and its path-ordered exponential is therefore always unitary. 

By contrast, the global condition
\begin{equation}
    \oint_C\mathrm{Tr} A^{(a)}(k')dk' \neq 0. \label{Sufficient}
\end{equation}
is \emph{sufficient} but not necessary to render the Wilson loop nonunitary, as it implies  $|\det \mathcal{W}_k| \neq 1$. For instance, pseudo-unitary Wilson loops---discussed in Sec. III---obey \eqref{NecessaryCondition} but break \eqref{Sufficient}. As a result, they have a unit determinant but non-unimodular eigenvalues.

The nonunitarity of the Wilson loop, parametrized by $\kappa$, is not merely a formal artifact of non-Hermiticity: it has a clear manifestation in the dynamics of WFs, as we shall now motivate with a heuristic argument and then demonstrate from first principles.

\subsection{Drifting Wannier functions}

To understand the physics of $\kappa$, consider a one-dimensional crystal with $N_{\mathrm{occ}}=1$ occupied band. In the Hermitian limit, a Wannier function $\ket{W_{x_0}}$ is a state exponentially localized and peaked at a point $x_0+\nu$ in space, where $x_0$ is a unit cell coordinate (which we take to be $x_0=0$ for simplicity) and $\nu$, the real WC, is the coordinate relative to the center of the unit cell. For each orbital component $\alpha=1,\dots,N$, we define the position-space amplitude
\begin{equation}
    F_\alpha(x)\equiv (\bra{x}\otimes\bra{\alpha})\ket{W_{0}},
\end{equation}
and its corresponding weight profile $\rho_\alpha(x)\equiv |F_\alpha(x)|^2$, typically smooth and symmetric about its maximum at $x=\nu$. Expanding near the peak, we may approximate the localized envelope by a Gaussian,
\begin{equation}
    F_{\alpha}(x)\approx F_{\alpha}(\nu)\exp\!\left[-\frac{(x-\nu)^2}{2\sigma^2}\right],
    \label{Eq:GaussianEnvelope}
\end{equation}
where the width $\sigma$ is fixed by the curvature of $\rho_\alpha(x)$ at $x=\nu$,
\begin{equation}
    \frac{\sigma^2}{2}=-\frac{\rho_\alpha(\nu)}{\rho_\alpha''(\nu)}.
    \label{Eq:SigmaCurvature}
\end{equation}

Now, imagine that we weakly break non-Hermiticity so that the Wilson loop (a Berry phase for $N_{\mathrm{occ}}=1$) acquires a small nonunitary part and the corresponding WC becomes complex,
\begin{equation}
    \nu \to \nu + \mathrm{i}\kappa.
\end{equation}
For $|\kappa|\ll 1$, one expects the WF to remain close to its Hermitian limit. In this regime, the leading effect of a small imaginary WC is the displacement of the centroid of the envelope \eqref{Eq:GaussianEnvelope} from the real axis $(x=\nu)$ onto the complex plane $(x=\nu+\mathrm{i}\kappa)$, giving us
\begin{equation}
    F_{\alpha}(x)\approx F_{\alpha}(\nu+\mathrm{i}\kappa)\exp\!\left[-\frac{(x-\nu-\mathrm{i}\kappa)^2}{2\sigma^2}\right]
    \label{Eq:GaussianEnvelope2}.
\end{equation}
Expanding the square in the expression above yields
\begin{equation}
    F_{\alpha}(x)
    \approx
    F_{\alpha}(\nu+\mathrm{i}\kappa)e^{-\frac{(x-\nu)^2}{2\sigma^2}}
    e^{\mathrm{i}\frac{\kappa}{\sigma^2}(x-\nu)}
    e^{\frac{\kappa^2}{2\sigma^2}},
\end{equation}
so that, up to an overall normalization factor, the non-Hermitian Wannier function differs from the Hermitian one only by a plane wave factor with an effective momentum
\begin{equation}
    k_{\mathrm{eff}}\approx \frac{\kappa}{\sigma^2}.
    \label{Eq:keffEstimateGaussian}
\end{equation}

In summary, our heuristic argument suggests that, to a first approximation, the imaginary part $\kappa$ of the Wannier center manifests as a linear phase gradient in space, i.e., a momentum shift, superimposed on the WF envelope.

To make our interpretation of $\kappa$ more rigorous, we must first carefully specify what we mean by the term ``Wannier function'', as it has two definitions that coincide in one-dimensional Hermitian Hamiltonians, but are not equivalent upon breaking Hermiticity.

\subsubsection{Bloch-Fourier vs projected-position\\ definitions of a Wannier function}

Wannier functions are usually defined as Fourier transforms of the Bloch energy eigenstates~\eqref{BlochState} in a smooth gauge~\cite{PhysRev.52.191,BLOUNT1962305}. 
Restricting for simplicity to one-dimensional systems and to a single occupied band ($N_{\mathrm{occ}}=1$), this construction reads
\begin{align}
    \ket{W_{x}} 
    &= \frac{1}{\sqrt{\ell}}\sum_{k} e^{-\mathrm{i} k x}\ket{\psi_{k}^{R}} \nonumber \\
    &= \frac{1}{\sqrt{\ell}}\sum_{k} e^{-\mathrm{i} k x}\ket{k} \otimes \ket{u_{k}^{R}},
    \label{FourierConstructed}
\end{align}
where the sum runs over the $\ell$ crystal momenta in the Brillouin zone. 
We refer to this construction as the \emph{Bloch--Fourier} definition of a Wannier function.

Wannier functions have also been treated~(see, for instance, Ref.~\cite{PhysRevB.26.4269}) as the maximally localized states belonging to the occupied bands. In one dimension, they are the (right) eigenstates of the projected position operator $X_p$ as defined in \eqref{Xp}, i.e., the \emph{projected-position} Wannier functions satisfy
\begin{equation}
    X_p \ket{W_{x}} = \lambda_x \ket{W_{x}},
    \label{PositionConstructed}
\end{equation}
with eigenvalues
\begin{equation}
    \lambda_x = \exp\left(-\frac{2 \pi \mathrm{i} x}{\ell}\right)\exp\left(-\frac{2 \pi \mathrm{i} z}{\ell}\right) \label{LambdaX}
\end{equation}
depending on the Wannier center $z$ (see Appendix \ref{AppendixEigenvaluesofProjectedPositionOperator}). In the definition of $X_p$ in \eqref{Xp}, $X$ is the unitary (periodic) position operator introduced by Resta~\cite{PhysRevLett.80.1800},
\begin{equation}
    X = \sum_{x=0}^{\ell-1} 
    \ket{x} \, e^{-\frac{2\pi \mathrm{i}}{\ell} x} \bra{x}
    \otimes \mathbb{I}_{N},
    \label{Resta}
\end{equation}
where $\mathbb{I}_{N}$ is the identity operator acting on the orbital degrees of freedom.

In one-dimensional Hermitian insulators, one can always find a gauge where the Bloch-Fourier construction~\eqref{FourierConstructed} coincides with the projected-position ~\eqref{PositionConstructed} definition. We now demonstrate that this correspondence breaks down in generic non-Hermitian Hamiltonians and argue that the projected-position construction is the physically relevant definition of Wannier functions; we henceforth adopt it.

\subsubsection{Nonuniform momentum weight distribution\\ of non-Hermitian Wannier functions}
\label{Sec:InequivalenceNH}

\begin{figure}
    \centering
    \includegraphics[width=\linewidth]{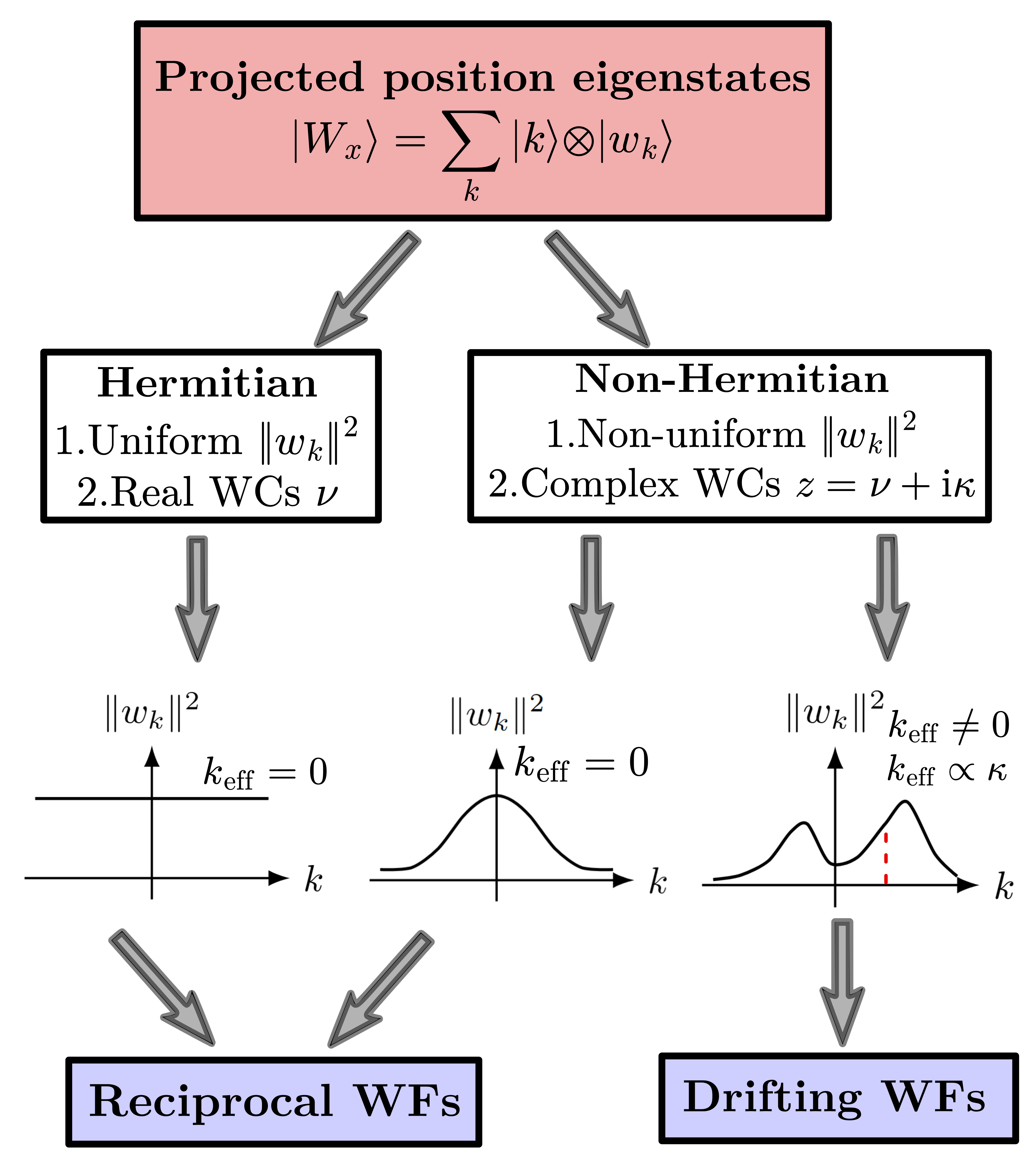}
    \caption{Flowchart summarizing key properties of Wannier functions (understood here as eigenstates of the projected position operator) in one dimension.
    In Hermitian Hamiltonians, the momentum-space distribution $\|w_k\|^2$ is uniform and consistent with the Bloch--Fourier construction, leading to reciprocal wave-packet dynamics.
    In non-Hermitian Hamiltonians, $\|w_k\|^2$ is generically non-uniform and Wannier centers can become complex, $z=\nu+\mathrm{i}\kappa$.
    The non-uniform momentum distribution may be symmetric or asymmetric about $k=0$, leading respectively to reciprocal or drifting (nonreciprocal) dynamics characterized by a nonzero average momentum $k_{\mathrm{eff}}$.
    In many cases of interest, and for weak non-Hermiticity, $k_{\mathrm{eff}}$ scales linearly with $\kappa$.}
    \label{Diagram}
\end{figure}

Let $\ket{W_x}$ be a right projected-position Wannier function satisfying the eigenvalue equation~\eqref{PositionConstructed}. 
One may always expand such a state in the Bloch basis as
\begin{equation}
    \ket{W_x} = \sum_{k}\ket{k} \otimes \ket{w_k},
\end{equation}
where $\ket{w_k}=a_k\ket{u^{R}_{k}}$ equals the right-occupied Bloch state $\ket{u^{R}_{k}}$ up to a $k$-dependent coefficient~\footnote{The coefficient $a_k=\langle u^{L}_{k}|w_{k}\rangle$ compensates the gauge dependence of $\ket{u^{R}_k}$.}. The (gauge-invariant) momentum-space weight distribution of $\ket{W_x}$ is then given by $\|w_k\|^2$ (see Appendix~\ref{Appendix:Inequivalence}).

If $\ket{W_x}$ could be written in the equal-weight Bloch--Fourier form~\eqref{FourierConstructed}, its momentum components would necessarily have uniform weight,
\begin{equation}
    \|w_k\|^2 = \frac{1}{\ell},
\end{equation}
independent of $k$. However, we show in Appendix~\ref{Appendix:Inequivalence} that the distribution $\|w_{k}\|^2$ for a single non-Hermitian band satisfies
\begin{equation}
    \|w_k\|^2
    =\|w_{k_0}\|^2
    \exp\!\left[
        2\int_{k_0}^{k}\big(\Gamma(k')-\overline{\Gamma}\big)\,dk'
    \right],
    \label{Eq:WeightDistributionGamma}
\end{equation}
where
\begin{equation}
    \Gamma(k)
    \equiv
    \mathrm{Im}\,A(k)
    +\frac{1}{2}\,\partial_k\log\|u_k^{R}\|^2,
    \quad
    A(k)=-\mathrm{i}\langle u_k^{L}|\partial_k u_k^{R}\rangle.
    \label{Eq:GammaDefMainText}
\end{equation}
The second term in $\Gamma(k)$ accounts for the possible non-unitarity of the right Bloch eigenvectors and vanishes identically in Hermitian Hamiltonians, while
\begin{equation}
    \overline{\Gamma}=\frac{1}{2\pi}\int_{\mathrm{BZ}}\Gamma(k)\,dk = \kappa \label{GammaAverage}
\end{equation}
The object $\Gamma(k)$ is gauge invariant and its Brillouin-zone average \eqref{GammaAverage} equals the imaginary part of the Wannier center $z=\nu+\mathrm{i}\kappa$ [see Appendix \ref{Appendix:Inequivalence}].

Thus, whenever $\Gamma(k)$ has non-uniform $k$ dependence, the weights $\|w_k\|^2$ are non-uniform across the Brillouin zone, and $\ket{W_x}$ does not have an equal-modulus superposition of Bloch eigenstates. Therefore, the Bloch-Fourier \eqref{FourierConstructed} and projected-position \eqref{PositionConstructed} definitions of Wannier functions are inequivalent in non-Hermitian Hamiltonians.

On physical grounds, this inequivalence singles out the projected-position construction as the appropriate definition of Wannier functions in non-Hermitian Hamiltonians.
The operator $X_p = PXP$ is a bona fide observable: its spectrum is gauge-invariant, directly tied to the Wilson loop, and its eigenstates are uniquely defined up to global phase redefinitions.
In contrast, the Bloch--Fourier construction does not faithfully capture the spatial structure of localized states selected by the physical position operator, and from now on, we reserve the term ``Wannier function'' for the projected-position construction \eqref{PositionConstructed}.

Crucially, the nonuniform momentum weight distribution $\|w_k\|^2$ is not merely a formal feature of Wannier functions but has salient dynamical consequences. If $\|w_k\|^2$ is not symmetric about $k=0$, then the average momentum of the projected-position Wannier function (in the $[-\pi,\pi)$ convention),
\begin{equation}
    k_{\mathrm{eff}} = \frac{\int_{-\pi}^{\pi}k\|w_k\|^2 \ dk}{\int_{-\pi}^{\pi}\|w_k\|^2 \ dk}, \label{AvgMomentum}
\end{equation}
is nonzero and produces a directional drift of \emph{Wannier wave packets}, i.e., wave packets that time evolve from Wannier functions. Using \eqref{Eq:WeightDistributionGamma}, we can relate the momentum weights at $k$ and $-k$ as
\begin{equation}
    \frac{1}{2}\log\left(\frac{\|w_k\|^2}{\|w_{-k}\|^2}\right) = \int_{-k}^{k}(\Gamma(k)-\overline{\Gamma})dk.
\end{equation}
Thus, an asymmetry is present unless $\Gamma(k)-\overline{\Gamma}$ is an odd function of $k$. In Appendix \ref{AppendixProtectionGamma}, we show that for a single occupied band, symmetries of the kind
\begin{equation}
    h(k)=Sh(-k)S^{-1},
\end{equation}
of which inversion symmetry is an example, impose the constraint $\Gamma(-k)=-\Gamma(k)$, and therefore imply $\|w_{-k}\|^2=\|w_{k}\|^2$ and $k_{\mathrm{eff}}=0$ (the $N_{\mathrm{occ}}>1$ case less restrictive). Meanwhile, symmetries of the form
\begin{equation}
    h(k)^{\dagger}=Sh(-k)S^{-1},
\end{equation}
of which pseudo-inversion---discussed in Sec. \ref{SecSymmetries}---is an example, enforce $\Gamma(-k)=\Gamma(k)$, thus protecting the asymmetry $\|w_{-k}\|^2\neq\|w_{k}\|^2$ and $k_{\mathrm{eff}} \neq 0$.

The distribution $\|w_k\|^2$ also allows us to determine the drift velocity $v_{\mathrm{drift}}=\frac{d}{dt}\langle x \rangle $ of a Wannier wave packet. For instance, when an energy band has a purely real dispersion $E(k)$, the drift velocity is given by [c.f. Appendix \ref{Appendix:Driftvelocity}]
\begin{align}
    v_{\mathrm{drift}} 
    &= \frac{1}{\hbar}\frac{\int_{-\pi}^{\pi} \partial_kE(k) \|w_k\|^2 \ dk}{\int_{-\pi}^{\pi} \|w_k\|^2 \ dk}. \label{DriftExact}
\end{align}
If $\|w_k\|^2$ is sufficiently narrow in $k$, we can estimate \eqref{DriftExact} as
\begin{equation}
    v_{\mathrm{drift}} \sim \frac{1}{\hbar}\partial_kE(k)\big|_{k_{\mathrm{eff}}}. \label{DriftApproximate}
\end{equation}

\subsubsection{Scaling of $k_{\mathrm{eff}}$ with $\kappa$ for weak nonunitarity of the Wilson loop}

While $k_{\mathrm{eff}}$ and the drift velocity in general depend nontrivially on $\Gamma(k)$ through \eqref{Eq:WeightDistributionGamma}, in many cases of interest [cf.\ model~\eqref{ToyModel1} in Sec.~\ref{SectionExamples1D}] we recover a linear scaling of $k_{\mathrm{eff}}$ with $\kappa$ for weak nonunitarity.
Assume $\Gamma(k)=\varepsilon f(k)$ for a real parameter $\varepsilon$, which implies $\kappa=\overline{\Gamma}=\varepsilon\overline{f}$.
Under these circumstances, the momentum-weight distribution~\eqref{Eq:WeightDistributionGamma} reads
\begin{equation}
    \|w_k\|^2
    =\|w_{k_0}\|^2
    e^{\varepsilon \Phi(k)},
\end{equation}
where
\begin{equation}
    \Phi(k) = 2\int_{k_0}^{k}\left[f(k')-\overline{f}\right]dk'.
\end{equation}
Using these definitions, Eq.~\eqref{AvgMomentum} becomes
\begin{equation}
    k_{\mathrm{eff}} = \frac{\int_{-\pi}^{\pi} k\, e^{\varepsilon \Phi(k)} \, dk}{\int_{-\pi}^{\pi} e^{\varepsilon\Phi(k)} \, dk}.
\end{equation}
For small $\varepsilon$, we expand $e^{\varepsilon \Phi(k)} = 1 + \varepsilon \Phi(k)+ O(\varepsilon^2)$ and obtain
\begin{equation}
    k_{\mathrm{eff}} = \frac{\varepsilon}{2\pi}\int_{-\pi}^{\pi}k\,\Phi(k)\,dk + O(\varepsilon^2).
\end{equation}
Finally, using $\varepsilon=\kappa/\overline{f}$ (for $\overline{f}\neq 0$)\footnote{The case $\overline{f}=0$ is a codimension-one constraint and is therefore nongeneric in parameter space unless enforced by symmetry.},
we find
\begin{equation}
    k_{\mathrm{eff}}
    =
    \frac{\kappa}{2\pi\,\overline{f}}
    \int_{-\pi}^{\pi}k\,\Phi(k)\,dk
    + O(\kappa^2), \label{KappaRigorous}
\end{equation}
showing that $k_{\mathrm{eff}}$ is linear in $\kappa$ to leading order, in agreement with the heuristic estimate~\eqref{Eq:keffEstimateGaussian}.
Figure~\ref{Diagram} presents a flowchart summarizing the main properties of Wannier functions developed in this section.

\subsubsection{Robustness of drifting Wannier functions\\ in the thermodynamic limit}

At this stage, it is natural to ask whether the dynamical consequences of a complex Wannier center persist in the thermodynamic limit. On the one hand, line-gapped non-Hermitian Hamiltonians generically exhibit a nonunitary Wilson loop and complex Wannier centers $z=\nu+\mathrm{i}\kappa$, which can produce an effective momentum shift $k_{\mathrm{eff}}$ and directional drift of Wannier wave packets.
On the other hand, the projected position operator is constructed from the Resta operator~\eqref{Resta}, and its eigenvalues obey [c.f. Eq.~\eqref{LambdaX}]
\begin{equation}
    |\lambda_x| = \exp\left(\frac{2\pi \kappa}{\ell}\right) = 1 + \frac{2\pi\kappa}{\ell} + \mathcal{O}(\ell^{-2}). \label{Leigenvalue}
\end{equation}

Note from \eqref{Leigenvalue} that the deviation from $|\lambda_x|=1$ scales as $1/\ell$ and thus vanishes in the thermodynamic limit. In other words, although the Wilson loop of a line-gapped non-Hermitian Hamiltonian is generically nonunitary even as $\ell \to \infty$, its biorthogonal projected position operator becomes asymptotically unitary---in the sense that its eigenvalues approach the unit circle---because the Resta operator \eqref{Resta} itself contains an explicit factor of $1/\ell$. 

However, this asymptotic unitarity of the projected position operator \emph{does not} imply that the imaginary part $\kappa$ of the WC becomes unphysical. The situation is fully analogous to the Hermitian case where the real Wannier center $\nu$ yields a finite bulk polarization even though $e^{2\pi\mathrm{i}\nu/\ell} \to 1$ as $\ell \to \infty$. Likewise, the effective momentum shift on Wannier functions depends on $\kappa$ but not on $\ell$ [cf. \eqref{Eq:keffEstimateGaussian} and \eqref{KappaRigorous}], so the asymptotic unitarity of the projected position operator has no bearing on the nonreciprocal dynamics encoded by the complex Wannier centers. Rather, $\kappa$ encodes a real dynamical effect that survives in the thermodynamic limit.

\subsection{Example of a complex Wannier center\\ and a nonreciprocal Wannier function}\label{SectionExamples1D}

\begin{figure}
    \centering
    \includegraphics[width=\linewidth]{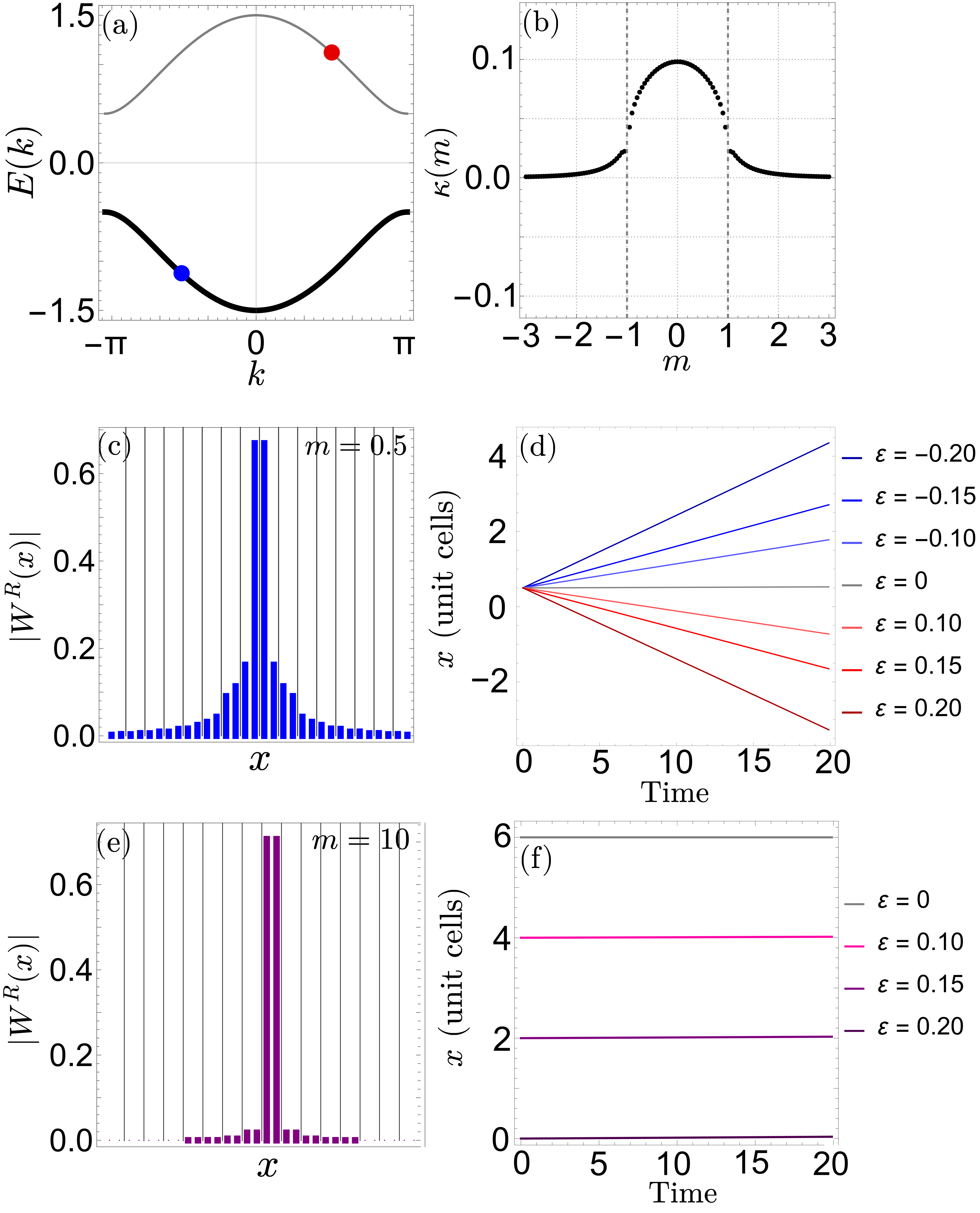}
    \caption{Complex Wannier centers and nonreciprocal Wannier function dynamics of the lowest band of Hamiltonian \eqref{ToyModel1}. (a) Energy spectrum \eqref{DispersionToyModel} highlighting the lower band $E_{k,-}$, from which panels (b-f) are extracted. (b) imaginary part $\kappa(m)$ of the Wannier center as a function of the mass parameter $m$ for fixed non-Hermiticity $\varepsilon=0.2$. (c) Absolute value of the components of a Wannier function in the topological phase $(m=0.5)$. Vertical gray lines delineate unit cells. (d) Center of mass of time-evolved Wannier functions for several $\varepsilon$ at fixed $m=0.5$; the finite $\kappa$ produces nonreciprocal drift. (e) Wannier function in the trivial phase $(m=10)$. (f) Corresponding center-of-mass evolution for several $\varepsilon$, showing the absence of drift in the trivial phase as $m\to \infty$, where $\kappa \to 0$. Initial Wannier functions are taken at different positions to avoid overlap. System size $\ell=400$ and PBC are used in dynamical simulations.}
    \label{Fig1}
\end{figure}

As an example of a system with a nonunitary Wilson loop, complex Wannier centers, and drifting Wannier functions, consider the two-band model
\begin{equation}
    h_\varepsilon(k) = S_{\varepsilon}(k)h_{0}(k)S_{\varepsilon}^{-1}(k), \label{ToyModel1}
\end{equation}
where $h_0(k)=\sin k \ \sigma_x + (m+\cos k)\sigma_z$ is a Hermitian operator and
\begin{equation}
    S_{\varepsilon}(k)=\begin{pmatrix}
        e^{\varepsilon\sin k} & 0 \\
        0 & e^{-\varepsilon \sin k}
    \end{pmatrix}
    \label{SimTrans}
\end{equation}
is a non-unitary similarity transformation that renders the Bloch Hamiltonian \eqref{ToyModel1} non-Hermitian. Its eigenvalues are purely real,
\begin{equation}
    E_{k,\pm} = \pm \sqrt{\sin^2k + (m+\cos k)^2}, \label{DispersionToyModel}
\end{equation}
and there is a line gap for $m\neq 1$, as Fig.~\ref{Fig1}(a) illustrates. We can express the eigenvectors of the Hermitian operator $h_0(k)$ in a smooth gauge as
\begin{align}
    \ket{u^{(0)}_{k,-}} &= e^{\mathrm{i}\theta_k/2}\begin{pmatrix}
        -\sin\frac{\theta_k}{2} \\
        \cos\frac{\theta_k}{2}
    \end{pmatrix}, \\
    \ket{u^{(0)}_{k,+}} &= e^{\mathrm{i}\theta_k/2}\begin{pmatrix}
        \cos\frac{\theta_k}{2} \\
        \sin\frac{\theta_k}{2}
    \end{pmatrix},
\end{align}
where $\theta_k$ is defined through
\begin{equation}
    \tan\theta_k = \frac{\sin k}{m+\cos k}.
\end{equation}
Moreover, we can find the right eigenkets of the non-Hermitian Hamiltonian \eqref{ToyModel1} as $\ket{u_{k,\pm}^{R}}=S_{k}(\varepsilon)\ket{u{(0)}_{k,\pm}}$, yielding
\begin{align}
    \ket{u^{R}_{k,-}} &= e^{\mathrm{i}\theta_k/2}\begin{pmatrix}
        -e^{\varepsilon\sin k}\sin\frac{\theta_k}{2} \\
        e^{-\varepsilon\sin k}\cos\frac{\theta_k}{2}
    \end{pmatrix}, \label{RightMinus} \\
    \ket{u^{R}_{k,+}} &= e^{\mathrm{i}\theta_k/2}\begin{pmatrix}
        e^{\varepsilon\sin k}\cos\frac{\theta_k}{2} \\
        e^{-\varepsilon\sin k}\sin\frac{\theta_k}{2}
    \end{pmatrix}.
\end{align}
Similarly, we can obtain the left eigenbras as $\bra{u^{L}_{k,\pm}}=\bra{u^{(0)}_{k,\pm}}S^{-1}_{k}(\varepsilon)$, giving us
\begin{align}
    \bra{u^{L}_{k,-}} &= e^{-\mathrm{i}\theta_k/2}\begin{pmatrix}
        -e^{-\varepsilon\sin k}\sin\frac{\theta_k}{2}, &
        e^{\varepsilon\sin k}\cos\frac{\theta_k}{2}
    \end{pmatrix}, \label{LeftMinus} \\
    \bra{u^{L}_{k,+}} &= e^{-\mathrm{i}\theta_k/2}\begin{pmatrix}
        e^{-\varepsilon\sin k}\cos\frac{\theta_k}{2}, &
        e^{\varepsilon\sin k}\sin\frac{\theta_k}{2}
    \end{pmatrix}.
\end{align}
Using equations \eqref{RightMinus} and \eqref{LeftMinus}, the Berry connection associated with the lower band is complex,
\begin{align*}
    A_k &= -\ii \bra{u^{L}_{k,-}}\partial_k\ket{u^{R}_{k,-}} \nonumber \\
    &= \frac{1}{2}\frac{d\theta_k}{dk}+
    \mathrm{i}\varepsilon\cos k \cos\theta_k. \label{BerryConnectionToyModel}
\end{align*}
and thus satisfies the necessary condition \eqref{NecessaryCondition} for the nonunitarity of the Wilson loop. 

When integrated over the BZ, the real part of the Berry connection gives rise to the real part of the WC (in units of the unit cell length),
\begin{align}
    \nu &= \frac{1}{4\pi}\int_{0}^{2\pi}\frac{d\theta_k} {dk}dk =\begin{cases}
        0, \quad |m|>1 \\
        \frac{1}{2}, \quad |m|<1.
    \end{cases}
\end{align}
The quantization of the real part of the WC is a consequence of chiral ($\sigma_y$) and pseudo-inversion~\footnote{We define pseudo-Inversion symmetry in Eq. \eqref{PInvCondition}.} ($\sigma_z$) symmetries, and its values distinguish two phases, one trivial $(|m|>1)$ and one topological $(|m|<1)$.

Meanwhile, the imaginary part of the Berry connection yields a nonzero integral over the BZ,
\begin{equation}
    \kappa = \frac{\varepsilon}{2\pi}\int_{0}^{2\pi}\cos k \cos\theta_k dk, \label{ImaginaryToyModel}
\end{equation}
and thus also satisfies the sufficient condition \eqref{Sufficient} for the nonunitarity of the Wilson loop. Fig. \ref{Fig1}(b) shows the imaginary part $\kappa$ of the WC from numerical calculations of the Wilson loop at multiple values of $m$ and at fixed $\varepsilon=0.2$. In particular, one can see that $\kappa$ reaches its maximum $\kappa = \frac{\varepsilon}{2}$ at $m=0$, and that $\kappa$ approaches zero as $m \to \infty$.

By employing \eqref{Eq:keffEstimateGaussian} and \eqref{DriftApproximate} to the band dispersions $E_{k,\pm}$ in \eqref{DispersionToyModel}, we obtain the following expression for the drift velocities of Wannier functions in both bands~\footnote{The fact that the group velocities are the same is model-dependent, and follows from the fact that, while having $\kappa$s with opposite signs, the energy bands $E_{\pm}$ have opposite derivatives at $k=0$.}:
\begin{align}
    v_{\mathrm{drift}}^{\pm}(\kappa) &\sim \frac{1}{\hbar}\frac{\sin (\frac{\kappa}{\sigma^2})}{\sqrt{\sin^2(\frac{\kappa}{\sigma^2})+[m+\cos(\frac{\kappa}{\sigma^2})]^2}} \nonumber \\
    &= \frac{1}{\hbar}\frac{m}{|m+1|\sigma^2}\kappa + \mathcal{O}(\kappa^2). \label{DriftVel}
\end{align}
Since $\kappa$ is proportional to $\varepsilon$ according to \eqref{ImaginaryToyModel}, we conclude that the drift velocity of nonreciprocal Wannier functions should scale linearly with $\varepsilon$.

We verify this dependence through dynamical simulations of Wannier functions. To obtain a representation of our model in real space, we approximate the similarity transformation~\eqref{SimTrans} by taking $e^{\varepsilon\sin k} \approx 1 + \varepsilon \sin k$, which amounts to truncating an exponentially decaying series of long-range hoppings up to next-nearest neighbors (NNN). The resulting expression preserves the same symmetries of the exact Hamiltonian~\eqref{ToyModel1} and accurately describes its physics for sufficiently small $\varepsilon$, the regime explored in our simulations. Moreover, we numerically obtain the Wannier functions by calculating the right eigenvectors of the periodic position operator (see~\cite{PhysRevLett.80.1800} and the definition~\eqref{Resta}) restricted to the occupied subspace (lower or upper band) according to the biorthogonal projection \eqref{Projection}.

Upon such definitions and under the NNN truncation of the Hamiltonian, Fig.~\ref{Fig1}(d) confirms the qualitative behavior of~\eqref{DriftVel}, showing how the centers of mass of Wannier functions at time $t=0$ drift according to the sign and magnitude of $\varepsilon$ as the wave packet evolves. However, as $\kappa$ becomes negligible deep enough in the trivial phase $\nu=0$, the centers of mass become stationary, and the states lose nonreciprocity, as Fig.~\ref{Fig1}(f) shows.

\begin{figure}[h]
    \centering
    \includegraphics[width=0.95\linewidth]{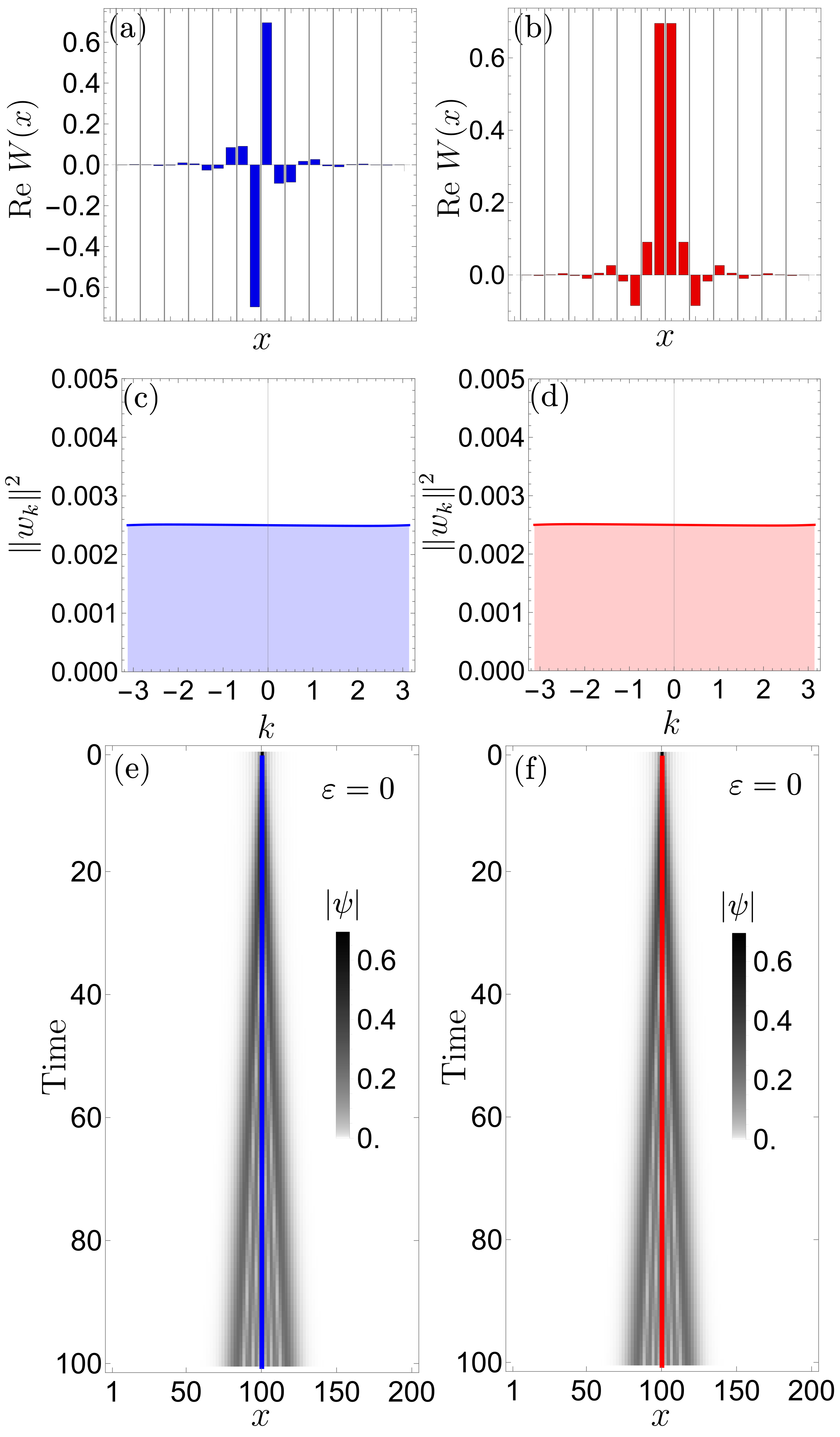}
    \caption{Dynamics of Wannier functions of the model \eqref{ToyModel1} in the Hermitian case $(\varepsilon=0)$. Position-space profiles (real parts) of the Wannier wave packets at $t=0$ for the lower (a) and upper (b) bands. There are two sites per unit cell, whose boundaries are indicated by the vertical gray lines. Momentum-weight distribution $\|w_{k}\|^2$ of the Wannier functions in the lower (c) and upper (d) band. Evolution of the Wannier wave packets for the lower (e) and upper (f) bands. The gray color scale indicates the amplitude $|\psi(x,t)|$ of the wave packet at site $x$, and the blue (e) and red (f) curves track the wave packets' center of mass. We employed $\ell=100$ unit cells and periodic boundary conditions (PBC) in the simulations.}
    \label{HermitianWannier}
\end{figure}

\begin{figure}[h]
    \centering
    \includegraphics[width=0.95\linewidth]{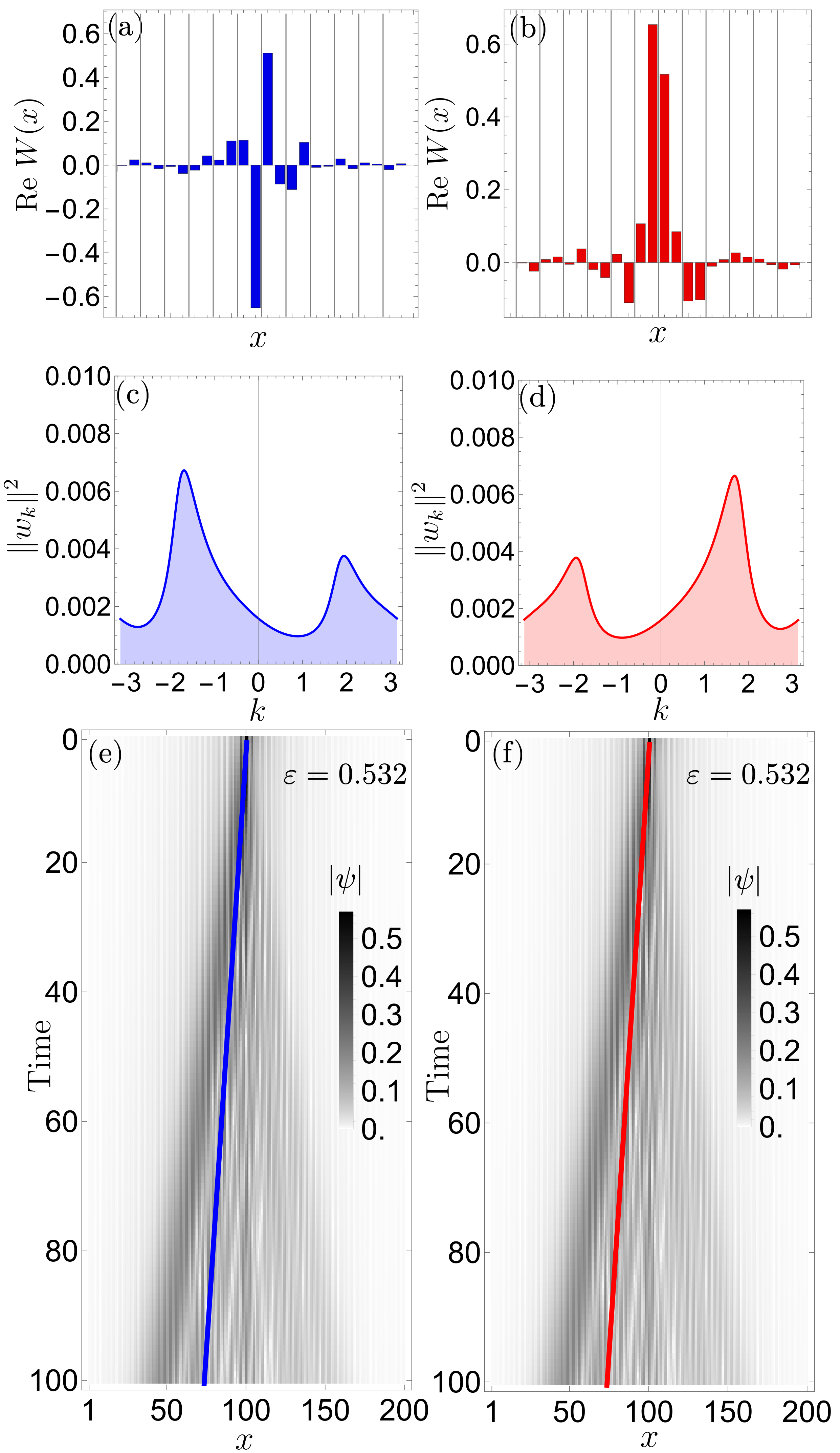}
    \caption{Dynamics of Wannier functions of the model \eqref{ToyModel1} in the non-Hermitian case $(\varepsilon=0.532)$. Position-space profiles (real parts) of the Wannier wave packets at $t=0$ for the lower (a) and upper (b) bands. There are two sites per unit cell, whose boundaries are indicated by the vertical gray lines. Momentum-weight distribution $\|w_{k}\|^2$ of the Wannier functions in the lower (c) and upper (d) band. Evolution of the Wannier wave packets for the lower (e) and upper (f) bands, showing asymmetric spreading and a net drift of the center of mass. The gray color scale indicates the amplitude $|\psi(x,t)|$ of the wave packet at site $x$, and the blue (e) and red (f) curves track the wave packets' center of mass. We employed $\ell=100$ unit cells and periodic boundary conditions (PBC) in the simulations.}
    \label{NHWannierMainText}
\end{figure}

Now let us focus on the topological phase at $m=0.5$ and discuss the wave-packet dynamics in more detail. Figure~\ref{HermitianWannier} serves as our control in the Hermitian limit $\varepsilon=0$, showing the profiles of the purely real Wannier functions of the lower and upper bands at $t=0$ in panels~\ref{HermitianWannier}(a) and~\ref{HermitianWannier}(b), respectively. These profiles closely follow the odd (even) site-symmetry representations of the energy eigenstates at $k=0$ in the lower (upper) bands. Panels~\ref{HermitianWannier}(c) and~\ref{HermitianWannier}(d) show the expected uniform momentum weights $\|w_k\|^2$, while ~\ref{HermitianWannier}(e) and ~\ref{HermitianWannier}(f) show the amplitude per site of the Wannier wave packets and their centers of mass as they evolve from the configurations (a) and (b) at $t=0$. One can see how these initial configurations spread over time while their centers of mass remain stationary, as expected for Wannier functions in Hermitian Hamiltonians.

Upon turning on $\varepsilon \sim 0.5$, we are well inside the non-Hermitian regime. The Wannier profiles in Fig.~\ref{NHWannierMainText}(a) and~\ref{NHWannierMainText}(b) are not purely real (imaginary components not shown). Panels~\ref{NHWannierMainText}(c) and~\ref{NHWannierMainText}(d) confirm that the non-Hermitian Wannier functions have nonuniform and asymmetric momentum-weight distributions $\|w_{k}\|^2$. The mirrored asymmetries of $\|w_{k}\|^2$ for the upper and lower bands reflect the proportionality of the momentum bias $k_{\mathrm{eff}}$ with $\kappa$. Finally, Figs.~\ref{NHWannierMainText}(e) and ~\ref{NHWannierMainText}(f) give us the position-space manifestation of the Wannier functions' nonreciprocal behavior. Both panels show an asymmetric spread of the wave packets (darker shades towards the left), together with a corresponding drift of their centers of mass indicated by the blue and red curves. Notice that, even though they have opposite $k_{\mathrm{eff}}$, the Wannier wave packets drift in the same direction because the slope of $E_{-}(k)$ [$E_{+}(k)]$ is negative for $k_{\mathrm{eff}}<0$ ($k_{\mathrm{eff}}>0$), as one can see in the blue (red) dot in Fig. \ref{Fig1} (a).

\begin{figure}
    \centering
    \includegraphics[width=\linewidth]{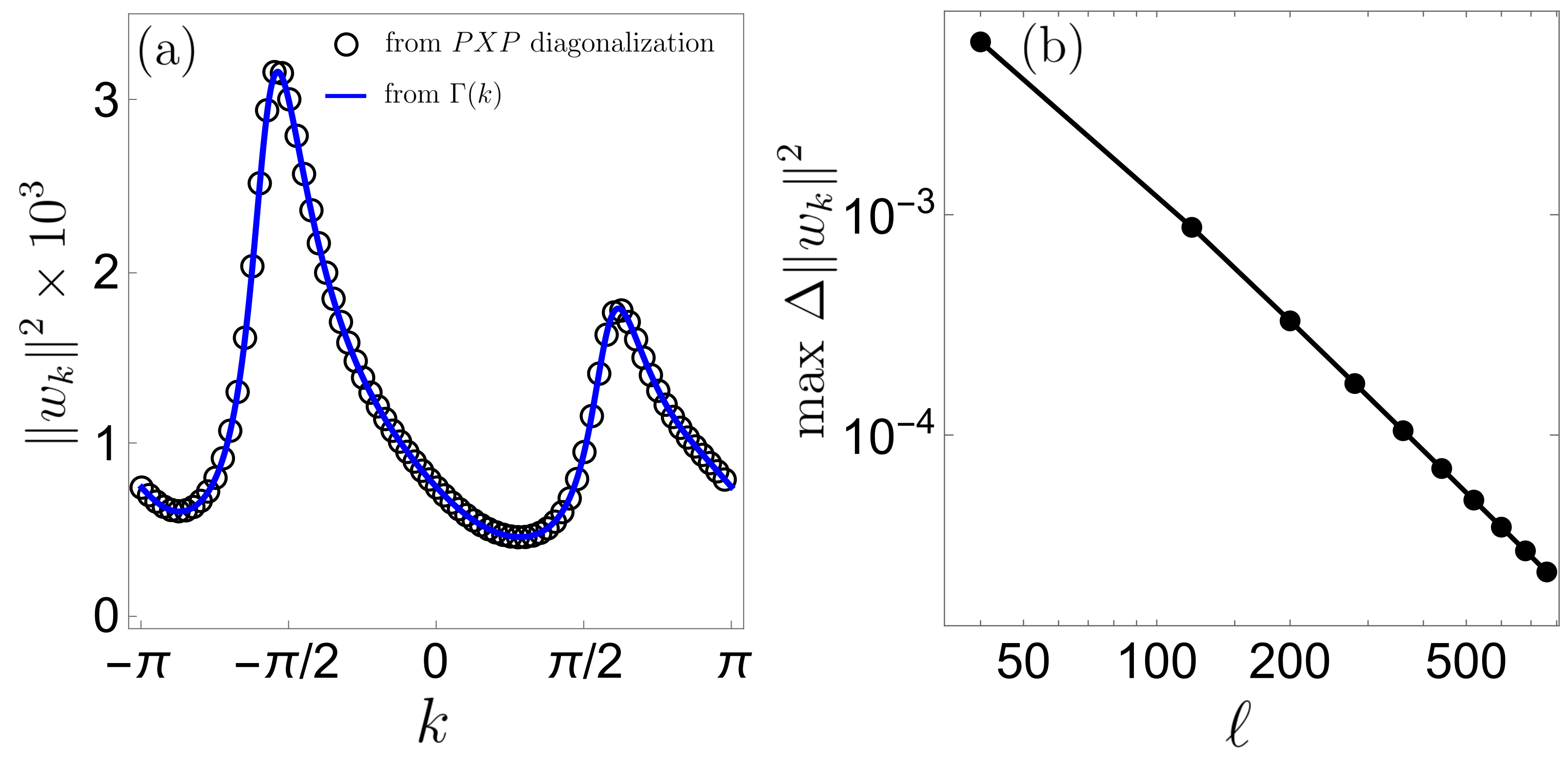}
    \caption{(a) Momentum distribution $\|w_{k}\|^2$ of Wannier functions from direct diagonalization of $PXP$ (circles) and from equation \eqref{Eq:WeightDistributionGamma} (red line) for system size $\ell=800$. (b) Maximum absolute deviation
$\max_k |\|w_k^{PXP}\|^2 - \|w_k^{\Gamma}\|^2|$
as a function of system size $\ell$.}
    \label{Verification}
\end{figure}

To further validate the analytical expression~\eqref{Eq:WeightDistributionGamma}, 
we directly compare the momentum-weight distribution of 
Fig.~\ref{NHWannierMainText}(c) with the prediction obtained from $\Gamma(k)$, 
computed independently from the left and right Bloch eigenvectors of the occupied band 
[see~\eqref{Eq:GammaDefMainText} and its discretized implementation in 
Appendix~\ref{Appendix:Inequivalence}]. 
Figure~\ref{Verification}(a) shows the excellent agreement between the 
direct diagonalization of $PXP$ and the $\Gamma$-dependent expression. 
The remaining finite-size deviations originate from the discretization of the Brillouin zone 
and decay exponentially with $\ell$, as demonstrated in Fig.~\ref{Verification}(b), 
thereby confirming the continuum-limit derivation.

After investigating the unconstrained spectra of the Wilson loop and the projected position operator in generic non-Hermitian Hamiltonians, we now focus on the special class of pseudo-Hermitian Hamiltonians. In this case, pseudo-Hermiticity of the Hamiltonian implies that the Wilson loop is pseudo-unitary (see Sec. III A), which in turn enforces eigenvalue pairing in the Wannier spectrum.

\section{Wannier functions and spectra of pseudo-Hermitian Hamiltonians}

A Hamiltonian $H$ is pseudo-Hermitian (pH) if there exists a Hermitian and invertible operator $\xi:\mathcal{H}\to\mathcal{H}$ (often called \emph{metric}) such that~\cite{Ali}
\begin{equation}
    H^{\dagger} = \xi \ H \ \xi^{-1}. \label{PseudoHermiticity}
\end{equation}
The special case $\xi=\mathbb{I}$ reduces to Hermiticity. Here, we focus on the cases where the pseudo-Hermitian metric acts nontrivially only on the internal (orbital) degrees of freedom,
\begin{equation}
    \xi = \mathbb{I}_L \otimes \eta,
\end{equation}
so that the Bloch Hamiltonian is also pH with a $k$-independent metric,
\begin{equation}
    h^{\dagger}(k) = \eta \  h(k) \ \eta^{-1}. \label{PseudoHermiticitybloch}
\end{equation}

An immediate consequence of \eqref{PseudoHermiticitybloch} is that the right and left eigenvectors of $h_k$ are not independent but related by $\eta$. Indeed, if $\ket{u^{R}_{k,n}}$ is a right eigenvector with eigenvalue $E_{k,n}$, then
\begin{equation}
    h^{\dagger}(k)(\eta\ket{u^{R}_{k,n}}) = \eta h(k)\ket{u^{R}_{k,n}} = E_{k,n}(\eta\ket{u^{R}_{k,n}}). \label{PseudoRelation}
\end{equation}
Eq.~\eqref{PseudoRelation} implies $\eta\ket{u^{R}_{k,n}}$ is a right eigenvector of $h(k)^{\dagger}$ with eigenvalue $E_{k,n}$ and, hence, a left eigenvector of $h(k)$ with energy $E_{k,n}^{*}$. Consequently, the spectrum of a pH Hamiltonian consists of real or complex-conjugate pairs of eigenvalues. 

Since the occupied subspace is defined as the span of the eigenvectors with $\mathrm{Re}[E_{k,n}] \leq E_0$ [c.f. \eqref{RightandLeftsubspace}] and the real part of an eigenvalue is invariant under complex conjugation, it follows that $\eta\ket{u^{R}_{k,n}}$ lies in the occupied subspace if and only if $\ket{u^{R}_{k,n}}$ does. Therefore, the matrix of occupied right eigenvectors $R_k$ [c.f. \eqref{RandL}] and the left set of eigenvectors $\tilde{L}_k=\eta R_k$ both span the same subspace, even though they are not necessarily biorthonormal:
\begin{equation}
    \tilde{L}_{k}^{\dagger}R_{k}= R_{k}^{\dagger}\eta R_{k}=M_k, \label{Def1}
\end{equation}
where $M_k$ is an $N_{\mathrm{occ}} \times N_{\mathrm{occ}}$ invertible Hermitian matrix. Defining a new set of left vectors,
\begin{equation}
    L_{k}=\tilde{L}_kM_{k}^{-1}=\eta R_k M^{-1}_k, \label{Lk}
\end{equation}
leads to biorthonormality,
\begin{equation}
    L_{k}^{\dagger}R_{k} = M_{k}^{-1}M_k = \mathbb{I}_{N_{\mathrm{occ}}}. \label{Def2}
\end{equation}

\subsection{Pseudo-unitary Wilson loop}\label{SubPseudoWL}

In Appendix \ref{AppendixPseudoUnitaryWilsonLoop}, we show that the relations \eqref{Lk} and \eqref{Def2} imply that the Wilson loop is \emph{pseudo-unitary},
\begin{align}
    \mathcal{W}^{-1}_k &=M_{k}^{-1}\mathcal{W}_k^{\dagger}M_k, \label{PseudoWL}
\end{align}
where $M_k$ plays the role of a metric in the occupied subspace.

Now, let $\ket{w^{R,L}_{\lambda}}$ denote the right/left eigenvectors of the Wilson loop with eigenvalue $\lambda$,
\begin{align}
    \mathcal{W}_k\ket{w^{R}_{\lambda}} &= \lambda\ket{w^{R}_{\lambda}}, \nonumber \\
    \mathcal{W}^{\dagger}_k\ket{w^{L}_{\lambda}} &= \lambda^{*}\ket{w^{L}_{\lambda}}.
\end{align}
The pseudo-unitary condition \eqref{PseudoWL} implies that $M_k\ket{w^{R}_{\lambda}}$ is a right eigenstate of $\mathcal{W}_k^{\dagger}$ with eigenvalue $1/\lambda$ and therefore a left eigenvalue of $\mathcal{W}_k$ with eigenvalue $1/\lambda^{*}$ (see Ref. \cite{mostafazadeh2004pseudounitary} and Appendix~\ref{AppendixSpectral}). As a result, the spectrum of pseudo-unitary Wilson loops consists of unimodular eigenvalues $|\lambda|=1$ or inverse-conjugate pairs $(\lambda,1/\lambda^{*})$. If $|\lambda|=1$, we know that the imaginary part $\kappa$ of the WC is zero and the corresponding Wannier function is reciprocal. Now, for a nontrivial pair $\lambda_2=1/\lambda_1^{*}$, we have
\begin{align}
    \kappa_2 = \frac{1}{2\pi}\log|\lambda_2|=\frac{1}{2\pi}\log\Big|\frac{1}{\lambda_1^{*}}\Big|=-\kappa_1.
\end{align}
Hence, for pseudo-unitary Wilson loops, the WCs are either real or come in complex-conjugate pairs $(\nu+\mathrm{i}\kappa,\nu-\mathrm{i}\kappa)$, the latter case generally corresponding to counterpropagating Wannier functions.

Taking determinants on both sides of \eqref{PseudoWL} gives
\begin{equation}
    |\mathrm{det}\mathcal{W}_k|=1.
\end{equation}
On the other hand, since $\det\mathcal{W}_k = \prod_{n=1}^{N_{\mathrm{occ}}}\lambda_n$,
we may write
\begin{equation}
    |\det\mathcal{W}_k|
    = \exp\!\left(2\pi\sum_{n=1}^{N_{\mathrm{occ}}}\kappa_n\right). \label{BoostNeutrality}
\end{equation}
Therefore, although individual Wannier functions may be nonreciprocal $(\kappa_n \neq 0)$, pseudo-unitarity enforces a global reciprocity condition,
\begin{equation}
    \sum_{n=1}^{N_{\mathrm{occ}}}\kappa_n=0.
\end{equation}
In other words, when the Wilson loop is pseudo-unitary---or more generally, when it does not satisfy \eqref{Sufficient}---nonreciprocity appears at the level of individual channels, not as a net flow.

\subsection{Krein signatures of Wannier centers}\label{Sec:KreinStability}

Beyond these spectral constraints, pseudo-unitarity endows the space of Wilson loop eigenvectors with an indefinite inner product determined by $M_k$. For two right eigenvectors $\ket{w^{R}_{\lambda}}$ and $\ket{w^{R}_{\mu}}$, we define the \emph{Krein overlap}
\begin{equation}
    [M_{k}]_{\lambda\mu}=\langle w^{R}_{\lambda}| M_{k}\ket{w^{R}_{\mu}}. \label{KreinOverlap}
\end{equation}
Using $M_{k}\mathcal{W}_{k}^{-1}=W^{\dagger}_{k}M_{k}$, which follows from \eqref{PseudoWL}, we obtain the condition
\begin{equation}
    (\lambda^{*}\mu - 1)[M_k]_{\lambda\mu}=0,
\end{equation}
which means that Wilson loop eigenvectors with nonvanishing Krein overlap necessarily have eigenvalues related by $\mu=1/\lambda^{*}$. In particular, if $\ket{w^{R}_{\lambda}}$ has a nontrivial \emph{Krein norm} (self-overlap), then its eigenvalue $\lambda$ is unitary and its WC is purely real,
\begin{equation}
    [M_k]_{\lambda\lambda} \neq 0 \Rightarrow |\lambda|=1 \ \ \text{and} \ \ \kappa=0. 
\end{equation}
The sign of the Krein norm defines the \emph{Krein signature} $s(\lambda)$ of a Wilson loop eigenvector $\ket{w^{R}_{\lambda}}$,
\begin{equation}
    s(\lambda)=\begin{cases}
        -1, \quad [M_k]_{\lambda\lambda}<0, \\
        +1, \quad [M_k]_{\lambda\lambda}>0. \label{KreinSignature}
    \end{cases}
\end{equation}

The Krein stability theorem (see, for instance, Refs.~\cite{geyer2025stability,chernyavsky2018krein,PhysRevB.111.064312}) states that the Krein signature of a vector is smooth under perturbations and can only change or become ill-defined if the corresponding eigenvalue becomes degenerate with another one of opposite Krein signature. In our setting, this implies that a unimodular Wilson loop eigenvalue $|\lambda_1|=1$ cannot leave the unit circle under deformations that preserve pseudo-unitarity unless it collides with another unimodular eigenvalue $|\lambda_2|=1$ carrying the opposite Krein signature. At such a ``Krein collision'', the Krein norms $[M_k]_{\lambda_1\lambda_1}$ and $[M_k]_{\lambda_2\lambda_2}$ cancel, and the pair of eigenvalues may move away from the unit circle as inverse-conjugate pairs $(\lambda,1/\lambda^{*})$, corresponding to complex-conjugate WCs $(\nu+\mathrm{i}\kappa,\nu-\mathrm{i}\kappa)$.

\subsection{Krein signature as a chiral charge}

The Krein signature has a natural interpretation in terms of the Wannier Hamiltonian $h_{W}(k) = \frac{\mathrm{i}}{2\pi}\log\mathcal{W}_k$, whose eigenvectors are the same as those of the Wilson loop, and its eigenvalues are simply the WCs,
\begin{equation}
    h_W(k)\ket{w^{R}_{\lambda}}=(\nu+\mathrm{i}\kappa)\ket{w^{R}_{\lambda}}.
\end{equation}
Since $\mathcal{W}_k$ is pseudo-unitary, $h_{W}(k)$ is automatically pH with respect to the same metric $M_k$,
\begin{equation}
    h_{W}^{\dagger}(k) = M_k^{-1} h_{W}(k)M_k. \label{PseudoHermitianWannierHamiltonian}
\end{equation}
As a consequence, the Hermitian part of the Wannier Hamiltonian, defined as $h^{(h)}_{W}(k)=\frac{1}{2}[h_W(k)+h^{\dagger}_W(k)]$, commutes with the metric,
\begin{equation}
    [M_k,h^{(h)}_{W}(k)]=0, \label{HermitianPart}
\end{equation}
while the anti-Hermitian part, defined as $h^{(a)}_{W}(k)=\frac{1}{2}[h_W(k)-h^{\dagger}_W(k)]$, \emph{anticommutes} with it,
\begin{equation}
    \{M_k,h_W^{(a)}\}=0. \label{AntiHermitianPart}
\end{equation}
Therefore, the metric $M_k$ plays the role of a chiral operator and maps the eigenstates of the Wannier Hamiltonian with $+\mathrm{i}\kappa$ to those with $-\mathrm{i}\kappa$. We can thus say that Wannier functions associated with complex-conjugate WCs are counter-propagating chiral partners related by $M_k$.

For $\kappa = 0$, the Wilson loop eigenvector $\ket{w^R_\lambda}$ lies in the kernel of the anti-Hermitian part $h_W^{(a)}(k)$. Within this subspace, Eq.~\eqref{HermitianPart} implies that $h_W(k)$ and the projected metric $M_k$ can be simultaneously diagonalized, so that the eigenvector can be chosen to have a definite chirality, i.e., a definite Krein signature defined in Eq.~\eqref{KreinSignature}. Since Krein collisions create or annihilate pairs of real Wannier centers with opposite signature, the net Krein signature in the occupied subspace is conserved as long as the pseudo-Hermitian structure is preserved. Consequently, it protects a minimum number of Wannier centers with $\kappa = 0$.
We now verify these constraints in concrete examples.

\subsection{One-dimensional pseudo-Hermitian examples}{\label{1DModelSection}}

We consider two minimal pseudo-Hermitian models that illustrate the pairing structure and Krein signatures of complex Wannier centers. The first model consists of two coupled Rice-Mele chains~\cite{PhysRevLett.49.1455} with alternating on-site gain and loss,
\begin{align}
    h_{1}(k) &= \tau_0 \otimes [(1+t\cos k)\sigma_x+t\sin k \sigma_y] + \alpha \  \tau_x \otimes \sigma_0 + \nonumber \\ 
    & \ \ +m \ \tau_0 \otimes \sigma_z+ \mathrm{i}\gamma \ \tau_z \otimes \sigma_z, \label{DoubleSSH}
\end{align}
where $\tau_0$ is the identity and $\tau_{x,y,z}$ are Pauli matrices acting in the copy space, while $\sigma_0$ and $\sigma_{x,y,z}$ act in the orbital space of each individual Rice-Mele layer. Here, $t$ controls the nearest-neighbor (NN) hopping dimerization, $m$ is a staggered on-site potential, $\alpha$ couples the two layers, and $\gamma$ sets the strength of the non-Hermitian gain/loss terms. The Hamiltonian \eqref{DoubleSSH} is pH with respect to the metric $\eta = \tau_x \otimes \sigma_0$.

We focus on two regions of parameter space. Both satisfy $\gamma \in (-0.5,0.5)$, $t = m = 0.3$ and exhibit a real bulk energy spectrum with a gap at half-filling, so the resulting Wilson loops and the projected metrics $M_k$ [c.f Eq.~\eqref{Def1}] are well-defined $2\times 2$ matrices. Regime (i) is characterized by strong coupling between the two layers $(\alpha=2)$, while regime (ii) has weak inter-layer coupling $(\alpha=0.2)$.

\begin{figure}
    \centering
    \includegraphics[width=\columnwidth]{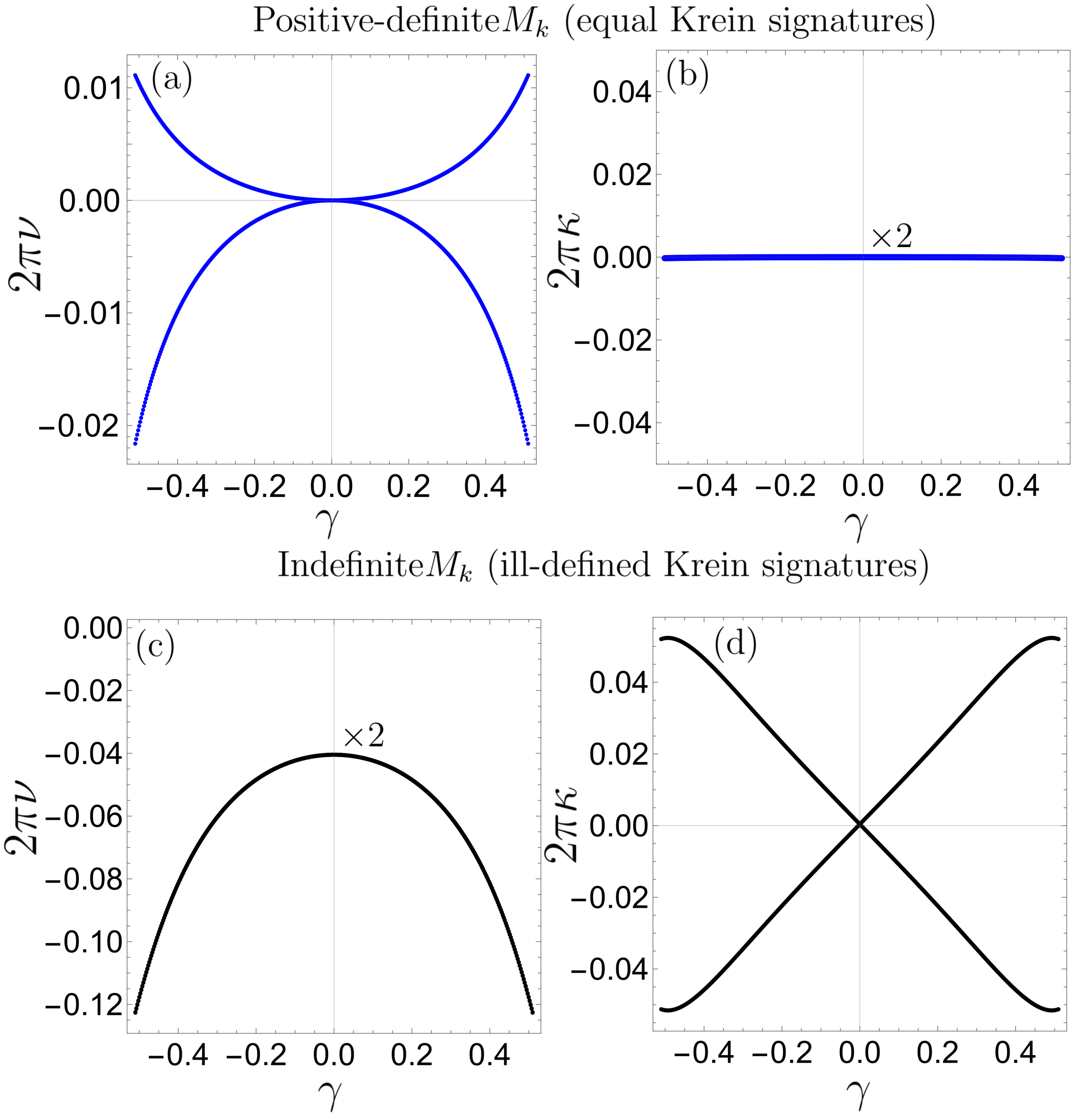}
    \caption{Protection of real WCs in the model \eqref{DoubleSSH} by the Krein signature. (a) Real and (b) imaginary parts of the WCs as a function of the gain/loss parameter $\gamma$ in regime (i). (c,d) are the regime (ii) equivalents.}
    \label{fig:KreinCollision}
\end{figure}

In regime (i), the eigenvalues of $M_k$ have the same sign for all $k$, so that $M_k$ is definite and the Wilson loop right eigenstates $\ket{w^{R}_{1}}$ and $\ket{w^{R}_{2}}$ necessarily carry the same Krein signature. As a result, we have $\kappa_{1}=\kappa_{2}=0$ for all $\gamma$---even when the real parts $\nu_{1}$, $\nu_2$ pass through each other at $\gamma=0$, as shown in Figs.~\ref{fig:KreinCollision}(a,b).

In regime (ii), the eigenvalues of $M_k$ have opposite signs, so $M_k$ is not positive-definite. Hence, the diagonal elements $[M_k]_{11}$ and $[M_k]_{22}$ are unconstrained in sign, allowing the WCs to move into the imaginary axis and forcing $\nu_1=\nu_{2}$, as seen in Figs.~\ref{fig:KreinCollision}(c,d).

A third and qualitatively different possibility occurs when $M_k$ is indefinite yet the WCs remain real and carry opposite, well-defined Krein signatures until they collide and split in the imaginary direction. We observe such Krein collisions in the model
\begin{align}
    h_{2}(k)&=\tau_{z}\otimes [(1+t\cos k)\sigma_x+t\sin k \sigma_y] + \alpha \  \tau_x \otimes \sigma_0 \nonumber \\
    & \ \ + \mathrm{i}\gamma \tau_{z} \otimes \sigma_z. \label{MinusOneModel}
\end{align}
The above expression is similar to \eqref{DoubleSSH} at $m=0$, but with $\tau_z$ on the first term instead of $\tau_0$. This endows the Hamiltonian~\eqref{MinusOneModel} with inversion $(I_s=\tau_x \otimes \sigma_y)$ and pseudo-inversion $(I_c = \tau_0 \otimes \sigma_x)$ symmetries—to be discussed in detail in Sec.~IV—as well as pseudo-Hermiticity, with $\eta=\tau_x \otimes \sigma_z$. As we show later, the interplay of these symmetries allows the nontrivial Krein collisions we report below.

\begin{figure}
    \centering
    \includegraphics[width=\columnwidth]{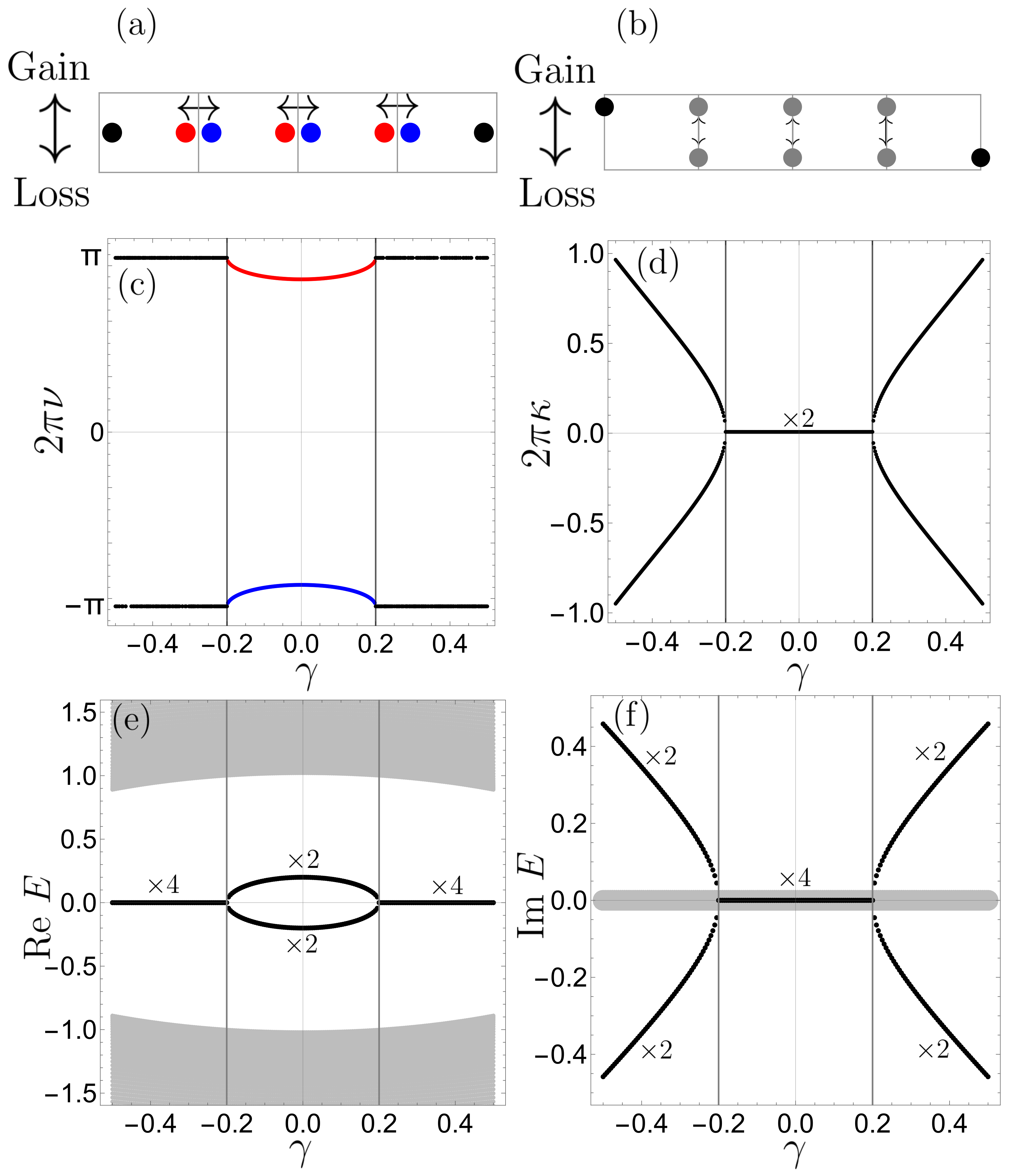}
    \caption{Bulk boundary correspondence in regime (i) of model \eqref{MinusOneModel}. 
    (a,b) Position-space schematic illustrating edge states and Wannier centers for $\gamma^2<\alpha^2$ (a) and $\gamma^2>\alpha^2$ (b).
Black disks denote edge states under open boundary conditions, while colored disks represent Wannier centers of the occupied bands (blue/red indicate negative/positive Krein signature, and gray indicates an ill-defined Krein signature).
The vertical position encodes different quantities for these objects: it represents the imaginary part of the energy, $\mathrm{Im}\,E$, for edge states, and the imaginary part $\kappa$ of the complex Wannier center $z=\nu+i\kappa$ for Wannier centers.
(c,d) Real and imaginary parts of the Wannier centers as functions of $\gamma$.
(e,f) Real and imaginary parts of the bulk (gray) and edge (black) energy spectra.
Vertical gray lines indicate $|\gamma|=\alpha$.}
    \label{fig:BBC}
\end{figure}

\begin{figure}
    \centering
    \includegraphics[width=\columnwidth]{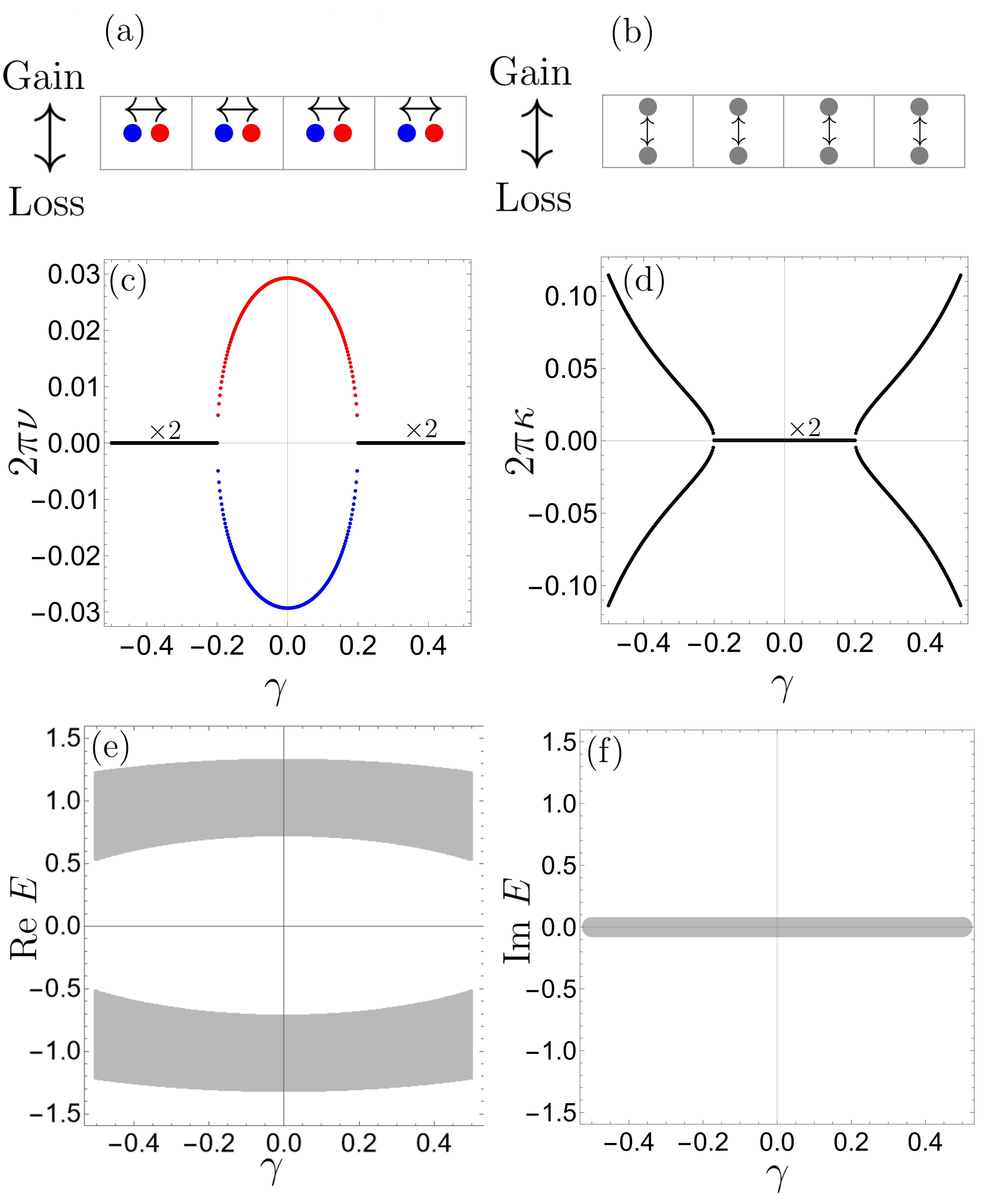}
    \caption{Bulk boundary correspondence in regime (ii) of model \eqref{MinusOneModel}. 
    (a,b) Position-space schematic illustrating edge states and Wannier centers for $\gamma^2<\alpha^2$ (a) and $\gamma^2>\alpha^2$ (b). Colored disks represent Wannier centers of the occupied bands, with blue (red) indicating negative (positive) Krein signature and gray indicating an ill-defined Krein signature. Their vertical displacement corresponds to the imaginary part $\kappa$ of the complex Wannier center $z=\nu+i\kappa$.
    (c) Real and imaginary (d) parts of the Wannier centers as functions of $\gamma$. (e) Real and (f) imaginary parts of the energy spectrum.}
    \label{fig:BBCTrivial}
\end{figure}

We again focus on two gapped regions of parameter space with real bulk spectra, satisfying $-0.5<\gamma<0.5$ and $\alpha=0.2$. Regimes (i) and (ii) are characterized by strong ($t=2$) and weak ($t=0.3$) inter-cell hoppings, respectively.

Regime (i) is connected to the obstructed atomic limit, so the real parts of the Wannier centers of the occupied bands lie near the boundaries between unit cells. Under open boundary conditions (OBC), this configuration leaves one unpaired Wannier center at each edge and produces a filling anomaly in the occupied bands~\cite{benalcazar_quantization_2019}. This anomaly requires the presence of a pair of midgap edge-localized states. The correspondence between Wannier centers and edge modes is illustrated schematically in Figs.~\ref{fig:BBC}(a,b). The behavior of the complex Wannier centers as functions of $\gamma$ is shown in Figs.~\ref{fig:BBC}(c,d), while Figs.~\ref{fig:BBC}(e,f) display the corresponding bulk (gray) and edge (black) energy spectra.

For $\gamma^2<\alpha^2$, the Wannier centers are real and carry opposite Krein signatures (indicated by the blue and red colors), and the edge-localized states likewise have real energies. At $\gamma^2=\alpha^2$, the two Wannier centers collide at $\nu=1/2$. For $\gamma^2>\alpha^2$, the centers split into complex-conjugate pairs and acquire imaginary parts. The edge spectrum follows the same transition: the edge states become degenerate at $\gamma^2=\alpha^2$ and subsequently develop opposite imaginary energies, opening an imaginary gap for $\gamma^2>\alpha^2$.

In region (ii), which is connected to the trivial atomic limit, the real parts of the WCs lie near the center of the unit cell, as illustrated schematically in Figs.~\ref{fig:BBCTrivial}(a,b). Opening the boundaries therefore does not leave unpaired WCs at the edges and does not produce a filling anomaly in the occupied bands. Consequently, no edge-localized states appear. Although the opposite Krein-signature WCs still undergo a Krein collision at $\gamma^2=\alpha^2$ and split into complex-conjugate pairs for $\gamma^2>\alpha^2$ [c.f. Figs.~\ref{fig:BBCTrivial}(c,d)], this transition has no counterpart in the energy spectrum [c.f. Figs.~\ref{fig:BBCTrivial}(e,f)] because no edge modes exist.

In summary, using only bulk information—Wilson loops and their complex Wannier centers—one can determine both the existence and the dynamical character of edge states. The real parts of the Wannier centers determine whether a filling anomaly occurs and thus whether edge states are present, while their imaginary parts predict whether those edge states exhibit gain or loss.

Pseudo-Hermiticity, together with crystalline symmetries such as inversion, can stabilize Wannier configurations that have no analogue in Hermitian Hamiltonians. More generally, this enrichment reflects the broader principle of \emph{symmetry ramification} in non-Hermitian systems, whereby a symmetry that is unique in the Hermitian limit splits into distinct symmetry branches that independently constrain the energy spectrum and the Wilson loop. In the next section, we analyze this mechanism for crystalline (pseudo-)inversion and show how its ramified symmetry branches enforce characteristic patterns of Wannier-center flows.

\section{Symmetry ramification and Wilson loops}\label{SecSymmetries}

In Hermitian Hamiltonians, the statements
\begin{align}
    h(k) &= S h(\pm k) S^{-1}, \quad 
    h(k)^{\dagger} =Sh(\pm k)S^{-1},
\end{align}
are equivalent as symmetry constraints, since $h(k)^{\dagger}=h(k)$. Both express that the Bloch Hamiltonian is symmetric under an invertible (linear or antilinear) operator $S$, which may either preserve ($+k$) or reverse $(-k)$ momentum.  In non-Hermitian Hamiltonians, however, these relations are no longer identical. The similarity condition
\begin{equation}
    h(k) = S h(\pm k) S^{-1} \label{Similarity}
\end{equation}
acts within the right (and separately within the left) eigenspaces and may connect states at different momenta, but it does not relate right and left eigenvectors. In contrast, conjugation-type symmetry
\begin{equation}
     h^{\dagger}(k) =Sh(\pm k)S^{-1}, \label{Conjugate}
\end{equation}
of which pseudo-Hermiticity is the special $S^{\dagger}=S$ and $+k$ case, does relate right and left eigenvectors. Consequently, the conditions \eqref{Similarity} and \eqref{Conjugate} impose distinct constraints on the biorthogonal Berry connection~ \eqref{BiorthogonalBerry} and Wilson loop~\eqref{BiorthW}. This branching---where a single Hermitian symmetry becomes two inequivalent non-Hermitian symmetry classes---is known as \emph{symmetry ramification}~\cite{kawabata_symmetry_2019}. Let us now take the concrete example of a crystalline symmetry and see how its ramification affects the Wannier spectrum.

\subsection{Inversion symmetry as a similarity constraint}

The most familiar definition of inversion symmetry (IS) is
\begin{equation}
    h(k)=I_{s}h(-k)I_{s}^{-1}, \label{InvCondition}
\end{equation}
where $I_s^{\dagger}=I_{s}$ is the spatial inversion operator, which also obeys $I_{s}^2=\mathbb{I}_N$ and has eigenvalues $\pm 1$. If $\ket{u^{R}_{k,n}}$ is a right energy eigenstate with eigenvalue $E_{k,n}$, we have
\begin{equation}
    h(-k)I_s\ket{u^{R}_{k,n}} = I_sh(k)\ket{\psi^{R}_{k,n}} = E_{k,n}I_s\ket{u^{R}_{k,n}},
\end{equation}
so $I_s\ket{u^{R}_{k,n}}$ is also a right eigenstate at $-k$ with the same eigenvalue $E_{k,n}$. Similarly, $I_s\ket{u^{L}_{k,n}}$ is a left eigenstate of $h(-k)$. In Appendix~\ref{Appendix:Inversion}, we define a map between the $k$ and $-k$ eigenspaces through the sewing matrix
\begin{equation}
    B^{s}_{k} = L^{\dagger}_{-k}I_sR_k, \label{SewingMatrix}
\end{equation}
which is unitary due to the biorthonormality of $L_k$ and $R_k$, and show that the Wilson loop obeys
\begin{equation}
    \mathcal{W}_{k} = (B^{s}_k)^{-1}\mathcal{W}_{-k}^{-1}B^{s}_k, \label{pInvolutory}
\end{equation}
and actually becomes \emph{pseudo-involutory} at high-symmetry points (HSPs) $k_{I}=\{0,\pi\}$,
\begin{equation}
    \mathcal{W}_{k_I} = (B^{s}_{k_I})^{-1}\mathcal{W}_{k_I}^{-1}B^{s}_{k_I}.
\end{equation}
Since the sewing matrix is invertible, $\mathcal{W}_{k_I}$ and $\mathcal{W}^{-1}_{{k_I}}$ must share the same set of eigenvalues. Thus, the Wannier spectrum is symmetric under
\begin{equation}
    \lambda \leftrightarrow 1/\lambda \Leftrightarrow e^{-2\pi\mathrm{i}z}\leftrightarrow e^{2\pi\mathrm{i}z},
\end{equation}
where $z=\nu+\mathrm{i}\kappa$ are the complex Wannier centers. Such symmetry can manifest in two ways: (i) isolated real Wannier centers with $\kappa=0$ and $\nu=\{0,\frac{1}{2}\}$, or (ii) inversion-related pairs $(\nu,\kappa)$ and $(-\nu,-\kappa)$. Case (i) occurs when the Wilson loop at $k_I$ commutes with the projected inversion operator $B^{s}_k$, so that its eigenvectors can be chosen as inversion eigenstates. In contrast, case (ii) arises when $B^{s}_{k_I}$ and $\mathcal{W}_{k_I}$ do not commute; in this situation inversion sends each Wilson loop eigenstate to a distinct partner, enforcing the paired structure $z \leftrightarrow -z$. Which of these two scenarios occurs is determined by the symmetry-protected phase encoded in the pattern of inversion irreps carried by the occupied energy eigenstates at the two HSPs. For $N_{\mathrm{occ}}=1,2$, all possible irrep patterns and the corresponding Wannier spectra are listed in Table~\ref{tab:table1}. For a derivation of the results, see Appendix~\ref{Appendix:Inversion}.

\begin{table}[t]
\caption{\label{tab:table1}%
Constraints on the complex  Wannier centers, $z=\nu+\ii \kappa$, due to the  eigenvalues of inversion symmetry $I_s$ at HSPs $\{0,\pi\}$. 
}
\begin{ruledtabular}
\begin{tabular}{cccc}
$N_\mathrm{occ}$ &\textrm{$I_s$ eigenvalues at $\{0,\pi\}$}&
$\nu$&
$\kappa$\\
\colrule
1 &$\{+,+\}$ &$0$ & $0$\\
 &$\{+,-\}$ &$\frac{1}{2}$ & $0$\\
\colrule
2 &$\{(++),(++)\}$ &$[0,0]$ & $[0,0]$\\
&$\{(++),(+-)\}$ & $[0,\frac{1}{2}]$ & $[0,0]$ \\
&$\{(++),(--)\}$ & $[\frac{1}{2},\frac{1}{2}]$ & $[0,0]$ \\
&$\{(+-),(+-)\}$ &  $\nu_1=-\nu_2$ & $\kappa_1=-\kappa_2$
\end{tabular}
\end{ruledtabular}
\end{table}

\subsection{Pseudo-inversion as a conjugate-type symmetry}

We define pseudo-inversion symmetry (pIS) as
\begin{equation}
    h(k)^{\dagger}=I_{c}h(-k)I_{c}^{-1}, \label{PInvCondition}
\end{equation}
where $I_c^{\dagger}=I_{c}$ and $I_{c}^2=\mathbb{I}_N$. If $\ket{u^{R}_{k,n}}$ is a right eigenvector of $h(k)$, then
\begin{equation}
    h(-k)^{\dagger}I_c\ket{u^{R}_{k,n}} = I_c h(k)\ket{u^{R}_{k,n}}=E_{k,n}I_c
    \ket{u^{R}_{k,n}},
\end{equation}
so that $I_c\ket{u^{R}_{k,n}}$ is a right eigenstate of $h(-k)^{\dagger}$ with eigenvalue $E_{k,n}$, and therefore a left eigestate of $h(-k)$ with eigenvalue $E^{*}_{k,n}$. Therefore, $I_cR_k$ lies in the subspace spanned by $L_{-k}$. Similarly, $I_c\ket{u^{L}_{k,n}}$ is a right eigenstate of $h(-k)$, so $I_cL_{k}$ lies in the subspace spanned by $R_{-k}$. From these facts, we can define the sewing matrix and its inverse as
\begin{equation}
    B^{c}_{k} = R^{\dagger}_{-k}I_cR_{k}, \quad (B^{c}_k)^{-1} = L^{\dagger}_{-k}I_cL_k.
\end{equation}
In Appendix \ref{Appendix:Inversion}, we show that the Wilson loop satisfies
\begin{equation}
    \mathcal{W}_{k} = (B^{c}_{k})^{-1}\mathcal{W}_{-k}^{\dagger}B^{c}_{k},
\end{equation}
and therefore it is pseudo-Hermitian at HSPs $k_I$:
\begin{equation}
    \mathcal{W}_{k} = (B^{c}_{k_I})^{-1}\mathcal{W}_{k_I}^{\dagger}B^{c}_{k_I} \label{pHermitian}.
\end{equation}
As a consequence, its spectrum must be symmetric under $\lambda \leftrightarrow \lambda^{*}$. In terms of Wannier centers, we may have (i) isolated WCs with $\nu=0$ and $\kappa \in \mathbb{R}$, which have well-defined representations under pseudo-inversion, or (ii) inverse-related pairs $(\nu,\kappa)$ and $(-\nu,\kappa)$, whose corresponding Wannier functions map onto one another under pseudo-inversion. Once again, whether we have case (i) or (ii) depends on the pattern or pseudo-inversion irreps carried by the occupied energy eigenstates at the two HSPs. The possibilities for $N_{\mathrm{occ}}=2$ are listed in Table~\ref{tab:table2} and derived in Appendix \ref{Appendix:Inversion}.

\begin{table}[t]
\caption{\label{tab:table2}%
Constraints on the complex Wannier centers, $z=\nu+\ii \kappa$, due to the eigenvalues of pseudo-inversion symmetry $I_c$ at HSPs $\{0,\pi\}$.
}
\begin{ruledtabular}
\begin{tabular}{cccc}
$N_{\mathrm{occ}}$ & \textrm{$I_c$ eigenvalues at $\{0,\pi\}$}&
$\nu$&
$\kappa$\\
\colrule
1 & $\{+,+\}$ & 0 & $\kappa \in \mathbb{R}$ \\
 & $\{+,-\}$ & $\frac{1}{2}$ & $\kappa \in \mathbb{R}$ \\
\colrule
2 & $\{(++),(++)\}$ &$[0,0]$ & $\kappa_1,\kappa_2 \in\mathbb{R}$\\
& $\{(++),(+-)\}$ & $[0, \frac{1}{2}]$ & $\kappa_1,\kappa_2 \in\mathbb{R}$ \\
& $\{(++),(--)\}$ & $[ \frac{1}{2}, \frac{1}{2}]$ & $\kappa_1,\kappa_2 \in\mathbb{R}$ \\
& $\{(+-),(+-)\}$ & $\nu_1=-\nu_2$ & $\kappa_1=\kappa_2$
\end{tabular}
\end{ruledtabular}
\end{table}

\subsection{Combined effects of inversion and pseudo-inversion}
\label{sec:combined-syms}

After discussing how inversion (IS) and pseudo-inversion (pIS) separately constrain the biorthogonal Wilson loop and the Wannier spectrum, we now analyze
their combined action. In non-Hermitian Hamiltonians, IS and pIS arise from the
ramification of Hermitian inversion into two distinct symmetry types: a
similarity constraint acting within the right and left eigenspaces, and a
conjugation-type constraint relating right to left eigenvectors. Although these two involutory operators need not coincide, one may consider models in which both symmetries are present simultaneously and represented by distinct operators. 

A natural question, then, is whether IS and pIS together impose additional
constraints on the Wilson loop beyond those enforced by each symmetry
individually. To address this, we first derive the algebraic relation that
follows whenever both symmetries hold. Inversion symmetry imposes the
similarity relation \eqref{Similarity}
while pseudo-inversion imposes the conjugation condition \eqref{Conjugate}. Combining the two yields
\begin{align}
    h(k)^{\dagger}
        &= I_c\, h(-k)\, I_c^{-1} \nonumber\\
        &= I_c\!\left(I_s h(k) I_s^{-1}\right)\! I_c^{-1} \nonumber\\
        &= (I_c I_s)\, h(k)\, (I_cI_s)^{-1}. \label{AlgebraicRelation}
\end{align}

\subsubsection{Commuting symmetries}

If the IS ans pIS symmetry operators commute, the object
\begin{equation}
    \eta \equiv I_c I_s \label{etaFromI}
\end{equation}
is a Hermitian and involutory operator,
\begin{align*}
    \eta^{\dagger}&=(I_sI_{c})^{\dagger}=(I_{c})^{\dagger}I_{s}^{\dagger}=I_{c}I_{s}=I_{s}I_{c}=\eta, \nonumber \\
    \eta^{2}&=I_sI_cI_sI_c = I_sI_{c}^{2}I_s = \mathbb{I}_N,
\end{align*}
which, according to \eqref{AlgebraicRelation}, implements pseudo-Hermiticity:
\begin{equation}
    h(k)^{\dagger} = \eta\, h(k)\, \eta^{-1}.
    \label{eq:gen-pH-from-IS-pIS}
\end{equation}
We therefore obtain a set $\{I_s,I_c,\eta\}$ of mutually commuting operators with eigenvalues $\pm 1$. At the HSPs $\{0,\pi\}$, they also commute with the Bloch Hamiltonian $h(k_I)$ so there is an energy eigenbasis $\{\ket{u}^{R}_{k,n}, n=1,\dots,N_{\mathrm{occ}}\}$ such that
\begin{align*}
    \eta\ket{u^{R}_{k_{I},n}} &= \sigma_{k_I}\ket{u^{R}_{k_{I},n}}, \nonumber \\
    I_s\ket{u^{R}_{k_{I},n}} &= \xi_{k_I}\ket{u^{R}_{k_{I},n}}, \\
    I_c\ket{u^{R}_{k_{I},n}} &= \chi_{k_I}\ket{u^{R}_{k_{I},n}}, \nonumber 
\end{align*}
where $\sigma_{k_I},\xi_{k_I},\chi_{k_I} \in \{-1,1\}$. In other words, the energy eigenstates at HSPs transform according to one-dimensional irreducible representations (irreps) of the three symmetry operators. Moreover, acting with \eqref{etaFromI} on $\ket{u^{R}_{k_{I}}}$ yields a constraint between the eigenvalues of $\eta$, $I_s$, and $I_c$,
\begin{equation}
    \sigma_{k_{I},n}=\xi_{k_{I}}\chi_{k_{I}}. \label{EigenvalueConstraint}
\end{equation}
Thus, specifying the irreps of all three symmetries at every HSP is redundant: from now on, we adopt $\xi_{k_{I}}$ and $\chi_{k_I}$ as our independent indices.

We can define the topological invariant
\begin{align}
    [\xi]=\#\xi^{(+)}_{0} - \#\xi^{(+)}_{\pi},
\end{align}
where $\#\xi^{(+)}_{k_I}$ is the number of $+1$ eigenvalues of $I_s$ in the occupied subspace at the HSP $k_{I}$. Accordingly, we may define the invariant $[\chi]$ for $I_c$, and classify topological phases by $([\xi],[\chi])$.

For $N_{\mathrm{occ}}=1$, the Krein signature associated with the $k$-independent pseudo-Hermitian metric cannot change continuously across the Brillouin zone. Therefore $\sigma_0=\sigma_\pi$, and by \eqref{EigenvalueConstraint}, we have $\xi_0\xi_\pi=\chi_0\chi_{\pi}$. That leaves us with only two topological phases, as shown in the $N_{\mathrm{occ}}=1$ rows of Tab. \ref{tab:ThreeSymetries}. For two occupied bands, a careful comparison of Tabs. \ref{tab:table1} and \ref{tab:table2} yields six combinations of irreps that are compatible with simultaneous IS and pIS, as laid out in the $N_{\mathrm{occ}}=2$ rows of Tab. \ref{tab:ThreeSymetries}.

\begin{table*}[t]
\caption{\label{tab:ThreeSymetries}
Constraints on complex Wannier centers, $z=\nu + \ii \kappa$, by the eigenvalues of commuting IS, pIS, and pH at HSPs for \(N_{\mathrm{occ}}=1,2\).}
\begin{ruledtabular}
\begin{tabular}{cccccc}
$N_{\mathrm{occ}}$ & $I_s$ irreps at $\{0,\pi\}$ & $I_c$ irreps at $\{0,\pi\}$ & ($[\xi],[\chi]$) & $\nu$ & $\kappa$ \\
\colrule
1 & \(\{+,+\}\) & \(\{+,+\}\) & $(0,0)$ & $0$ & $0$ \\
  & \(\{+,-\}\) & \(\{+,-\}\) & $(1,1)$ & $\frac{1}{2}$ & $0$ \\
\colrule
2 & \(\{(++),(++)\}\) & \(\{(++),(++)\}\) & $(0,0)$ & $[0,0]$ & $[0,0]$ \\
  & \(\{(++),(+-)\}\) & \(\{(++),(+-)\}\) & $(1,1)$ & $[0,\tfrac{1}{2}]$ & $[0,0]$ \\
  & \(\{(++),(--)\}\) & \(\{(++),(--)\}\) & $(2,2)$ & $[\tfrac{1}{2},\tfrac{1}{2}]$ & $[0,0]$ \\
  & \(\{(+-),(+-)\}\) & \(\{(+-),(+-)\}\) & $(0,0)$ & $\nu_1=-\nu_2$ & $[0,0]$ \\
  & \(\{(+-),(+-)\}\) & \(\{(++),(++)\}\) & $(0,1)$ & $[0,0]$ & $\kappa_1=-\kappa_2\neq 0$ \\
  & \(\{(+-),(+-)\}\) & \(\{(++),(--)\}\) & $(0,2)$ & $[\tfrac{1}{2},\tfrac{1}{2}]$ & $\kappa_1=-\kappa_2\neq 0$ \\
\end{tabular}
\end{ruledtabular}
\end{table*}

Lastly, we note that if a Bloch Hamiltonian is pH with metric $\eta$ and possesses IS with an inversion operator $I_s$ such that $[\eta,I_s]=0$, then it automatically satisfies pIS with the operator $I_c = \eta I_s$. Therefore, pseudo-Hermiticity and IS+pIS---when the two corresponding operators commute and can therefore be simultaneously diagonalized---are not independent symmetries. The presence of any two of them implies the third.

\subsubsection{Anticommuting symmetries}

If the IS and pIS symmetry operators anticommute, then the operator $\eta$ defined in \eqref{etaFromI} is anti-Hermitian and squares to minus one,
\begin{align*}
    \eta^{\dagger}&=(I_sI_{c})^{\dagger}=(I_{c})^{\dagger}I_{s}^{\dagger}=I_{c}I_{s}=-I_{s}I_{c}=-\eta, \nonumber \\
    \eta^{2}&=I_sI_cI_sI_c = -I_sI_{c}^{2}I_s = -\mathbb{I}_N.
\end{align*}
We therefore introduce
\begin{equation}
    \tilde{\eta} \equiv \mathrm{i}\eta = \mathrm{i}I_sI_c,
\end{equation}
which \emph{anticommutes} with $I_s$ and $I_c$, satisfies $\tilde{\eta}^{\dagger}=\tilde{\eta}$ and $\tilde{\eta}^2=\mathbb{I}_{N}$, and implements pseudo-Hermiticity,
\begin{equation}
    h(k)^{\dagger} = \tilde{\eta}\, h(k)\, \tilde{\eta}^{-1}.
\end{equation}
The three symmetry generators are thus Hermitian involutions that mutually anticommute,
$A_i^\dagger=A_i$, $A_i^2=\mathbb I_N$, and $\{A_i,A_j\}=0$ for $i\neq j$,
with $A_i\in\{I_s,I_c,\tilde{\eta}\}$. They furnish a representation of the complex Clifford algebra
\begin{equation}
    \{A_i,A_j\}=2\delta_{ij}\mathbb I_N .
    \label{Clifford}
\end{equation}
At inversion-invariant momenta $k_I\in\{0,\pi\}$ the occupied subspace is invariant under the symmetry
actions, so the induced representation on the occupied subspace is well-defined. In a biorthonormal
occupied basis $\{|u^R_{k_I,n}\rangle,\langle u^L_{k_I,n}|\}_{n=1}^{N_{\rm occ}}$ we define
\begin{equation}
    [A_i^{\rm occ}(k_I)]_{mn}\equiv
    \langle u^L_{k_I,m}|\,A_i\,|u^R_{k_I,n}\rangle .
\end{equation}
These $N_{\rm occ}\times N_{\rm occ}$ matrices satisfy the same Clifford relations
$\{A_i^{\rm occ}(k_I),A_j^{\rm occ}(k_I)\}=2\delta_{ij}\mathbb I_{N_{\rm occ}}$.

In contrast to the commuting case, the Clifford algebra admits no one-dimensional irreducible representations. The smallest representation therefore occurs at $N_{\rm occ}=2$, where the occupied subspace furnishes the unique two-dimensional irreducible representation. Up to a unitary change of basis one may choose
\begin{equation}
A_i^{\rm occ}(k_I)=\sigma_i, \qquad i=x,y,z.
\end{equation}
In particular, the three symmetry generators cannot be simultaneously diagonalized, and the occupied subspace cannot be decomposed into independent symmetry sectors.

At high-symmetry momenta $k_I$, the Wilson loop $\mathcal{W}_{k_I}$ satisfies
\begin{align}
\mathcal{W}_{k_I} &= (B^c_{k_I})^{-1} \mathcal{W}_{k_I}^\dagger B^c_{k_I}, \nonumber \\
\mathcal{W}_{k_I} &= (B^s_{k_I})^{-1} \mathcal{W}_{k_I}^{-1} B^s_{k_I}, \nonumber \\
\mathcal{W}_{k_I}^{-1} &= M_{k_I}^{-1} \mathcal{W}_{k_I}^\dagger M_{k_I},
\end{align}
as per Eqs.~\eqref{PseudoWL}, \eqref{pInvolutory}, and \eqref{pHermitian}, with sewing matrices acting within the same two-dimensional irreducible space. These relations imply that the Wilson-loop spectrum is invariant as a set under Hermitian conjugation and inversion,
\begin{equation}
\mathrm{Spec}(\mathcal{W}_{k_I})
=
\mathrm{Spec}(\mathcal{W}_{k_I}^\dagger)
=
\mathrm{Spec}(\mathcal{W}_{k_I}^{-1}).
\label{SpectrumConstraints}
\end{equation}

For $N_{\mathrm{occ}}=2$, let the Wilson-loop eigenvalues be $\lambda_1=r_1e^{\mathrm{i}\theta_1}$ and $\lambda_2=r_2e^{\mathrm{i}\theta_2}$, where $r_{1,2}>0$ and $\theta_{1,2}\in(-\pi,\pi]$. The first equality in \eqref{SpectrumConstraints} implies that the spectrum is invariant under complex conjugation, $\{\lambda_1,\lambda_2\}=\{\lambda_1^*,\lambda_2^*\}$. Since the spectrum contains only two elements, this can occur in two ways. Either each eigenvalue is individually invariant under complex conjugation, $\lambda_1=\lambda_1^*$ and $\lambda_2=\lambda_2^*$, which implies $\theta_{1,2}\in\{0,\pi\}$ and therefore real eigenvalues; or the two eigenvalues are exchanged, $\lambda_1=\lambda_2^*$, which implies $r_1=r_2$ and $\theta_2=-\theta_1$.

We now impose the second equality in \eqref{SpectrumConstraints}, $\{\lambda_1,\lambda_2\}=\{1/\lambda_1,1/\lambda_2\}$. In the first case, where the eigenvalues are real, this requires them to form a reciprocal pair, $\lambda_2=1/\lambda_1$. Writing $\lambda_1=\pm r$ with $r>0$, we obtain $\lambda_2=\pm 1/r$ with the same sign, implying $r_2=1/r_1$. In the second case, where $\lambda_1=\lambda_2^*$ and $r_1=r_2$, the reciprocal condition forces $r_1=1/r_2$, which implies $r_1=r_2=1$. The eigenvalues are therefore unimodular and form a complex-conjugate pair, $\lambda_1=e^{\mathrm{i}\theta}$ and $\lambda_2=e^{-\mathrm{i}\theta}$.

In terms of the complex Wannier centers
\begin{equation}
z=\frac{i}{2\pi}\log\lambda \equiv \nu + i\kappa ,
\end{equation}
the two possibilities become
\begin{align}
&\mathrm{(i)}:\quad \nu_1=\nu_2\in\{0,1/2\}, \qquad \kappa_1=-\kappa_2\in\mathbb{R},\\
&\mathrm{(ii)}:\quad \nu_1=-\nu_2, \qquad \kappa_1=\kappa_2=0 .
\end{align}

Thus, the Wannier centers are either pinned to $\nu=0$ or $\nu=1/2$ while moving along the imaginary axis with opposite $\kappa$, or confined to the real axis while moving along it in opposite directions. Transitions between these behaviors can therefore only occur through Krein collisions at $\nu=0$ or $\nu=1/2$. This is precisely what we observe in Figs.~\ref{fig:BBC} and \ref{fig:BBCTrivial} on the model \eqref{MinusOneModel}, which has anti-commuting IS, pIS, and pH with $I_s=\tau_x \otimes \sigma_y$, $I_c = \tau_0 \otimes \sigma_x$, and $\tilde{\eta}=\tau_x \otimes \sigma_z$.

\begin{figure}
    \centering
    \includegraphics[width=\columnwidth]{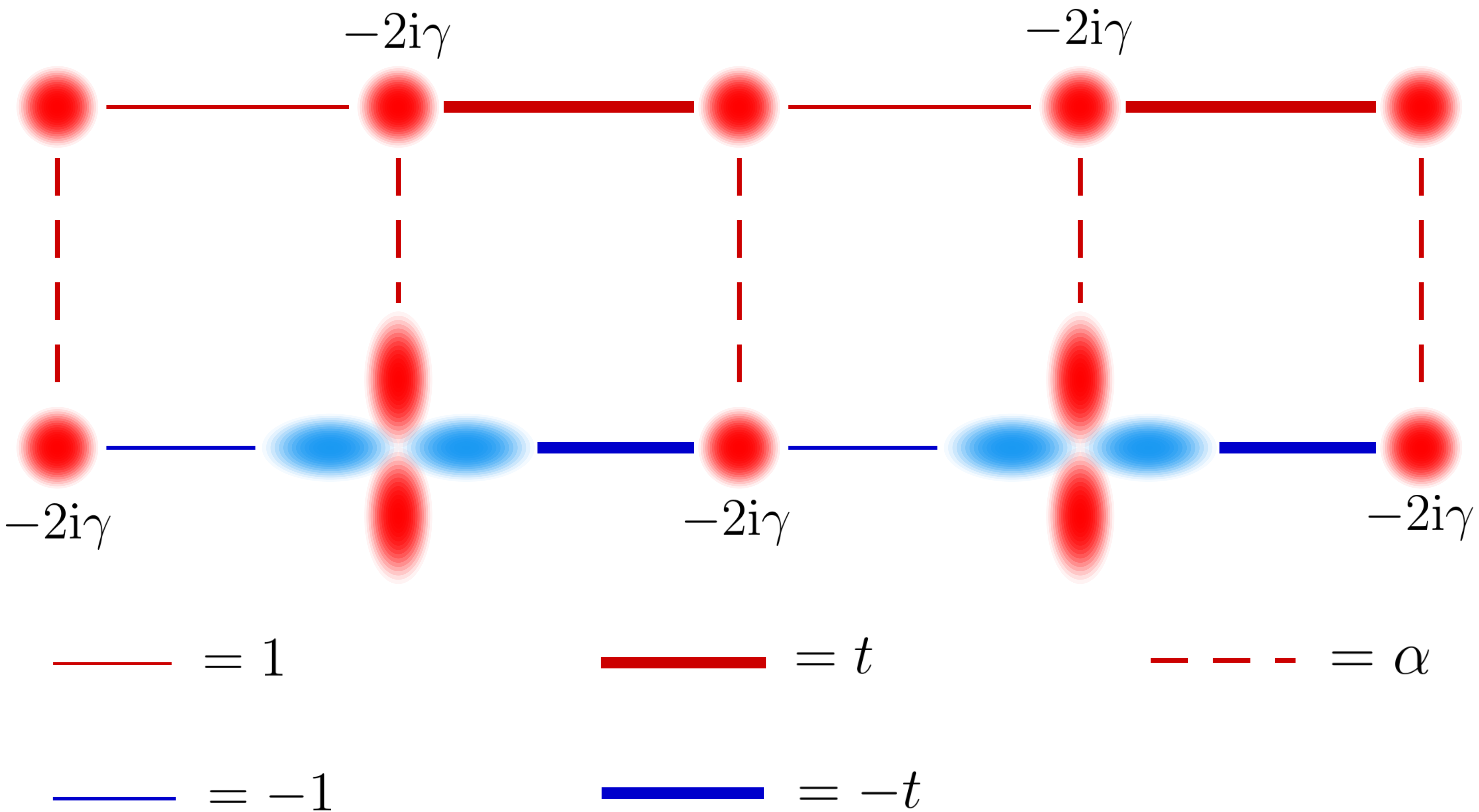}
    \caption{Schematic photonic implementation of model \eqref{MinusOneModel}. The upper leg consists of $s$-orbital resonators with alternating couplings $1$ and $t$, with on-site loss $-2\mathrm{i}\gamma$ at even sites. The lower leg alternates between $s$ and $d$ orbitals, whose symmetry produces sign-changing overlaps that generate couplings $-1$ and $-t$. On-site loss $-2\mathrm{i}\gamma$ at every $s$ orbital in the lower leg. Dashed bonds denote inter-leg couplings of strength $\alpha$.}
    \label{Schematic}
\end{figure}

Figure \eqref{Schematic} schematically shows a ladder-shaped photonic waveguide array that can 
implement model \eqref{MinusOneModel} and test its bulk-boundary correspondence. The upper leg consists of resonators supporting $s$-orbital modes with alternating couplings $1$ and $t$. The lower leg alternates between resonators whose relevant modes have $s$- and $d$-orbital symmetry. While the $s$-orbital field is isotropic and positive, the $d$-orbital lobes are positive (negative) along the vertical (horizontal) direction. As a result, the overlap of neighboring $s$ and $d$ modes produces a positive inter-leg coupling $\alpha$ and negative couplings $-1$ and $-t$ along the lower leg. Non-Hermiticity enters through on-site loss in the even (odd) resonators of the upper (lower) leg, giving a contribution $\mathrm{diag}(0,-2\mathrm{i}\gamma,-2\mathrm{i}\gamma,0)=-\mathrm{i}\gamma\mathbb{I}+i\gamma\tau_{z}\otimes\sigma_z$ to the Hamiltonian. The resulting Bloch Hamiltonian is that of \eqref{MinusOneModel}
\begin{equation}
    h_{\mathrm{array}}(k) = h_{2}(k)-\mathrm{i}\gamma\mathbb{I},
\end{equation}
up to a rigid shift of the spectrum. All the ingredients for this implementation (orbital-mode engineering, tunable couplings, and site-selective losses) have been demonstrated in multimode photonic platforms~\cite{NohSchulzBenalcazarJorg+2025+4273+4283}. The lattice geometry of Fig.~\ref{Schematic} closely resembles that used in the photonic Mobius topological insulator~\cite{jiang2023photonic}.

\section{Discussion}

We have shown necessary and sufficient conditions for biorthogonal Wilson loops to break unitarity, yielding complex Wannier centers in non-Hermitian bands. More importantly, we demonstrated that the imaginary parts of such complex Wannier centers are associated with an imbalance in the momentum-weight distribution of Wannier functions and often act as their effective average momenta. These momentum shifts lead to observable drifts of Wannier functions in real space, thus establishing a direct dynamical manifestation of complex Wannier centers and providing a practical route for probing non-Hermitian Wannier configurations.

In pseudo-Hermitian Hamiltonians, such as those in models~\eqref{ToyModel1} and~\eqref{DoubleSSH}, preparing localized Wannier functions and adiabatically tuning a parameter would reveal whether a pair of Wannier centers carries equal or opposite Krein signatures: only the latter undergo Krein collisions and acquire imaginary parts, producing a measurable drift, whereas the former remain stationary. The relationship between non-Hermitian Wannier Hamiltonians with complex spectra and the edge Hamiltonians of these systems has remained largely unexplored. Here we show that, in a system with anticommuting pseudo-Hermiticity and (pseudo) inversion symmetries, the Krein structure of the Wilson loop establishes a bulk–boundary correspondence between complex Wannier centers and the spectrum of edge states. In addition, we present a possible experimental realization of that system in a well-known platform.

Wannier functions also govern the emergence and symmetry character of solitons bifurcating from linear bands in weakly nonlinear lattices~\cite{PhysRevB.111.064312}. This suggests an avenue to engineer topologically nonreciprocal solitons. Existing proposals rely on gain--loss mechanisms that generically introduce amplification or acceleration~\cite{6hy3-jmk9,PhysRevB.104.L020303,PhysRevB.105.125421,YUCE2021127484,PhysRevB.110.L180302,PhysRevB.109.094308,veenstra2024non}. In contrast, solitons that inherit the profile of non-Hermitian Wannier functions would propagate unidirectionally with constant amplitude and velocity, offering a more robust, stable, and spectrally cleaner mechanism for realizing nonreciprocal wave packets.

Finally, by clarifying the structure and transitions of complex Wannier centers, this work lays the foundation for a systematic program of non-Hermitian Wannier topology. Our results indicate that symmetry constraints—--particularly pseudo-Hermiticity and inversion-type symmetries—--organize non-Hermitian bands into distinct Wannier symmetry classes with well-defined invariants, suggesting a non-Hermitian analogue of the Wannier-based classification familiar from Hermitian band theory~\cite{kawabata_symmetry_2019}.

\bibliography{NH}

@article{Benalcazar,
  title = {Electric multipole moments, topological multipole moment pumping, and chiral hinge states in crystalline insulators},
  author = {Benalcazar, Wladimir A. and Bernevig, B. Andrei and Hughes, Taylor L.},
  journal = {Phys. Rev. B},
  volume = {96},
  issue = {24},
  pages = {245115},
  numpages = {59},
  year = {2017},
  month = {Dec},
  publisher = {American Physical Society},
  doi = {10.1103/PhysRevB.96.245115},
  url = {https://link.aps.org/doi/10.1103/PhysRevB.96.245115}
}

@misc{tanaka2024exceptionalsecondordertopologicalinsulators,
      title={Exceptional Second-Order Topological Insulators}, 
      author={Yutaro Tanaka and Daichi Nakamura and Ryo Okugawa and Kohei Kawabata},
      year={2024},
      eprint={2411.06898},
      archivePrefix={arXiv},
      primaryClass={cond-mat.mes-hall},
      url={https://arxiv.org/abs/2411.06898}, 
}

@article{Heiss_2012, title={The physics of exceptional points}, volume={45}, url={https://arxiv.org/pdf/1210.7536}, DOI={10.1088/1751-8113/45/44/444016}, number={44}, journal={Journal of Physics A}, publisher={IOP Publishing}, author={Heiss, W. D.}, year={2012}, month=oct, pages={444016} }

@article{Kawabata_2017_Exceptional,
  title = {Classification of Exceptional Points and Non-Hermitian Topological Semimetals},
  author = {Kawabata, Kohei and Bessho, Takumi and Sato, Masatoshi},
  journal = {Phys. Rev. Lett.},
  volume = {123},
  issue = {6},
  pages = {066405},
  numpages = {7},
  year = {2019},
  month = {Aug},
  publisher = {American Physical Society},
  doi = {10.1103/PhysRevLett.123.066405},
  url = {https://link.aps.org/doi/10.1103/PhysRevLett.123.066405}
}

@article{jiang2023photonic,
  title={Photonic M{\"o}bius topological insulator from projective symmetry in multiorbital waveguides},
  author={Jiang, Chuang and Song, Yiling and Li, Xiaohong and Lu, Peixiang and Ke, Shaolin},
  journal={Optics Letters},
  volume={48},
  number={9},
  pages={2337--2340},
  year={2023},
  publisher={Optica Publishing Group}
}

@article{NohSchulzBenalcazarJorg+2025+4273+4283,
url = {https://doi.org/10.1515/nanoph-2025-0492},
title = {Orbital frontiers: harnessing higher modes in photonic simulators},
title = {},
author = {Jiho Noh and Julian Schulz and Wladimir Benalcazar and Christina Jörg},
pages = {4273--4283},
volume = {14},
number = {24},
journal = {Nanophotonics},
doi = {doi:10.1515/nanoph-2025-0492},
year = {2025},
lastchecked = {2026-03-15}
}

@article{Bergholtz_2021,
  title = {Exceptional topology of non-Hermitian systems},
  author = {Bergholtz, Emil J. and Budich, Jan Carl and Kunst, Flore K.},
  journal = {Rev. Mod. Phys.},
  volume = {93},
  issue = {1},
  pages = {015005},
  numpages = {31},
  year = {2021},
  month = {Feb},
  publisher = {American Physical Society},
  doi = {10.1103/RevModPhys.93.015005},
  url = {https://link.aps.org/doi/10.1103/RevModPhys.93.015005}
}

@article{Ding_Fang_Ma_2022, title={Non-Hermitian topology and exceptional-point geometries}, volume={4}, url={http://arxiv.org/pdf/2204.11601}, DOI={10.1038/s42254-022-00516-5}, number={12}, journal={Nature Reviews Physics}, author={Ding, Kun and Fang, Chen and Ma, Guancong}, year={2022}, month=apr, pages={745–760} }

@article{PhysRevLett.49.1455,
  title = {Elementary Excitations of a Linearly Conjugated Diatomic Polymer},
  author = {Rice, M. J. and Mele, E. J.},
  journal = {Phys. Rev. Lett.},
  volume = {49},
  issue = {19},
  pages = {1455--1459},
  numpages = {0},
  year = {1982},
  month = {Nov},
  publisher = {American Physical Society},
  doi = {10.1103/PhysRevLett.49.1455},
  url = {https://link.aps.org/doi/10.1103/PhysRevLett.49.1455}
}

@article{masuda2022relationship,
  title={Relationship between the electronic polarization and the winding number in non-hermitian systems},
  author={Masuda, Shohei and Nakamura, Masaaki},
  journal={Journal of the Physical Society of Japan},
  volume={91},
  number={4},
  pages={043701},
  year={2022},
  publisher={The Physical Society of Japan}
}

@article{PhysRevLett.123.073601,
  title = {Higher-Order Topological Corner States Induced by Gain and Loss},
  author = {Luo, Xi-Wang and Zhang, Chuanwei},
  journal = {Phys. Rev. Lett.},
  volume = {123},
  issue = {7},
  pages = {073601},
  numpages = {8},
  year = {2019},
  month = {Aug},
  publisher = {American Physical Society},
  doi = {10.1103/PhysRevLett.123.073601},
  url = {https://link.aps.org/doi/10.1103/PhysRevLett.123.073601}
}

@article{mostafazadeh2004pseudounitary,
  title={Pseudounitary operators and pseudounitary quantum dynamics},
  author={Mostafazadeh, Ali},
  journal={Journal of mathematical physics},
  volume={45},
  number={3},
  pages={932--946},
  year={2004},
  publisher={American Institute of Physics}
}

@article{Hu,
  title = {Electric polarization and its quantization in one-dimensional non-Hermitian chains},
  author = {Hu, Jinbing and Perroni, Carmine Antonio and De Filippis, Giulio and Zhuang, Songlin and Marrucci, Lorenzo and Cardano, Filippo},
  journal = {Phys. Rev. B},
  volume = {107},
  issue = {12},
  pages = {L121101},
  numpages = {6},
  year = {2023},
  month = {Mar},
  publisher = {American Physical Society},
  doi = {10.1103/PhysRevB.107.L121101},
  url = {https://link.aps.org/doi/10.1103/PhysRevB.107.L121101}
}

@article{Chen,
  title = {Comparative study of Hermitian and non-Hermitian topological dielectric photonic crystals},
  author = {Chen, Menglin L. N. and Jiang, Li Jun and Zhang, Shuang and Zhao, Ran and Lan, Zhihao and Sha, Wei E. I.},
  journal = {Phys. Rev. A},
  volume = {104},
  issue = {3},
  pages = {033501},
  numpages = {6},
  year = {2021},
  month = {Sep},
  publisher = {American Physical Society},
  doi = {10.1103/PhysRevA.104.033501},
  url = {https://link.aps.org/doi/10.1103/PhysRevA.104.033501}
}

@article{Kunst,
  title = {Biorthogonal Bulk-Boundary Correspondence in Non-Hermitian Systems},
  author = {Kunst, Flore K. and Edvardsson, Elisabet and Budich, Jan Carl and Bergholtz, Emil J.},
  journal = {Phys. Rev. Lett.},
  volume = {121},
  issue = {2},
  pages = {026808},
  numpages = {6},
  year = {2018},
  month = {Jul},
  publisher = {American Physical Society},
  doi = {10.1103/PhysRevLett.121.026808},
  url = {https://link.aps.org/doi/10.1103/PhysRevLett.121.026808}
}

@article{Edvardsson,
  title = {Phase transitions and generalized biorthogonal polarization in non-Hermitian systems},
  author = {Edvardsson, Elisabet and Kunst, Flore K. and Yoshida, Tsuneya and Bergholtz, Emil J.},
  journal = {Phys. Rev. Res.},
  volume = {2},
  issue = {4},
  pages = {043046},
  numpages = {10},
  year = {2020},
  month = {Oct},
  publisher = {American Physical Society},
  doi = {10.1103/PhysRevResearch.2.043046},
  url = {https://link.aps.org/doi/10.1103/PhysRevResearch.2.043046}
}

@article{OrtegaTaberner,
  title = {Polarization and entanglement spectrum in non-Hermitian systems},
  author = {Ortega-Taberner, Carlos and R\o{}dland, Lukas and Hermanns, Maria},
  journal = {Phys. Rev. B},
  volume = {105},
  issue = {7},
  pages = {075103},
  numpages = {12},
  year = {2022},
  month = {Feb},
  publisher = {American Physical Society},
  doi = {10.1103/PhysRevB.105.075103},
  url = {https://link.aps.org/doi/10.1103/PhysRevB.105.075103}
}

@article{Fidkowski,
  title = {Model Characterization of Gapless Edge Modes of Topological Insulators Using Intermediate Brillouin-Zone Functions},
  author = {Fidkowski, Lukasz and Jackson, T. S. and Klich, Israel},
  journal = {Phys. Rev. Lett.},
  volume = {107},
  issue = {3},
  pages = {036601},
  numpages = {4},
  year = {2011},
  month = {Jul},
  publisher = {American Physical Society},
  doi = {10.1103/PhysRevLett.107.036601},
  url = {https://link.aps.org/doi/10.1103/PhysRevLett.107.036601}
}

@article{mostafazadeh2002pseudo,
  title={Pseudo-Hermiticity versus PT symmetry: The necessary condition for the reality of the spectrum of a non-Hermitian Hamiltonian},
  author={Mostafazadeh, Ali},
  journal={Journal of Mathematical Physics},
  volume={43},
  number={1},
  pages={205--214},
  year={2002},
  publisher={American Institute of Physics}
}

@article{mostafazadeh2001pseudo,
  title={Pseudo-Hermiticity versus PT-Symmetry II: A complete characterizatio n of non-Hermitian Hamiltonians with a real spectrum},
  author={Mostafazadeh, Ali},
  journal={arXiv preprint math-ph/0110016},
  year={2001}
}

@article{mostafazadeh2002pseudo3,
  title={Pseudo-Hermiticity versus PT-symmetry III: Equivalence of pseudo-Hermiticity and the presence of antilinear symmetries},
  author={Mostafazadeh, Ali},
  journal={Journal of Mathematical Physics},
  volume={43},
  number={8},
  pages={3944--3951},
  year={2002},
  publisher={American Institute of Physics}
}

@article{Ali,
   title={PSEUDO-HERMITIAN REPRESENTATION OF QUANTUM MECHANICS},
   volume={07},
   ISSN={1793-6977},
   url={http://dx.doi.org/10.1142/S0219887810004816},
   DOI={10.1142/s0219887810004816},
   number={07},
   journal={International Journal of Geometric Methods in Modern Physics},
   publisher={World Scientific Pub Co Pte Lt},
   author={Mostafazadeh, Ali},
   year={2010},
   month=nov, pages={1191–1306} }

@misc{edvardsson2023biorthogonalrenormalization,
      title={Biorthogonal Renormalization}, 
      author={Elisabet Edvardsson and J Lukas K König and Marcus Stålhammar},
      year={2023},
      eprint={2212.06004},
      archivePrefix={arXiv},
      primaryClass={quant-ph},
      url={https://arxiv.org/abs/2212.06004}, 
}

@article{Curtright_2007,
   title={Biorthogonal quantum systems},
   volume={48},
   ISSN={1089-7658},
   url={http://dx.doi.org/10.1063/1.2196243},
   DOI={10.1063/1.2196243},
   number={9},
   journal={Journal of Mathematical Physics},
   publisher={AIP Publishing},
   author={Curtright, Thomas and Mezincescu, Luca},
   year={2007},
   month=sep }

@article{Pati_2009,
   title={Entanglement in non-Hermitian quantum theory},
   volume={73},
   ISSN={0973-7111},
   url={http://dx.doi.org/10.1007/s12043-009-0101-0},
   DOI={10.1007/s12043-009-0101-0},
   number={3},
   journal={Pramana},
   publisher={Springer Science and Business Media LLC},
   author={Pati, Arun K.},
   year={2009},
   month=sep, pages={485–498} }

@article{Brody_2013,
   title={Biorthogonal quantum mechanics},
   volume={47},
   ISSN={1751-8121},
   url={http://dx.doi.org/10.1088/1751-8113/47/3/035305},
   DOI={10.1088/1751-8113/47/3/035305},
   number={3},
   journal={Journal of Physics A: Mathematical and Theoretical},
   publisher={IOP Publishing},
   author={Brody, Dorje C},
   year={2013},
   month=dec, pages={035305} }

@article{Ashida_2020,
   title={Non-Hermitian physics},
   volume={69},
   ISSN={1460-6976},
   url={http://dx.doi.org/10.1080/00018732.2021.1876991},
   DOI={10.1080/00018732.2021.1876991},
   number={3},
   journal={Advances in Physics},
   publisher={Informa UK Limited},
   author={Ashida, Yuto and Gong, Zongping and Ueda, Masahito},
   year={2020},
   month=jul, pages={249–435} }

@article{Hwang,
  title = {Fragile topology protected by inversion symmetry: Diagnosis, bulk-boundary correspondence, and Wilson loop},
  author = {Hwang, Yoonseok and Ahn, Junyeong and Yang, Bohm-Jung},
  journal = {Phys. Rev. B},
  volume = {100},
  issue = {20},
  pages = {205126},
  numpages = {38},
  year = {2019},
  month = {Nov},
  publisher = {American Physical Society},
  doi = {10.1103/PhysRevB.100.205126},
  url = {https://link.aps.org/doi/10.1103/PhysRevB.100.205126}
}

@Book{vanderbilt_2018,
  author    = {Vanderbilt, David},
  publisher = {Cambridge University Press},
  title     = {Berry Phases in Electronic Structure Theory: Electric Polarization, Orbital Magnetization and Topological Insulators},
  year      = {2018},
  doi       = {10.1017/9781316662205},
  place     = {Cambridge},
}

@misc{wang2024higherordertopologicalknotsnonreciprocal,
      title={Higher-order Topological Knots and Nonreciprocal Dynamics in non-Hermitian lattices}, 
      author={Yifan Wang and Wladimir A. Benalcazar},
      year={2024},
      eprint={2412.05809},
      archivePrefix={arXiv},
      primaryClass={cond-mat.mes-hall},
      url={https://arxiv.org/abs/2412.05809}, 
}

@article{Alexandradinata,
  title = {Wilson-loop characterization of inversion-symmetric topological insulators},
  author = {Alexandradinata, A. and Dai, Xi and Bernevig, B. Andrei},
  journal = {Phys. Rev. B},
  volume = {89},
  issue = {15},
  pages = {155114},
  numpages = {18},
  year = {2014},
  month = {Apr},
  publisher = {American Physical Society},
  doi = {10.1103/PhysRevB.89.155114},
  url = {https://link.aps.org/doi/10.1103/PhysRevB.89.155114}
}

@article{Jiang_2024,
  title = {Tunable non-Hermitian skin effect via gain and loss},
  author = {Jiang, Wen-Cheng and Wu, Hong and Li, Qing-Xu and Li, Jian and Zhu, Jia-Ji},
  journal = {Phys. Rev. B},
  volume = {110},
  issue = {15},
  pages = {155144},
  numpages = {9},
  year = {2024},
  month = {Oct},
  publisher = {American Physical Society},
  doi = {10.1103/PhysRevB.110.155144},
  url = {https://link.aps.org/doi/10.1103/PhysRevB.110.155144}
}

@article{Yu,
  title = {Equivalent expression of $z_{2}$ topological invariant for band insulators using the non-Abelian Berry connection},
  author = {Yu, Rui and Qi, Xiao Liang and Bernevig, Andrei and Fang, Zhong and Dai, Xi},
  journal = {Phys. Rev. B},
  volume = {84},
  issue = {7},
  pages = {075119},
  numpages = {12},
  year = {2011},
  month = {Aug},
  publisher = {American Physical Society},
  doi = {10.1103/PhysRevB.84.075119},
  url = {https://link.aps.org/doi/10.1103/PhysRevB.84.075119}
}

@article{GARRISON1988177,
title = {Complex geometrical phases for dissipative systems},
journal = {Physics Letters A},
volume = {128},
number = {3},
pages = {177-181},
year = {1988},
issn = {0375-9601},
doi = {https://doi.org/10.1016/0375-9601(88)90905-X},
url = {https://www.sciencedirect.com/science/article/pii/037596018890905X},
author = {J.C. Garrison and E.M. Wright}
}

@article{PhysRevA.98.053833,
  title = {Complex Berry phase dynamics in $\mathcal{PT}$-symmetric coupled waveguides},
  author = {Hayward, Rosie and Biancalana, Fabio},
  journal = {Phys. Rev. A},
  volume = {98},
  issue = {5},
  pages = {053833},
  numpages = {8},
  year = {2018},
  month = {Nov},
  publisher = {American Physical Society},
  doi = {10.1103/PhysRevA.98.053833},
  url = {https://link.aps.org/doi/10.1103/PhysRevA.98.053833}
}

@article{Zhang_Yang_Fang_2022, title={Universal non-hermitian skin effect in two and higher dimensions}, volume={13}, DOI={10.1038/s41467-022-30161-6}, number={1}, journal={Nature Communications}, author={Zhang, Kai and Yang, Zhesen and Fang, Chen}, year={2022}, month={May}}

@misc{wang2024classifyingordertwospatialsymmetries,
      title={Classifying Order-Two Spatial Symmetries in Non-Hermitian Hamiltonians: Point-gapped AZ and AZ$^\dag$ Classes}, 
      author={Yifan Wang},
      year={2024},
      eprint={2411.03410},
      archivePrefix={arXiv},
      primaryClass={cond-mat.mes-hall},
      url={https://arxiv.org/abs/2411.03410}, 
}

@article{PhysRevB.100.195135,
  title = {Wilson loop approach to fragile topology of split elementary band representations and topological crystalline insulators with time-reversal symmetry},
  author = {Bouhon, Adrien and Black-Schaffer, Annica M. and Slager, Robert-Jan},
  journal = {Phys. Rev. B},
  volume = {100},
  issue = {19},
  pages = {195135},
  numpages = {25},
  year = {2019},
  month = {Nov},
  publisher = {American Physical Society},
  doi = {10.1103/PhysRevB.100.195135},
  url = {https://link.aps.org/doi/10.1103/PhysRevB.100.195135}
}

@article{PhysRevLett.124.056802,
  title = {Non-Hermitian Boundary Modes and Topology},
  author = {Borgnia, Dan S. and Kruchkov, Alex Jura and Slager, Robert-Jan},
  journal = {Phys. Rev. Lett.},
  volume = {124},
  issue = {5},
  pages = {056802},
  numpages = {6},
  year = {2020},
  month = {Feb},
  publisher = {American Physical Society},
  doi = {10.1103/PhysRevLett.124.056802},
  url = {https://link.aps.org/doi/10.1103/PhysRevLett.124.056802}
}

@article{PhysRevLett.80.5243,
  title = {Real Spectra in Non-Hermitian Hamiltonians Having $PT$ Symmetry},
  author = {Bender, Carl M. and Boettcher, Stefan},
  journal = {Phys. Rev. Lett.},
  volume = {80},
  issue = {24},
  pages = {5243--5246},
  numpages = {0},
  year = {1998},
  month = {Jun},
  publisher = {American Physical Society},
  doi = {10.1103/PhysRevLett.80.5243},
  url = {https://link.aps.org/doi/10.1103/PhysRevLett.80.5243}
}

@article{Bender_2015,
doi = {10.1088/1742-6596/631/1/012002},
url = {https://doi.org/10.1088/1742-6596/631/1/012002},
year = {2015},
month = {jul},
publisher = {IOP Publishing},
volume = {631},
number = {1},
pages = {012002},
author = {Bender, Carl M},
title = {PT-symmetric quantum theory},
journal = {Journal of Physics: Conference Series}
}

@article{Xiong_2018, title={Why does bulk boundary correspondence fail in some non-hermitian topological models}, journal = {J. Phys. Commun.},volume={2}, url={https://robots.iopscience.iop.org/article/10.1088/2399-6528/aab64a/pdf}, DOI={10.1088/2399-6528/AAB64A}, number={3}, publisher={IOP Publishing}, author={Xiong, Ye}, year={2018}, month=mar, pages={035043} }

@article{Song_2019,
  title = {Non-Hermitian Skin Effect and Chiral Damping in Open Quantum Systems},
  author = {Song, Fei and Yao, Shunyu and Wang, Zhong},
  journal = {Phys. Rev. Lett.},
  volume = {123},
  issue = {17},
  pages = {170401},
  numpages = {8},
  year = {2019},
  month = {Oct},
  publisher = {American Physical Society},
  doi = {10.1103/PhysRevLett.123.170401},
  url = {https://link.aps.org/doi/10.1103/PhysRevLett.123.170401}
}

@article{altland1997nonstandard,
  title={Nonstandard symmetry classes in mesoscopic normal-superconducting hybrid structures},
  author={Altland, Alexander and Zirnbauer, Martin R},
  journal={Physical Review B},
  volume={55},
  number={2},
  pages={1142},
  year={1997},
  publisher={APS}
}

@article{benalcazar_quantization_2019,
	title = {Quantization of fractional corner charge in $C_n$-symmetric higher-order topological crystalline insulators},
	volume = {99},
	url = {https://link.aps.org/doi/10.1103/PhysRevB.99.245151},
	doi = {10.1103/PhysRevB.99.245151},
	pages = {245151},
	number = {24},
	journal = {Phys. Rev. B},
	shortjournal = {Phys. Rev. B},
	author = {Benalcazar, Wladimir A. and Li, Tianhe and Hughes, Taylor L.},
	urldate = {2024-01-11},
	date = {2019-06-26},
    year = {2019},
}

@article{kawabata_symmetry_2019,
	title = {Symmetry and Topology in Non-Hermitian Physics},
	volume = {9},
	url = {https://link.aps.org/doi/10.1103/PhysRevX.9.041015},
	doi = {10.1103/PhysRevX.9.041015},
	pages = {041015},
	number = {4},
	journal = {Phys. Rev. X},
	shortjournal = {Phys. Rev. X},
	author = {Kawabata, Kohei and Shiozaki, Ken and Ueda, Masahito and Sato, Masatoshi},
	urldate = {2024-01-07},
	date = {2019-10-21},
}

@article{schindler_hermitian_2023,
	title = {Hermitian Bulk – Non-Hermitian Boundary Correspondence},
	volume = {4},
	issn = {2691-3399},
	url = {https://link.aps.org/doi/10.1103/PRXQuantum.4.030315},
	doi = {10.1103/PRXQuantum.4.030315},
	pages = {030315},
	number = {3},
	journal = {{PRX} Quantum},
	shortjournal = {{PRX} Quantum},
	author = {Schindler, Frank and Gu, Kaiyuan and Lian, Biao and Kawabata, Kohei},
	urldate = {2023-12-17},
	date = {2023-08-02},
	langid = {english},
}

@article{Zhang_2020,
  title = {Correspondence between Winding Numbers and Skin Modes in Non-Hermitian Systems},
  author = {Zhang, Kai and Yang, Zhesen and Fang, Chen},
  journal = {Phys. Rev. Lett.},
  volume = {125},
  issue = {12},
  pages = {126402},
  numpages = {6},
  year = {2020},
  month = {Sep},
  publisher = {American Physical Society},
  doi = {10.1103/PhysRevLett.125.126402},
  url = {https://link.aps.org/doi/10.1103/PhysRevLett.125.126402}
}

@article{Yokomizi_2019,
  title = {Non-Bloch Band Theory of Non-Hermitian Systems},
  author = {Yokomizo, Kazuki and Murakami, Shuichi},
  journal = {Phys. Rev. Lett.},
  volume = {123},
  issue = {6},
  pages = {066404},
  numpages = {6},
  year = {2019},
  month = {Aug},
  publisher = {American Physical Society},
  doi = {10.1103/PhysRevLett.123.066404},
  url = {https://link.aps.org/doi/10.1103/PhysRevLett.123.066404}
}

@article{Yang_2020,
  title = {Non-Hermitian Bulk-Boundary Correspondence and Auxiliary Generalized Brillouin Zone Theory},
  author = {Yang, Zhesen and Zhang, Kai and Fang, Chen and Hu, Jiangping},
  journal = {Phys. Rev. Lett.},
  volume = {125},
  issue = {22},
  pages = {226402},
  numpages = {6},
  year = {2020},
  month = {Nov},
  publisher = {American Physical Society},
  doi = {10.1103/PhysRevLett.125.226402},
  url = {https://link.aps.org/doi/10.1103/PhysRevLett.125.226402}
}

@article{PhysRevLett.80.1800,
  title = {Quantum-Mechanical Position Operator in Extended Systems},
  author = {Resta, Raffaele},
  journal = {Phys. Rev. Lett.},
  volume = {80},
  issue = {9},
  pages = {1800--1803},
  numpages = {0},
  year = {1998},
  month = {Mar},
  publisher = {American Physical Society},
  doi = {10.1103/PhysRevLett.80.1800},
  url = {https://link.aps.org/doi/10.1103/PhysRevLett.80.1800}
}

@article{Okugawa_2021,
  title = {Non-Hermitian band topology with generalized inversion symmetry},
  author = {Okugawa, Ryo and Takahashi, Ryo and Yokomizo, Kazuki},
  journal = {Phys. Rev. B},
  volume = {103},
  issue = {20},
  pages = {205205},
  numpages = {11},
  year = {2021},
  month = {May},
  publisher = {American Physical Society},
  doi = {10.1103/PhysRevB.103.205205},
  url = {https://link.aps.org/doi/10.1103/PhysRevB.103.205205}
}

@article{Vecsei_2021,
  title = {Symmetry indicators for inversion-symmetric non-Hermitian topological band structures},
  author = {Vecsei, Pascal M. and Denner, M. Michael and Neupert, Titus and Schindler, Frank},
  journal = {Phys. Rev. B},
  volume = {103},
  issue = {20},
  pages = {L201114},
  numpages = {5},
  year = {2021},
  month = {May},
  publisher = {American Physical Society},
  doi = {10.1103/PhysRevB.103.L201114},
  url = {https://link.aps.org/doi/10.1103/PhysRevB.103.L201114}
}

@article{okuma_topological_2020,
	title = {Topological Origin of Non-Hermitian Skin Effects},
	volume = {124},
	url = {https://link.aps.org/doi/10.1103/PhysRevLett.124.086801},
	doi = {10.1103/PhysRevLett.124.086801},
	pages = {086801},
	number = {8},
	journal = {Phys. Rev. Lett.},
	shortjournal = {Phys. Rev. Lett.},
	author = {Okuma, Nobuyuki and Kawabata, Kohei and Shiozaki, Ken and Sato, Masatoshi},
	urldate = {2023-12-17},
	date = {2020-02-25},
}

@article{PhysRevB.26.4269,
  title = {Wannier functions in one-dimensional disordered systems: Application to fractionally charged solitons},
  author = {Kivelson, S.},
  journal = {Phys. Rev. B},
  volume = {26},
  issue = {8},
  pages = {4269--4277},
  numpages = {0},
  year = {1982},
  month = {Oct},
  publisher = {American Physical Society},
  doi = {10.1103/PhysRevB.26.4269},
  url = {https://link.aps.org/doi/10.1103/PhysRevB.26.4269}
}

@article{yao_edge_2018,
	title = {Edge States and Topological Invariants of Non-Hermitian Systems},
	volume = {121},
	issn = {0031-9007, 1079-7114},
	url = {https://link.aps.org/doi/10.1103/PhysRevLett.121.086803},
	doi = {10.1103/PhysRevLett.121.086803},
	pages = {086803},
	number = {8},
	journal = {Phys. Rev. Lett.},
	shortjournal = {Phys. Rev. Lett.},
	author = {Yao, Shunyu and Wang, Zhong},
	urldate = {2023-12-17},
	date = {2018-08-21},
	langid = {english},
}

@article{konig_braid-protected_2023,
	title = {Braid-protected topological band structures with unpaired exceptional points},
	volume = {5},
	url = {https://link.aps.org/doi/10.1103/PhysRevResearch.5.L042010},
	doi = {10.1103/PhysRevResearch.5.L042010},
	pages = {L042010},
	number = {4},
	journal = {Phys. Rev. Res.},
	shortjournal = {Phys. Rev. Res.},
	author = {König, J. Lukas K. and Yang, Kang and Budich, Jan Carl and Bergholtz, Emil J.},
	urldate = {2023-12-17},
	date = {2023-10-11},
}

@article{shen_topological_2018,
	title = {Topological Band Theory for Non-Hermitian Hamiltonians},
	volume = {120},
	url = {https://link.aps.org/doi/10.1103/PhysRevLett.120.146402},
	doi = {10.1103/PhysRevLett.120.146402},
	pages = {146402},
	number = {14},
	journal = {Phys. Rev. Lett.},
	shortjournal = {Phys. Rev. Lett.},
	author = {Shen, Huitao and Zhen, Bo and Fu, Liang},
	urldate = {2023-12-17},
	date = {2018-04-06},
}

@article{Shiozaki_2021,
  title = {Symmetry indicator in non-Hermitian systems},
  author = {Shiozaki, Ken and Ono, Seishiro},
  journal = {Phys. Rev. B},
  volume = {104},
  issue = {3},
  pages = {035424},
  numpages = {35},
  year = {2021},
  month = {Jul},
  publisher = {American Physical Society},
  doi = {10.1103/PhysRevB.104.035424},
  url = {https://link.aps.org/doi/10.1103/PhysRevB.104.035424}
}

@book{geyer2025stability,
  title={Stability of nonlinear waves in Hamiltonian dynamical systems},
  author={Geyer, Anna and Pelinovsky, Dmitry E},
  volume={288},
  year={2025},
  publisher={American Mathematical Society}
}

@article{PhysRevB.111.064312,
  title = {Nonlinear breathers with crystalline symmetries},
  author = {Schindler, Frank and Bulchandani, Vir B. and Benalcazar, Wladimir A.},
  journal = {Phys. Rev. B},
  volume = {111},
  issue = {6},
  pages = {064312},
  numpages = {20},
  year = {2025},
  month = {Feb},
  publisher = {American Physical Society},
  doi = {10.1103/PhysRevB.111.064312},
  url = {https://link.aps.org/doi/10.1103/PhysRevB.111.064312}
}

@article{chernyavsky2018krein,
  title={Krein signature for instability of PT-symmetric states},
  author={Chernyavsky, Alexander and Pelinovsky, Dmitry E},
  journal={Physica D: Nonlinear Phenomena},
  volume={371},
  pages={48--59},
  year={2018},
  publisher={Elsevier}
}

@article{6hy3-jmk9,
  title = {Solitons with self-induced topological nonreciprocity},
  author = {de Castro, Pedro Fittipaldi and Benalcazar, Wladimir A.},
  journal = {Phys. Rev. B},
  volume = {112},
  issue = {12},
  pages = {L121107},
  numpages = {6},
  year = {2025},
  month = {Sep},
  publisher = {American Physical Society},
  doi = {10.1103/6hy3-jmk9},
  url = {https://link.aps.org/doi/10.1103/6hy3-jmk9}
}

@article{PhysRevA.87.012118,
  title = {Topological invariance and global Berry phase in non-Hermitian systems},
  author = {Liang, Shi-Dong and Huang, Guang-Yao},
  journal = {Phys. Rev. A},
  volume = {87},
  issue = {1},
  pages = {012118},
  numpages = {6},
  year = {2013},
  month = {Jan},
  publisher = {American Physical Society},
  doi = {10.1103/PhysRevA.87.012118},
  url = {https://link.aps.org/doi/10.1103/PhysRevA.87.012118}
}

@article{PhysRevB.107.085122,
  title = {Complex Berry phase and imperfect non-Hermitian phase transitions},
  author = {Longhi, Stefano and Feng, Liang},
  journal = {Phys. Rev. B},
  volume = {107},
  issue = {8},
  pages = {085122},
  numpages = {13},
  year = {2023},
  month = {Feb},
  publisher = {American Physical Society},
  doi = {10.1103/PhysRevB.107.085122},
  url = {https://link.aps.org/doi/10.1103/PhysRevB.107.085122}
}

@article{lane2025complex,
  title={Complex Berry phase and steady-state geometric amplification in non-Hermitian systems},
  author={Lane, JR and Guria, C and H{\"o}ller, J and Montalvo, TD and Patil, YSS and Harris, JGE},
  journal={arXiv preprint arXiv:2503.23197},
  year={2025}
}

@article{PhysRevResearch.5.L032026,
  title = {Measuring the adiabatic non-Hermitian Berry phase in feedback-coupled oscillators},
  author = {Singhal, Yaashnaa and Martello, Enrico and Agrawal, Shraddha and Ozawa, Tomoki and Price, Hannah and Gadway, Bryce},
  journal = {Phys. Rev. Res.},
  volume = {5},
  issue = {3},
  pages = {L032026},
  numpages = {7},
  year = {2023},
  month = {Aug},
  publisher = {American Physical Society},
  doi = {10.1103/PhysRevResearch.5.L032026},
  url = {https://link.aps.org/doi/10.1103/PhysRevResearch.5.L032026}
}

@article{PhysRev.52.191,
  title = {The Structure of Electronic Excitation Levels in Insulating Crystals},
  author = {Wannier, Gregory H.},
  journal = {Phys. Rev.},
  volume = {52},
  issue = {3},
  pages = {191--197},
  numpages = {0},
  year = {1937},
  month = {Aug},
  publisher = {American Physical Society},
  doi = {10.1103/PhysRev.52.191},
  url = {https://link.aps.org/doi/10.1103/PhysRev.52.191}
}

@incollection{BLOUNT1962305,
title = {Formalisms of Band Theory},
editor = {Frederick Seitz and David Turnbull},
series = {Solid State Physics},
publisher = {Academic Press},
volume = {13},
pages = {305-373},
year = {1962},
issn = {0081-1947},
doi = {https://doi.org/10.1016/S0081-1947(08)60459-2},
url = {https://www.sciencedirect.com/science/article/pii/S0081194708604592},
author = {E.I. Blount}
}

@article{PhysRevB.104.L020303,
  title = {Non-Hermitian topological end breathers},
  author = {Lang, Li-Jun and Zhu, Shi-Liang and Chong, Y. D.},
  journal = {Phys. Rev. B},
  volume = {104},
  issue = {2},
  pages = {L020303},
  numpages = {6},
  year = {2021},
  month = {Jul},
  publisher = {American Physical Society},
  doi = {10.1103/PhysRevB.104.L020303},
  url = {https://link.aps.org/doi/10.1103/PhysRevB.104.L020303}
}

@article{PhysRevB.105.125421,
  title = {Dynamical nonlinear higher-order non-Hermitian skin effects and topological trap-skin phase},
  author = {Ezawa, Motohiko},
  journal = {Phys. Rev. B},
  volume = {105},
  issue = {12},
  pages = {125421},
  numpages = {9},
  year = {2022},
  month = {Mar},
  publisher = {American Physical Society},
  doi = {10.1103/PhysRevB.105.125421},
  url = {https://link.aps.org/doi/10.1103/PhysRevB.105.125421}
}

@article{YUCE2021127484,
title = {Nonlinear non-Hermitian skin effect},
journal = {Physics Letters A},
volume = {408},
pages = {127484},
year = {2021},
issn = {0375-9601},
doi = {https://doi.org/10.1016/j.physleta.2021.127484},
url = {https://www.sciencedirect.com/science/article/pii/S0375960121003480},
author = {Cem Yuce}
}

@article{PhysRevB.110.L180302,
  title = {Insensitive edge solitons in a non-Hermitian topological lattice},
  author = {Many Manda, Bertin and Achilleos, Vassos},
  journal = {Phys. Rev. B},
  volume = {110},
  issue = {18},
  pages = {L180302},
  numpages = {6},
  year = {2024},
  month = {Nov},
  publisher = {American Physical Society},
  doi = {10.1103/PhysRevB.110.L180302},
  url = {https://link.aps.org/doi/10.1103/PhysRevB.110.L180302}
}

@article{PhysRevB.109.094308,
  title = {Skin modes in a nonlinear Hatano-Nelson model},
  author = {Many Manda, Bertin and Carretero-Gonz\'alez, Ricardo and Kevrekidis, Panayotis G. and Achilleos, Vassos},
  journal = {Phys. Rev. B},
  volume = {109},
  issue = {9},
  pages = {094308},
  numpages = {12},
  year = {2024},
  month = {Mar},
  publisher = {American Physical Society},
  doi = {10.1103/PhysRevB.109.094308},
  url = {https://link.aps.org/doi/10.1103/PhysRevB.109.094308}
}

@article{veenstra2024non,
  title={Non-reciprocal topological solitons in active metamaterials},
  author={Veenstra, Jonas and Gamayun, Oleksandr and Guo, Xiaofei and Sarvi, Anahita and Meinersen, Chris Ventura and Coulais, Corentin},
  journal={Nature},
  volume={627},
  number={8004},
  pages={528--533},
  year={2024},
  publisher={Nature Publishing Group UK London}
}

\newpage

\appendix

\section{Construction of a biorthonormal basis}\label{AppendixBiorthonormal}

Let $H$ be a diagonalizable non-Hermitian operator acting on a Hilbert space $\mathcal{H}$ of dimension $D$ and let $\{\ket{\psi^{R}_{j}}\}$ where $j=1,\dots,D$ be its right eigenvectors with eigenvalues $\{E_{j}\}$:
\begin{equation}
    H\ket{\psi^{R}_j} = E_{j}\ket{\psi^{R}_j}. \label{Eigenvalue}
\end{equation}
Now, consider the operator
\begin{equation}
    R = \sum_{j=1}^{D}\ket{\psi^{R}_{j}}\bra{j},
\end{equation}
where $\{\ket{j}\}$ is an orthonormal basis of $\mathcal{H}$. Note that
\begin{equation}
    R_{ij} = \bra{i}R\ket{j}=\langle{i}|{\psi^{R}_j}\rangle,
\end{equation}
are the components of $\ket{\psi^{R}_j}$ in the canonical basis. By taking \eqref{Eigenvalue} and summing over all eigenvectors, we have
\begin{align}
    HR&=\sum_{j}E_j\ket{\psi^{R}_j}\bra{j} \nonumber \\
    &= \left(\sum_{j}\ket{\psi^{R}_j}\bra{j}\right)\left(\sum_{i}E_i\ket{i}\bra{i}\right) \label{Separate} \nonumber \\
    &= R \Lambda,
\end{align}
where
\begin{equation}
    \Lambda = E_{i}\ket{i}\bra{i}
\end{equation}
is the matrix $\mathrm{diag}(E_1,\dots,E_D)$.

Multiplying \eqref{Separate} on the left by $r^{-1}$ ($r$ is invertible as long as $H$ is not defective) yields
\begin{equation}
    R^{-1}HR = \Lambda.
\end{equation}
Taking the Hermitian conjugate of the expression above gives us
\begin{equation}
    R^{\dagger}H^{\dagger}(R^{-1})^{\dagger}=\Lambda^{\dagger}=\Lambda^{*}.
\end{equation}
Multiplying the expression above by $(R^{\dagger})^{-1}=(R^{-1})^{\dagger}$ on the left yields
\begin{equation}
    H^{\dagger}L = L\Lambda^{*},
\end{equation}
where
\begin{equation}
    L \equiv (R^{-1})^{\dagger} \label{DefinitionL}
\end{equation}
is an operator that allows us to define the left eigenstates $\ket{\psi^{L}_{j}}$ through
\begin{equation}
    L = \sum_{j=1}^{D}\ket{\psi^{L}_{j}}\bra{j},
\end{equation}
or
\begin{equation}
    L_{ij} = \langle i|\psi^{L}_{j} \rangle.
\end{equation}

Note that the definition \eqref{DefinitionL} implies
\begin{equation}
    L^{\dagger}R=R^{-1}R=\mathbb{I},
\end{equation}
which is equivalent to the biorthonormality condition
\begin{equation}
    \langle\psi^{L}_i|\psi^{R}_{j}\rangle = \delta_{ij}.
\end{equation}

\section{Gauge freedom in non-Hermitian Hamiltonians}\label{Appendix:GaugeFreedom}

Let $R_k$ and $L_k$ be the $N \times N_{\mathrm{occ}}$ operators collecting the occupied right and left
Bloch eigenvectors, as defined in Eq.~\eqref{RandL}, such that
\begin{equation}
L_k^\dagger R_k = \mathbb{I}_{N_{\mathrm{occ}}}.
\label{BiorthGaugeConstraint}
\end{equation}
The choice of basis within the occupied subspace is not unique:
one may perform a $k$-dependent change of basis
\begin{equation}
R_k \to R_k' = R_k\,U_k,
\qquad
L_k \to L_k' = L_k\,(U_k^{-1})^\dagger,
\label{GaugeTransformRL}
\end{equation}
where $U_k \in GL(N_{\mathrm{occ}},\mathbb{C})$ is an invertible matrix.
This transformation preserves biorthonormality,
since $L_k'^\dagger R_k'=(U_k^{-1})(L_k^\dagger R_k)U_k=\mathbb{I}_{N_{\mathrm{occ}}}$.

Importantly, the biorthogonal projector onto the occupied subspace,
\begin{equation}
P_k = R_k L_k^\dagger,
\end{equation}
is invariant under Eq.~\eqref{GaugeTransformRL}, as
\begin{equation}
P_k' = R_k' L_k'^\dagger = R_k U_k \left[(U_k^{-1})^\dagger\right]^\dagger L_k^\dagger
      = R_k U_k (U_k^{-1}) L_k^\dagger = P_k.
\end{equation}

\subsection{Wilson line elements and Wilson loops}

The biorthogonal Wilson line element connecting neighboring momenta is
\begin{equation}
G_k \equiv G_{k+\Delta,k}=L_{k+\Delta}^\dagger R_k,
\label{WilsonElementGaugeAppendix}
\end{equation}
where $\Delta=2\pi/\ell$ for a discretization of the Brillouin zone into $\ell$ points.
Under the gauge transformation \eqref{GaugeTransformRL}, $G_k$ transforms as
\begin{equation}
G_k \to G_k' = L_{k+\Delta}'^\dagger R_k'
= (U_{k+\Delta}^{-1})\,G_k\,U_k,
\label{GaugeTransformG}
\end{equation}
i.e., it is gauge-covariant.

The biorthogonal Wilson loop based at $k$ is defined as the ordered product
\begin{equation}
\mathcal{W}_k = \prod_{j=0}^{\ell-1} G_{k+j\Delta},
\label{WilsonLoopDiscreteAppendix}
\end{equation}
where the product is ordered along increasing momentum.
Using Eq.~\eqref{GaugeTransformG}, we obtain
\begin{align}
\mathcal{W}_k'
&= \prod_{j=0}^{\ell-1} G_{k+j\Delta}'
= \prod_{j=0}^{\ell-1}\left[(U_{k+(j+1)\Delta}^{-1})\,G_{k+j\Delta}\,U_{k+j\Delta}\right] \nonumber \\
&= U_{k+\ell\Delta}^{-1}\left(\prod_{j=0}^{\ell-1}G_{k+j\Delta}\right)U_k.
\end{align}
Since $k+\ell\Delta=k+2\pi$ corresponds to the same crystal momentum, we may choose a periodic gauge
such that $U_{k+2\pi}=U_k$, implying
\begin{equation}
\boxed{
\mathcal{W}_k' = U_k^{-1}\,\mathcal{W}_k\,U_k.
}
\label{WilsonLoopSimilarity}
\end{equation}
Therefore, the Wilson loop transforms by similarity, and its eigenvalues $\lambda$ are gauge invariant.
Thus, the associated (possibly complex) Wannier centers
$z=\nu+i\kappa$ defined through $\lambda=e^{-2\pi i z}$ are also gauge-invariant quantities.

\subsection{Gauge dependence of the Berry connection}

In the continuum limit, the Wilson line element may be written as
$G_k=\mathbb{I}-i\Delta A(k)+\mathcal{O}(\Delta^2)$, where
$A(k)=-iL_k^\dagger\partial_k R_k$ is the biorthogonal Berry connection.
Under Eq.~\eqref{GaugeTransformRL}, one finds the standard non-Abelian transformation law
\begin{equation}
A(k)\to A'(k)=U_k^{-1}A(k)U_k-iU_k^{-1}\partial_k U_k. \label{GaugeCovariantA}
\end{equation}
Hence, $A(k)$ is gauge dependent pointwise, while the Wilson loop spectrum remains gauge invariant
as shown above.

\section{Conditions for nonunitarity of the biorthogonal Wilson loop}\label{AppendixNonUnitaryWL}

Let $R_k$ $(L_k)$ be the operator defined in \eqref{RandL} containing the right (left) eigenvectors of the Bloch Hamiltonian $h_k$ in the occupied bands. The biorthogonal Wilson line is defined as
\begin{equation}
    G_{k+\Delta,k}=L^{\dagger}_{k+\Delta}R_{k},
\end{equation}
where $\Delta = \frac{2\pi}{\ell}$. In the thermodynamic limit $\ell \to \infty$ ($\Delta \to 0$), we have
\begin{equation}
    G_{k+\Delta,k} = L^{\dagger}_{k}R_{k}+\mathcal{O}(\Delta) = \mathbb{I}_{N_{\mathrm{occ}}}+\mathcal{O}(\Delta), \label{FirstOrder}
\end{equation}
which naively may induce one to think that the Wilson loop,
\begin{equation}
    \mathcal{W}=\lim_{\Delta \to 0}\prod_{k}G_{k+\Delta,k},
\end{equation}
is unitary. However, we shall now see that the $\mathcal{O}(\Delta)$ terms in \eqref{FirstOrder} give non-unitary contributions that remain finite in the thermodynamic limit. Concretely, we have
\begin{align}
    G_{k+\Delta,k} &= \mathbb{I} + \Delta(\partial_kL^{\dagger}_k)R_k + \mathcal{O}(\Delta^2) \nonumber \\
    &=\mathbb{I} - \Delta (L^{\dagger}_k\partial_kR_k) + \mathcal{O}(\Delta^2) \nonumber \\
    &= \mathbb{I} - \mathrm{i}\Delta A_k+\mathcal{O}(\Delta^2),
\end{align}
where $A_k=-\mathrm{i}L_{k}^{\dagger}\partial_kR_k$ is the biorthogonal Berry connection. Thus, in the thermodynamic limit, the Wilson loop from $k$ to $k+2\pi$ is given by
\begin{align}
    \mathcal{W}_k &= \mathcal{P}\exp\left(-\mathrm{i}\int_{k}^{k+2\pi}A(k')dk'\right).
\end{align}
Therefore, we have
\begin{equation}
    \frac{d\mathcal{W}_k}{dk}=-\mathrm{i}A_k\mathcal{W}_k, \label{DifferentialEquation}
\end{equation}
where we have used $A_{k+2\pi}=A_k$.

Now, consider $Q=\mathcal{W}_k^{\dagger}\mathcal{W}_k$. If $\mathcal{W}$ is unitary, then $Q=\mathbb{I}_{N_{\mathrm{occ}}}$ and thus $\frac{dQ}{dk}=0$. We have
\begin{align}
    \frac{dQ}{dk} &= \frac{d\mathcal{W}_k^{\dagger}}{dk}\mathcal{W}_k+\mathcal{W}_k^{\dagger}\frac{d\mathcal{W}_k}{dk} \nonumber \\
    &= \mathrm{i}\mathcal{W}_k^{\dagger}(A^{\dagger}_{k}-A_{k})\mathcal{W}_k \nonumber \\
    &= -2\mathrm{i}\mathcal{W}_k^{\dagger}A^{(a)}_{k}\mathcal{W}_k,
\end{align}
where 
\begin{equation}
    A^{(a)}_{k} = \frac{A_k-A^{\dagger}_k}{2}
\end{equation}
is the anti-Hermitian part of the Berry connection. In particular, for Hermitian Hamiltonians, the Berry connection is also Hermitian, implying $\frac{dQ}{dk}=0$ and a unitary Wilson loop. But this is still not a sufficient condition for the Wilson loop to be nonunitary. By using the formula
\begin{equation}
    \frac{d}{dk}\log\det \mathcal{W}=\mathrm{Tr}\left[\mathcal{W}^{-1}\frac{d\mathcal{W}}{dk}\right]
\end{equation}
and employing equation \eqref{DifferentialEquation}, we obtain
\begin{equation}
    \frac{d}{dk}\log\det \mathcal{W} = \mathrm{i}\mathrm{Tr} A(k),
\end{equation}
and
\begin{equation}
    \log\det \mathcal{W} = \mathrm{i}\int_{k}^{k+2\pi}\mathrm{Tr} A(k')dk'. \label{TraceFormula}
\end{equation}
Note that the expression above is gauge invariant. Although $A(k)$ transforms according to \eqref{GaugeCovariantA}, the cyclic property of the trace yields
\begin{align}
    \mathrm{Tr} A'(k) &= \mathrm{Tr} A(k) - \mathrm{i}\mathrm{Tr}(U^{-1}_{k}\partial_kU_k) \nonumber \\
    &= \mathrm{Tr} A(k) - \partial_k\log\det U_k,
\end{align}
so
\begin{align}
    \int_{k}^{k+2\pi}\mathrm{Tr} [A'(k')-A(k')]dk' &= \log\det U_{k+2\pi} - \log\det U_{k} \nonumber \\
    &=0.
\end{align}

Now, noting that
\begin{equation}
    \log \det \mathcal{W}=\log|\det\mathcal{W}|+\mathrm{i}\ \mathrm{arg}\det \mathrm{W},
\end{equation}
and using \eqref{TraceFormula}, we have
\begin{align}
    \log|\det\mathcal{W}| &= \mathrm{Re}\log \det \mathcal{W} \nonumber \\
    &= \mathrm{Re}\left[\mathrm{i}\int_{k}^{k+2\pi}\mathrm{Tr} A(k')dk'\right].
\end{align}
But since the imaginary component of the trace must come solely from the anti-Hermitian part and unitary matrices satisfy $|\det\mathcal{W}|=1$ and $\log|\det\mathcal{W}|=0$, we see that the Wilson loop cannot be unitary if
\begin{equation}
    \int_{k}^{k+2\pi}\mathrm{Tr} A^{(a)}(k')dk' \neq 0.
\end{equation}

\section{Relationship between the eigenvalues of the Wilson loop and the projected position operator}\label{AppendixEigenvaluesofProjectedPositionOperator}

In the momentum basis, defined by the Fourier transform
\begin{equation}
    \ket{k} = \frac{1}{\sqrt{\ell}}\sum_{x=0}^{\ell-1}e^{\mathrm{i}kx}\ket{x},
\end{equation}
the Resta position operator~\eqref{Resta} reads
\begin{equation}
    X = \sum_{k}\ket{k+\Delta}\bra{k} \otimes \mathbb{I}_{N},
\end{equation}
where again $\Delta=2\pi/\ell$ and $\mathbb{I}_{N}$ is the identity in the internal degrees of freedom. Meanwhile, the right and left projectors onto the occupied subspace~\eqref{OccupiedSubspaces} can be written as
\begin{align}
    R &= \sum_{k}\ket{k}\bra{k} \otimes R_{k} \nonumber \\
    L &= \sum_{k}\ket{k}\bra{k}\otimes L_{k},
\end{align}
where $R_{k}$ and $L_{k}$ are defined in \eqref{RandL} and act on the internal degrees of freedom. Therefore, the biorthonormal projection position operator~\eqref{Projection} is
\begin{align}
    X_{\mathrm{occ}} &= \sum_{k}\ket{k+\Delta}\bra{k} \otimes G_{k},
\end{align}
where $G_{k}=L_{k+\Delta}^{\dagger}R_{k}$ is the biorthogonal Wilson line element. Its matrix elements in the momentum basis are
\begin{equation}
    \bra{k}X_{\mathrm{occ}}\ket{k'}=\delta_{k+\Delta,k'} \otimes G_{k'},
\end{equation}
which we may gather in the following matrix representation:
\begin{equation}
    X_{\mathrm{occ}} = \begin{pmatrix}
        0 & G_{k_1} & 0 & \cdots & 0 \\
        0 & 0 & G_{k_2} & \cdots & 0 \\
        0 & 0 & 0 & \cdots & 0 \\
        \vdots & \vdots & \vdots & \ddots & \vdots \\
        G_{k_L} & 0 & 0 & \cdots & 0
    \end{pmatrix}, \label{WilsonLoopMatrix}
\end{equation}
where $k_j=j \Delta$. From \eqref{WilsonLoopMatrix} we get
\begin{equation}
    (X_{\mathrm{occ}})^{\ell} = \mathrm{diag}(\mathcal{W}_{k_1},\mathcal{W}_{k_2},\dots,\mathcal{W}_{k_L}), \label{ChainofWL}
\end{equation}
where $\mathcal{W}_{k}=\mathcal{W}_{k} = G_{k+2\pi-\Delta}G_{k+2\pi-2\Delta}\cdots G_{k+\Delta}G_{k}$ is the Wilson loop at base point $k$. If $\lambda_{n}$ is an eigenvalue of \eqref{WilsonLoopMatrix} and $e^{-2\pi \mathrm{i}z_{n}}$ is an eigenvalue of the Wilson loop where $n \in \{1,\dots,N_{\mathrm{occ}}\}$, the expression \eqref{ChainofWL} implies $(\lambda_n)^{\ell}=e^{-2\pi \mathrm{i}z_{n}}$. But since $z_{n}=\nu_{n}+\mathrm{i}\kappa_n$, we have
\begin{align}
    \lambda_{j,n} = r_n \exp\left(-\frac{2\pi i}{L}(\nu_n+j)\right),
\end{align}
In the expression above, the complex phase encodes the position of a particle (unit cell coordinate $j=0,\dots,\ell-1$ and $\nu_{n}$ relative to the unit-cell origin) and the absolute value
\begin{equation}
    r_n = |\lambda_{j,n}| = \exp\left(\frac{2\pi\kappa_n}{L}\right)
\end{equation}
deviates from unity according to the imaginary part $\kappa_n$ of the Wilson loop. Note that, even though $r_{n} \to 1$ in the thermodynamic limit, the momentum boost of the Wannier functions does not vanish, as it depends solely on $\kappa_n$ [see Eqs. \eqref{Eq:keffEstimateGaussian} and \eqref{KappaRigorous}], and not directly on the eigenvalues of the projected position operator.

\section{Inequivalence of projected-position and Bloch-Fourier definitions of Wannier functions in non-Hermitian Hamiltonians}
\label{Appendix:Inequivalence}

In this Appendix, we prove that, in a generic non-Hermitian band, Wannier functions defined as eigenstates of the projected position operator cannot be written as equal-weight Bloch--Fourier superpositions of Bloch eigenstates.
The obstruction is a gauge-invariant non-uniform distribution of momentum-space weights.

Let $\ket{W_x}$ be a projected-position Wannier function of a non-Hermitian Hamiltonian, defined as an eigenstate of the projected position operator,
\begin{equation}
    X_p \ket{W_x}=\lambda_x\ket{W_{x}},
    \label{Xequation}
\end{equation}
where $X_p = PXP$ is the position operator restricted to a single occupied band, and
\begin{equation}
    P = RL^{\dagger}
    = \sum_{k}\ket{k}\bra{k} \otimes \ket{u^{R}_k}\bra{u^{L}_k}
    \label{ProjectorAppendix}
\end{equation}
is the biorthogonal projector onto the occupied band. We assume the standard biorthonormal convention
$\langle u_k^{L}|u_k^{R}\rangle=1$.
The periodic (unitary) position operator in the crystal-momentum basis acts as a discrete shift,
\begin{equation}
    X = \sum_{k}\ket{k+\Delta}\bra{k} \otimes \mathbb{I}_N,
    \qquad
    \Delta=\frac{2\pi}{\ell},
    \label{XinMomentumBasis}
\end{equation}
where $k+\Delta$ is understood modulo the Brillouin zone.

Since $P\ket{W_x}=\ket{W_x}$, we may expand the Wannier function in the (right) Bloch basis as
\begin{align}
    \ket{W_x} &= \sum_k \, \ket{k} \otimes \ket{w_k}  \nonumber \\
    &=\sum_{k} \, \ket{k} \otimes a_k \ket{u^{R}_k},
    \label{BlochExpansion}
\end{align}
where $\ket{w_k}$ is the orbital content of $\ket{W^{R}_x}$ at momentum $k$, and
\begin{equation}
    a_k = \langle u^{L}_{k}|w_k\rangle
    \label{akDefinition}
\end{equation}
for the case of $N_{\mathrm{occ}}=1$ occupied band.
For the time being, let us assume that the right eigenstates of the Bloch Hamiltonian are all normalized, i.e.,
$\|u^{R}_k\|^2=\langle u^{R}_k|u^{R}_k\rangle = 1$ for all $k$. This is always possible by a (real) non-unitary rescaling $\ket{u_k^{R}}\to e^{f(k)}\ket{u_k^{R}}$ that preserves the biorthonormality condition. Our goal here is to determine whether $|a_k|$ is $k$-independent, like in Hermitian Hamiltonians. We will consider the effect of general non-unitary gauge transformations later on.

Using Eqs.~\eqref{XinMomentumBasis} and~\eqref{BlochExpansion}, we obtain
\begin{align}
    X_p\ket{W_x}
    &= PXP\ket{W_x}
     = PX\ket{W_x} \nonumber \\
    &= P\sum_{k} a_k \, \ket{k+\Delta} \otimes \ket{u^{R}_{k}} \nonumber \\
    &= \sum_{k} a_{k-\Delta}\,
    \langle u^{L}_{k}|u^{R}_{k-\Delta}\rangle\;
    \ket{k} \otimes \ket{u^{R}_{k}} .
    \label{IntResultW}
\end{align}
On the other hand, Eq.~\eqref{Xequation} implies
\begin{equation}
    X_p \ket{W_x}=\lambda_x \ket{W_x}
    =\sum_{k} \lambda_x a_k \, \ket{k} \otimes \ket{u^{R}_k}.
    \label{ExpandedLambda}
\end{equation}
Comparing Eqs.~\eqref{IntResultW} and~\eqref{ExpandedLambda}, we find the recurrence relation
\begin{equation}
    a_{k} = \frac{\langle u^{L}_{k}|u^{R}_{k-\Delta}\rangle}{\lambda_x}\,a_{k-\Delta}.
    \label{Recurrence}
\end{equation}

In the thermodynamic limit ($\Delta\to 0$),
\begin{align}
    \langle u^{L}_k|u^{R}_{k-\Delta}\rangle
    &= \langle u^{L}_k|u^{R}_{k}\rangle - \Delta \langle u^{L}_k|\partial_k u^{R}_{k}\rangle
    + \mathcal{O}(\Delta^2) \nonumber \\
    &\approx 1 - \mathrm{i}\Delta A_k
    \approx \exp\!\left[-\mathrm{i}\Delta A_k\right],
\end{align}
where $A_k=-\mathrm{i}\langle u^{L}_k|\partial_k u^{R}_k\rangle$ is the biorthogonal Berry connection.
Taking the absolute value then yields, to leading order in $\Delta$,
\begin{equation}
    \left|\langle u^{L}_k|u^{R}_{k-\Delta}\rangle\right|
    \approx \exp\!\left[\Delta \,\mathrm{Im}\,A_k\right].
    \label{AbsOverlap}
\end{equation}
We emphasize that $\mathrm{Im}\,A(k)$ is gauge dependent under non-unitary rescalings; however, the final statement below will be expressed in terms of a gauge-invariant momentum-space weight distribution.

Meanwhile, the eigenvalues of the projected position operator can be written as
\begin{equation}
    \lambda_x = \exp\!\left(-\frac{2\pi\mathrm{i}}{\ell}x\right)
               \exp\!\left(-\frac{2\pi\mathrm{i}}{\ell}z\right),
\end{equation}
where $z=\nu+\mathrm{i}\kappa$ is the (generally complex) Wannier center. Using $\Delta=2\pi/\ell$ yields
\begin{equation}
    |\lambda_x| = \exp(\Delta \kappa).
    \label{AbsLambda}
\end{equation}

Taking absolute values in Eq.~\eqref{Recurrence} and using Eqs.~\eqref{AbsOverlap} and~\eqref{AbsLambda}, we obtain
\begin{align}
    \frac{|a_k|}{|a_{k-\Delta}|}
    &\approx \exp\!\left[\Delta\,\mathrm{Im}\,A_k - \Delta \kappa\right], \nonumber \\
    \frac{\log|a_k|-\log|a_{k-\Delta}|}{\Delta}
    &\approx \mathrm{Im}\,A_k - \kappa .
\end{align}
In the continuum limit, this becomes
\begin{equation}
    \partial_k \log|a(k)|
    = \mathrm{Im}\,A(k) - \kappa .
    \label{DiffEqForak}
\end{equation}
Finally, for a single occupied band, one has (cf.~Eq.~\eqref{BiorthW} in the main text)
\begin{equation}
    \kappa = \frac{1}{2\pi}\int_{\mathrm{BZ}} \mathrm{Im}\,A(k)\,dk
    \equiv \overline{\mathrm{Im}\,A},
\end{equation}
so Eq.~\eqref{DiffEqForak} can be written as
\begin{equation}
    \partial_k \log|a(k)|
    = \mathrm{Im}\,A(k) - \overline{\mathrm{Im}\,A}.
\end{equation}
Thus $|a(k)|$ is generically \emph{not} constant in $k$ when $\mathrm{Im}\,A(k)$ is not uniform, implying that $\ket{W_x^R}$ cannot be written as an equal-weight Fourier superposition of Bloch eigenstates.
Solving \eqref{DiffEqForak} explicitly for $a(k)$ yields
\begin{equation}
    |a(k)| = |a(k_0)|\exp\left[\int_{k_0}^{k}(\mathrm{Im}\,A(k')-\kappa)\,dk'\right].
\end{equation}
In the normalized gauge $\|u^{R}_k\|^2=1$, the norm (or ``weight'') of the orbital part $\ket{w_k}=a(k)\ket{u_k^R}$ as a function of $k$ is therefore
\begin{align}
    \|w_k\|^2 &= |a(k)|^2 \cdot \| u^{R}_k\|^2 \nonumber \\
    &= |a(k_0)|^2 \exp\left[2\int_{k_0}^{k}(\mathrm{Im}\,A(k')-\kappa)\,dk'\right],
    \label{WeightFunction}
\end{align}
which is generically non-uniform in $k$.

At this point, one may wonder if the expression \eqref{WeightFunction} is gauge invariant, since any rescaling of the eigenstates of the Bloch Hamiltonian
\begin{equation}
    \ket{u'^{R}_k} = e^{f(k)}\ket{u^{R}_k},
    \quad
    \bra{u'^{L}_k} = e^{-f(k)}\bra{u^{L}_k}, \label{GaugeTrans}
\end{equation}
for any periodic $f(k) \in \mathbb{C}$, preserves the biorthonormality relations. Using the definition of $a_k$ in \eqref{akDefinition}, we see that
\begin{equation}
    a_k'=e^{-f(k)} a_k.
\end{equation}
Therefore,
\begin{align}
    \|w_k'\|^2
    &= |a_k'|^2 \cdot \| u'^{R}_k\|^2 \nonumber \\
    &= |e^{-f(k)}|^2 |a_k|^2 \cdot |e^{f(k)}|^2 \| u^{R}_k\|^2 \nonumber \\
    &= \|w_k\|^2.
\end{align}

Therefore, the non-uniform ``weight'' distribution of the momentum components of projected-position Wannier functions of non-Hermitian Hamiltonians is gauge invariant.
Since $\|w_k\|^2$ is non-uniform in the normalized gauge discussed above, it must be non-uniform in any gauge.
We thus conclude that such functions cannot be expressed in Bloch--Fourier form \eqref{FourierConstructed} under any choice of gauge.

Finally, let us re-express \eqref{WeightFunction} in a manifestly gauge-invariant form. To do that, note
\begin{align}
    \| u_k^{R}\|^2 = \| u_{k_0}^{R}\|^2 \exp\left(\int_{k_0}^{k}\partial_{k'}\log\|u^{R}_{k'}\|^2\,dk'\right).
\end{align}
Therefore, \eqref{WeightFunction} in a generic gauge reads
\begin{align}
    \|w_k\|^2  &= |a(k_0)|^2 \|u_k^{R}\|^2\exp\left[2\int_{k_0}^{k}(\mathrm{Im}\,A(k')-\kappa)\,dk'\right] \nonumber \\
    &= |a(k_0)|^2  \|u_{k_0}^R\|^2 \exp\Big[2\int_{k_0}^{k}\Big(\mathrm{Im}\,A(k') \nonumber \\
    & \qquad +\frac{1}{2}\partial_{k'}\log\|u^{R}_{k'}\|^2-\kappa\Big)\,dk'\Big] \nonumber \\
    &= \|w_{k_0}\|^2\exp\left[2\int_{k_0}^{k}(\Gamma(k')-\kappa)dk'\right]
    ,
    \label{WeightFunctionInvariant}
\end{align}
where $\|w_{k_0}\|^2 = |a(k_0)|^2  \|u_{k_0}\|^2$ and
\begin{equation}
    \Gamma(k)=\mathrm{Im}\,A(k)+\frac{1}{2}\partial_{k}\log\|u^{R}_{k}\|^2
\end{equation}
are gauge-invariant quantities, as we prove below.

Under the non-unitary gauge transformation \eqref{GaugeTrans}, the biorthogonal Berry connection
$A(k)=-\mathrm{i}\langle u_k^{L}|\partial_k u_k^{R}\rangle$ transforms as
\begin{align}
A'(k)
&=-\mathrm{i}\langle u_k^{L}e^{-f(k)}|\partial_k\!\left(e^{f(k)}u_k^{R}\right)\rangle
= A(k)-\mathrm{i}\,\partial_k f(k),
\end{align}
so that
\begin{equation}
\mathrm{Im}\,A'(k)=\mathrm{Im}\,A(k)-\partial_k\mathrm{Re}\,f(k).
\label{ImAtrans}
\end{equation}
Moreover, $\|u_k^{R}\|^2=\langle u_k^{R}|u_k^{R}\rangle$ transforms as
$\|u_k^{R\,\prime}\|^2=e^{2\mathrm{Re}\,f(k)}\|u_k^{R}\|^2$, implying
\begin{equation}
\frac{1}{2}\partial_k\log\|u_k^{R\,\prime}\|^2
=
\frac{1}{2}\partial_k\log\|u_k^{R}\|^2
+\partial_k\mathrm{Re}\,f(k).
\label{NormTrans}
\end{equation}
Combining Eqs.~\eqref{ImAtrans} and \eqref{NormTrans}, we find that
\begin{equation}
\Gamma(k)\equiv \mathrm{Im}\,A(k)+\frac{1}{2}\partial_k\log\|u_k^{R}\|^2 \label{GammaDef}
\end{equation}
is gauge invariant: $\Gamma'(k)=\Gamma(k)$.
Finally, in the normalized gauge $\|u_k^{R}\|^2=1$ one has $\Gamma(k)=\mathrm{Im}\,A(k)$, and hence
\begin{equation}
\overline{\Gamma}\equiv \frac{1}{2\pi}\int_{\mathrm{BZ}}\Gamma(k)\,dk
=
\frac{1}{2\pi}\int_{\mathrm{BZ}}\mathrm{Im}\,A(k)\,dk
\equiv \kappa.
\end{equation}

\subsection{Numerical confirmation of the analytical expression}

We verify the accuracy of the expression \eqref{WeightFunctionInvariant} by comparing its discrete version to the momentum weight distribution obtained directly from the diagonalization of the projected position operator \eqref{Xp}. 

\paragraph{Discrete $\Gamma$-propagation of the momentum-weight distribution.}

Let the Brillouin zone be sampled on a uniform grid
\begin{equation}
k_j = k_0 + j\Delta,
\qquad
j=0,1,\dots,\ell-1,
\qquad
\Delta = \frac{2\pi}{\ell},
\end{equation}
with periodic identification $k_{j-L}\equiv k_j$.

Expanding a projected-position Wannier eigenstate in the momentum basis,
\begin{equation}
|W_x\rangle = \sum_{k_j}^{\ell-1} |k_j\rangle \otimes |w_{k_j}\rangle,
\end{equation}
we define the momentum-space weight
\begin{equation}
\|w_{k_j}\|^2 \equiv \langle w_{k_j}|w_{k_j}\rangle.
\end{equation}

Denote by $|u^R_{k}\rangle$ and $\langle u^L_{k}|$ the right and left
Bloch eigenvectors of the occupied band, biorthonormalized such that
$\langle u^L_k|u^R_k\rangle = 1$.
A gauge-invariant discrete estimator of $\Gamma(k)$ in \eqref{GammaDef} is
\begin{equation}
\Gamma_j =
\frac{1}{\Delta}
\log \left|
\langle u^L_{k_j} \mid u^R_{k_{j-1}} \rangle
\right|
+
\frac{1}{2\Delta}
\log \left(
\frac{\|u^R_{k_j}\|^2}
     {\|u^R_{k_{j-1}}\|^2}
\right),
\label{eq:GammaDiscrete}
\end{equation}
and its Brillouin-zone average is
\begin{equation}
\overline{\Gamma}
=
\frac{\Delta}{2\pi}
\sum_{j=0}^{\ell-1}
\Gamma_j .
\label{eq:GammaBarDiscrete}
\end{equation}

The discretized version of equation \eqref{WeightFunctionInvariant} for the momentum-weight distribution reads
\begin{equation}
\|w_{k_j}\|^2
=
\|w_{k_0}\|^2
\exp\!\left[
2\Delta
\sum_{m=1}^{j}
\big(\Gamma_m - \overline{\Gamma}\big)
\right].
\label{eq:DiscreteWeight}
\end{equation}

Now, let us apply \eqref{eq:DiscreteWeight} to the lowest energy band of model \eqref{ToyModel1} with parameters $m=\varepsilon=0.5$ and compare it to a Wannier function obtained as an eigenvector of the projected position operator. Figure \ref{Verification}(a) shows an excellent agreement between the two methods at system size $\ell=800$, while Figure \ref{Verification}(b) show how \eqref{eq:DiscreteWeight} becomes more accurate as $\ell$ grows.

\section{Effect of symmetries on the momentum-weight distributions of Wannier functions}\label{AppendixProtectionGamma}

In this appendix we analyze how linear symmetries that map $k \to -k$ constrain the biorthogonal Berry connection
\begin{equation}
    A(k) = -\mathrm{i}\bra{u^{L}_k}\partial_k\ket{u^{R}_k},
\end{equation}
and consequently the gauge-invariant quantity
\begin{equation}
    \Gamma(k)=\mathrm{Im} A(k)+\frac{1}{2}\partial_k\log\|u^{R}_k\|^2,
\end{equation}
which controls the $k$-space distribution of projected-position Wannier functions.
We restrict to a single isolated and nondegenerate band.

\subsection{Similarity-Type Linear Symmetry}

Consider a Bloch Hamiltonian satisfying
\begin{equation}
    h(k) = S h(-k) S^{-1},
\end{equation}
where $S$ is a $k$-independent unitary operator.

Let $\ket{u_k^R}$ be a right eigenvector,
\begin{equation}
    h(k)\ket{u_k^R} = E(k)\ket{u_k^R}.
\end{equation}
Using the symmetry, $h(k)S = S h(-k)$, we find
\begin{equation}
    h(k)(S\ket{u_{-k}^R})
    =
    S h(-k)\ket{u_{-k}^R}
    =
    E(-k)(S\ket{u_{-k}^R}).
\end{equation}
Since the band is nondegenerate and isolated, $E(-k)=E(k)$ and therefore
\begin{equation}
    \ket{u_k^R} = e^{f(k)} S \ket{u_{-k}^R},
\end{equation}
with $f(k)\in\mathbb{C}$.

A similar argument for left eigenvectors yields
\begin{equation}
    \bra{u_k^L} = \bra{u_{-k}^L} S^{-1} e^{-f(k)},
\end{equation}
where biorthonormality enforces the inverse exponent.

\subsubsection*{Berry Connection Transformation}

We compute
\begin{align}
    A(k)
    &= -\mathrm{i}\bra{u_k^L}\partial_k\ket{u_k^R} \nonumber \\
    &= -\mathrm{i}\bra{u_{-k}^L}S^{-1}e^{-f(k)}
       \partial_k\!\left[e^{f(k)}S\ket{u_{-k}^R}\right].
\end{align}

Since $S$ is $k$-independent,
\begin{align}
    \partial_k\!\left[e^{f(k)}S\ket{u_{-k}^R}\right]
    &=
    (\partial_k f(k)) e^{f(k)}S\ket{u_{-k}^R}
    +
    e^{f(k)}S\,\partial_k\ket{u_{-k}^R}.
\end{align}

Using biorthonormality $\braket{u_{-k}^L|u_{-k}^R}=1$, we obtain
\begin{equation}
    A(k)
    =
    -\mathrm{i}\partial_k f(k)
    -
    \mathrm{i}\bra{u_{-k}^L}\partial_k\ket{u_{-k}^R}.
\end{equation}

By definition,
\begin{equation}
    A(-k)
    =
    -\mathrm{i}\bra{u_{-k}^L}\partial_{(-k)}\ket{u_{-k}^R}.
\end{equation}

Since $\partial_{(-k)}=-\partial_k$, we have
\begin{equation}
    A(-k)
    =
    +\mathrm{i}\bra{u_{-k}^L}\partial_k\ket{u_{-k}^R}.
\end{equation}

Therefore,
\begin{equation}
    \boxed{
    A(-k)
    =
    -A(k)
    +
    \mathrm{i}\partial_k f(k)
    }.
\end{equation}

Taking imaginary parts,
\begin{equation}
    \boxed{
    \mathrm{Im} A(-k)
    =
    -\mathrm{Im} A(k)
    +
    \partial_k \mathrm{Re} f(k)
    }.
\end{equation}

\subsubsection*{Norm Transformation}

From the sewing relation,
\begin{equation}
    \|u_{-k}^R\|^2
    =
    e^{2\mathrm{Re} f(k)}
    \|u_k^R\|^2.
\end{equation}

Taking logarithms and differentiating with respect to $k$,
\begin{equation}
    \frac{1}{2}\partial_k\log\|u_{-k}^R\|^2
    =
    -\frac{1}{2}\partial_k\log\|u_k^R\|^2
    -
    \partial_k \mathrm{Re} f(k). \label{NormContribution}
\end{equation}

\subsubsection*{Transformation of $\Gamma(k)$}

Combining the previous results,
\begin{align}
    \Gamma(-k)
    &=
    \mathrm{Im} A(-k)
    +
    \frac{1}{2}\partial_k\log\|u_{-k}^R\|^2
    \nonumber \\
    &=
    -\Gamma(k).
\end{align}

Thus,
\begin{equation}
    \boxed{\Gamma(-k)=-\Gamma(k)}.
\end{equation}

It follows that
\begin{equation}
    \overline{\Gamma}
    =
    \frac{1}{2\pi}
    \int_{\mathrm{BZ}} dk\,\Gamma(k)
    =
    0,
\end{equation}
and therefore
\begin{equation}
    \|w_k\|^2=\|w_{-k}\|^2.
\end{equation}

Unitary similarity (inversion-type) symmetry forbids nonreciprocal drift in a single-band system.

\subsection{Linear Conjugation Symmetry}

Now consider a symmetry of the form
\begin{equation}
    S h(k) S^{-1} = h^\dagger(-k),
\end{equation}
with $S$ unitary and $k$-independent. Acting on a right eigenvector,
\begin{equation}
    h^\dagger(-k)\, S\ket{u_k^R}
    =
    E(k)\, S\ket{u_k^R}.
\end{equation}

Thus $S\ket{u_k^R}$ is proportional to a left eigenvector at $-k$,
\begin{equation}
    \ket{u_{-k}^L} = e^{f(k)} S\ket{u_k^R}.
\end{equation}
The symmetry therefore maps right and left eigenvectors onto one another. A direct computation yields
\begin{equation}
    A(-k)
    =
    A(k)^*
    +
    \mathrm{i}\partial_k f(k),
\end{equation}
so that
\begin{equation}
    \boxed{
    \mathrm{Im} A(-k)
    =
    \mathrm{Im} A(k)
    +
    \partial_k \mathrm{Re} f(k)
    }.
\end{equation}

Combining with the norm contribution \eqref{NormContribution} gives
\begin{equation}
    \boxed{\Gamma(-k)=\Gamma(k)}.
\end{equation}

Thus $\Gamma(k)$ is generically even and its Brillouin-zone average need not vanish. In contrast to the similarity case, this symmetry enforces an asymmetric momentum distribution
\begin{equation}
    \|w_{-k}\|^2 \neq \|w_{k}\|^2
\end{equation}
whenever $\Gamma(k)$ has a nontrivial dependence in $k$.

\subsection*{Summary}

For a single isolated band:

\begin{itemize}
    \item A unitary similarity symmetry $S h(k) S^{-1}=h(-k)$ enforces $\Gamma$ odd and forbids nonreciprocal drift.
    \item A linear conjugation symmetry $S h(k) S^{-1}=h^\dagger(-k)$ makes $\Gamma$ even and allows nonzero drift.
\end{itemize}

\section{Drift velocity of wave packets in non-Hermitian Hamiltonians with real energies}\label{Appendix:Driftvelocity}

Consider the state
\begin{align}
    \ket{\psi(0)} &= \int dk \ket{k} \otimes \ket{w_k}
\end{align}
belonging to a single occupied band in the continuum limit. Let such a state evolve under a non-Hermitian, time-independent Hamiltonian $H$:
\begin{align}
    \ket{\psi(t)} &= e^{-\mathrm{i}Ht/\hbar}\ket{\psi(0)} \nonumber \\
    &= \int dk e^{-\mathrm{i}E(k)t/\hbar} \ket{k} \otimes \ket{w_k},
\end{align}
where $E(k)$ is the dispersion of the occupied band and the integration is over the entire Brillouin zone. If $E(k) \in \mathbb{R}$ for all $k$, then the norm of $\ket{\psi(t)}$ is constant over time, 
\begin{equation}
    \langle\psi(t)|\psi(t)\rangle = \int dk \|w_k\|^2,   \label{Normwavepacket}
\end{equation}
where $\|w(k)\|^2$ is the initial state's momentum-weight distribution. 

In this representation, the position operator acts as $x \to \mathrm{i}\partial_k$. Hence, its expectation value reads
\begin{align}
    \langle x\rangle (t) &= \frac{\bra{\psi(t)}x\ket{\psi(t)}}{\bra{\psi(t)}\psi(t) \rangle}\nonumber \\
    &=\frac{\int dk \ e^{\mathrm{i}E(k)t/\hbar} \bra{w_k}\mathrm{i}\partial_k\Big(e^{-\mathrm{i}E(k)t/\hbar}\ket{w_k}\Big)}{\int dk \|w_k\|^2} \nonumber \\
    &= \frac{\frac{t}{\hbar}\int dk \|w_k\|^2\partial_kE(k) + \int dk \bra{w_k}\mathrm{i}\partial_k\ket{w_k}}{\int dk \|w_k\|^2}.
\end{align}
Note that the first term on the right-hand side of the expression above is linear in $t$ while the second is time-independent. Thus, the wave packet's drift velocity is constant and given by
\begin{align}
    v_{\mathrm{drift}} &\equiv \frac{d}{dt}\langle x\rangle \nonumber \\
    &= \frac{1}{\hbar}\frac{\int dk \|w_k\|^2\partial_kE(k)}{\int dk \|w_k\|^2} \nonumber \\
    &= \frac{1}{\hbar}\big\langle \partial_kE(k)\big\rangle.
\end{align}

\section{Pseudo-unitary Wilson loop and projected position operator}\label{AppendixPseudoUnitaryWilsonLoop}

From the definition \eqref{WilsonLineElement} and the relations \eqref{Def1}, \eqref{Lk}, and \eqref{Def2} between right/left eigenstates of pseudo-Hermitian Hamiltonians, we may express the Hermitian conjugate of the biorthogonal Wilson line element as
\begin{align}
    G_k^{\dagger}=R_{k}^{\dagger}L_{k+\Delta} &= R^{\dagger}_{k}\eta R_{k+\Delta}M_{k+\Delta}^{-1}, \label{LineConjugate}
\end{align}
where we have employed the definition \eqref{Lk}. At the same time, the inverse of the biorthogonal Wilson line element in the thermodynamic limit is
\begin{align}
    G_{k}^{-1} &= L_{k}^{\dagger}r_{k+\Delta} = M_{k}^{-1}R_{k}^{\dagger}\eta r_{k+\Delta}. \label{LineInverse}
\end{align}
By comparing \eqref{LineConjugate} and \eqref{LineInverse}, we immediately obtain 
\begin{equation}
    G_{k}^{-1}=M_{k}^{-1}G^{\dagger}_{k}M_{k+\Delta}. \label{WLelementRelation}
\end{equation}

Now, discretizing the Wilson loop as defined in \eqref{BiorthW} and applying \eqref{WLelementRelation} leads to
\begin{align}
    \mathcal{W}^{-1}_k &= G_{k}^{-1}G_{k+\Delta}^{-1}\cdots G^{-1}_{k+2\pi-\Delta} \nonumber \\
    &=(M^{-1}_kG^{\dagger}_kM_{k+\Delta})(M^{-1}_{k+\Delta}G^{\dagger}_{k+\Delta}M_{k+2\Delta})\cdots \nonumber \\
    & \ \times (M^{-1}_{k+2\pi-\Delta}G^{\dagger}_{k+2\pi-\Delta})M_{k+2\pi} \nonumber \\
    &= M_{k}^{-1}G^{\dagger}_kG^{\dagger}_{k+\Delta}\cdots G_{k+2\pi-\Delta}M_{k+2\pi} \nonumber \\
    &=M_{k}^{-1}\mathcal{W}_k^{\dagger}M_k. \label{WLpseudoUnitary}
\end{align}
The result \eqref{WLpseudoUnitary} means that the Wilson loop is \emph{pseudo-unitary}.

\section{Spectral properties of pseudo-unitary operators}\label{AppendixSpectral}

Let $W$ be a pseudo-unitary operator with respect to the metric $M$,
\begin{equation}
    W^{-1} = M^{-1}W^{\dagger}M \label{PseudoUnitarity}
\end{equation}
As always, we can define a bi-orthonormal set of vectors $\ket{r_{\lambda}},\ket{l_{\lambda}}$ such that
\begin{align}
    W\ket{r_{\lambda}}&=\lambda\ket{r_{\lambda}}, \nonumber \\
    W^{\dagger}\ket{l_{\lambda}} &= \lambda^{*}\ket{l_{\lambda}}, \label{BiOrthonormalWL}
\end{align}
where $\braket{l_{\lambda}}{r_{\lambda'}}=\delta_{r_{\lambda,\lambda'}}$, and we say $\ket{r_{\lambda}}$ ($\ket{l_{\lambda}}$) are the right (left) eigenvectors of $W$.

From Eqs. \eqref{PseudoUnitarity} and \eqref{BiOrthonormalWL}, it follows

\begin{enumerate}
    \item \emph{$W\ket{r_{\lambda}}$ is a left eigenvector of $W$ with eigenvalue $\frac{1}{\lambda^{*}}$}:
    \begin{align}
    W^{\dagger}M\ket{r_{\lambda}} &= MW^{-1}\ket{r_{\lambda}} = \lambda^{-1}M\ket{r_{\lambda}}\nonumber \\
    &\Rightarrow \bra{r_{\lambda}}MW = \frac{1}{\lambda^{*}}\bra{r_{\lambda}}M, \label{First}
    \end{align}
    \item \emph{$M^{-1}\ket{l_{\lambda}}$ is a right eigenvector of $W$ with eigenvalue $\frac{1}{\lambda^{*}}$}:
    \begin{align}
        W^{-1}M^{-1}\ket{l_{\lambda}} &= M^
        {-1}W^{\dagger}\ket{l_{\lambda}} = \lambda^{*}M^{-1}\ket{l_{\lambda}} \nonumber \\
        & \Rightarrow WM^{-1}\ket{l_{\lambda}} = \frac{1}{\lambda^{*}}M^{-1}\ket{l_{\lambda}}.
        \end{align}
        \item From items 1 and 2, we see that the spectrum of $W$ consists of unimodular eigenvalues $|\lambda|=1$ or inverse-conjugate pairs $(\lambda,\frac{1}{\lambda^{*}})$.
\end{enumerate}
If the spectrum of $W$ is nondegenerate, result 1 implies $M\ket{r_{\lambda}} \propto \ket{l_{\frac{1}{\lambda^{*}}}}$. And since $\langle{l_{\frac{1}{\lambda^{*}}}}|{r_{\lambda}}\rangle=\delta_{\frac{1}{\lambda^{*}},\lambda}$, we have
\begin{equation}
    \langle{r_{\lambda}}|{M}|{r_{\lambda}}\rangle = 0
\end{equation}
unless $|\lambda|= 1$. Analogously, result 2 implies $M\ket{l_{\lambda}} \propto \ket{r_{\frac{1}{\lambda^{*}}}}$ and, therefore,
\begin{equation}
    \langle{l_{\lambda}}|{M}|{l_{\lambda}}\rangle = 0
\end{equation}
unless $|\lambda|=1$.

Moreover, from the fact that $M\ket{r_{1/\lambda^{*}}} \propto \ket{l_{\lambda}}$, we find that
\begin{equation}
    1=\braket{l_{\lambda}}{r_{\lambda}}\propto\bra{r_{1/\lambda^{*}}}M\ket{r_{\lambda}} \neq 0.
\end{equation}
Similarly, from $M^{-1}\ket{l_{1/\lambda^{*}}}\propto \ket{r_{\lambda}}$, we have
\begin{equation}
    1=\braket{l_{\lambda}}{r_{\lambda}}\propto\bra{l_{\lambda}}M^{-1}\ket{l_{1/\lambda^{*}}} \neq 0. \label{Last}
\end{equation}

\section{Chiral symmetry of the anti-Hermitian part of pseudo-Hermitian operators}\label{AppendixChiral}

Let $H$ be a pseudo-Hermitian operator. Its eigenvalues are either real or come in complex-conjugate pairs~\cite{mostafazadeh2002pseudo}. Let us denote the real eigenvalues by $E_{\alpha} \in \mathbb{R}$ such that
\begin{equation}
    H\ket{\psi^{R}_{\alpha}}=E_{\alpha}\ket{\psi^{R}_{\alpha}},
\end{equation}
and the complex conjugate pairs by $(E_{\beta},E_{\beta}^{*}) =(\epsilon_{\beta}+\mathrm{i}\gamma_{\beta},\epsilon_{\beta}-\mathrm{i}\gamma_{\beta})$, $\epsilon_{\beta},\gamma_{\beta} \in \mathbb{R}$,
such that
\begin{align}
    H\ket{\psi^{R}_{\beta}} &= E_{\beta}\ket{\psi^{R}_{\beta}}, \nonumber \\
    H\ket{\overline{\psi}^{R}_{\beta}} &= E_{\beta}^{*}\ket{\overline{\psi}^{R}_{\beta}}.
\end{align}

Now, for every real eigenvalue $E_a$, define the operator
\begin{equation}
    \eta_{\alpha} = \ket{\psi^{L}_{\alpha}}\bra{\psi^{L}_{\alpha}} \label{etaa},
\end{equation}
where $\ket{\psi^{L}_{\alpha}}$ are biorthonormal left eigenvectors of $H$ satisfying $\langle \psi^{L}_{\alpha}|\psi^{R}_{\alpha'}\rangle=\delta_{\alpha,\alpha'}$. We have
\begin{equation}
    \eta_{\alpha}\ket{\psi^{R}_{\alpha'}} = \ket{\psi^{L}_{\alpha}}\bra{\psi^{L}_{\alpha}}\psi^{R}_{\alpha'} \rangle = \delta_{\alpha,\alpha'}\ket{\psi^{L}_{\alpha'}}\label{Relationshipa}.
\end{equation}
Meanwhile, for every pair $(E_{\beta},E^{*}_{\beta})$, define
\begin{equation}
    \eta_{\beta} = \ket{\overline{\psi}^{L}_{\beta}}\bra{\psi^{L}_{\beta}} + \ket{\psi^{L}_{\beta}}\bra{\overline{\psi}^{L}_{\beta}} \label{etab},
\end{equation}
such that
\begin{align}
    \eta_{\beta}\ket{\psi^{R}_{\beta'}} &= \delta_{\beta,\beta'}\ket{\overline{\psi}^{L}_{\beta'}}, \nonumber \\
    \eta_{\beta}\ket{\overline{\psi}^{R}_{\beta'}} &= \delta_{\beta,\beta'}\ket{\psi^{L}_{\beta'}}\label{Relationshipb}
\end{align}

Using the definitions \eqref{etaa} and \eqref{etab}, we can build the global metric operator
\begin{equation}
    \eta = \sum_{\alpha}\eta_{\alpha}+\sum_{\beta}\eta_{\beta} \label{etaunitarydefinition}.
\end{equation}
One can see that $H^{\dagger}=\eta H \eta^{-1}$ through the spectral decomposition of $H$,
\begin{equation}
    H = \sum_{\alpha}E_{\alpha}\ket{\psi^{R}_{\alpha}}\bra{\psi^{L}_{\alpha}}+\sum_{\beta}\left(E_{\beta}\ket{\psi^{R}_{\beta}}\bra{\psi^{L}_{\beta}}+E^{*}_{\beta}\ket{\overline{\psi}^{R}_{\beta}}\bra{\overline{\psi}^{L}_{\beta}}\right).
\end{equation}
From Eqs. \eqref{Relationshipa} and \eqref{Relationshipb}, and noting that
\begin{align}
    \eta_{\alpha}^{-1}\ket{\psi^{L}_{\alpha}}&=\ket{\psi^{R}_{\alpha}} \nonumber \\
    \eta_{\beta}^{-1}\ket{\psi^{L}_{\beta}}&=\ket{\overline{\psi}^{R}_{\beta}}, \quad \eta_{\beta}^{-1}\ket{\overline{\psi}^{L}_{\beta}}=\ket{\psi^{R}_{\beta}},
\end{align}
we obtain
\begin{align}
    \eta H \eta^{-1} &= \sum_{\alpha}E_{\alpha}\ket{\psi^{L}_{\alpha}}\bra{\psi^{R}_{\beta}} \nonumber \\
    & \ \ +\sum_{\beta}\left(E_{\beta}\ket{\overline{\psi}^{L}_{\beta}}\bra{\overline{\psi}^{R}_{\beta}}+E^{*}_{\beta}\ket{\psi^{L}_{\beta}}\bra{\psi^{R}_{\beta}}\right) \nonumber \\
    &= H^{\dagger} \label{etacondition}.
\end{align}
To the best of our knowledge, the construction of the metric as in Eq.~\eqref{etaunitarydefinition} first appears in Ref. ~\cite{mostafazadeh2002pseudo,mostafazadeh2001pseudo,mostafazadeh2002pseudo3}. Note that such a metric is a full-rank operator and all its eigenvalues are 1 or $-1$, so $\eta^{2}=\mathbb{I} \Rightarrow \eta^{-1}=\eta$ and we can write Eq. \eqref{etacondition} as
\begin{equation}
    H^{\dagger} = \eta H \eta \label{etaother}.
\end{equation}

Now, one may decompose any operator into its Hermitian and anti-Hermitian parts,
\begin{equation}
    H = H_h+\mathrm{i}H_a,
\end{equation}
where $H_{h}^{\dagger}=H_{h}$ and $H_a^{\dagger}=H_a$. Conversely, we have
\begin{align}
    H_h &= \frac{H+H^{\dagger}}{2}, \nonumber \\
    H_a &= \frac{H-H^{\dagger}}{2\mathrm{i}}.
\end{align}
Using Eq. \eqref{etaother}, we obtain
\begin{align}
    \eta H_{a} &= \frac{\eta H- \eta H^{\dagger}}{2\mathrm{i}} \nonumber \\
    &= \frac{H^{\dagger}\eta- H\eta}{2\mathrm{i}} \nonumber \\
    &= -H_{a}\eta.
\end{align}
Therefore, we show that the anti-Hermitian part of a pseudo-Hermitian operator anti-commutes with the metric $\eta$,
\begin{equation}
    \{H_a,\eta\}=0,
\end{equation}
that is, it has chiral symmetry. Similarly, one can show that $\eta$ commutes with the Hermitian part of the operator,
\begin{equation}
    [H_h,\eta]=0.
\end{equation}

\section{Constraints of (pseudo) inversion symmetry on the Wannier spectrum}\label{Appendix:Inversion}

\subsection{Hermitian case}

In this subsection, we review the constraints on the Wilson loop due to the inversion symmetry first described in Ref.~\cite{Alexandradinata}. As we will see, Wannier centers can be restricted to specific fixed values under inversion symmetry. An inversion-symmetric crystal has a Bloch Hamiltonian that obeys
\begin{equation}
    I h(\kv)I^{-1}=h(-\kv),
    \label{eq:InversionSymmetryHermitian}
\end{equation}
where $I$ is the inversion operator that obeys $I^{-1}=I^\dag$ and $I^2=1$. Hamiltonians with inversion symmetry \eqref{eq:InversionSymmetryHermitian} have Wilson loops that satisfy
\begin{equation}
    \mathcal{W}_{+,k}=B_{-k}\mathcal{W}_{+,-k}^\dag B_k,
    \label{eq:WilsonLoopUnderInversionSymmetry_H}
\end{equation}
where $B_k$ is known as the \emph{sewing matrix}, with components $[B_k]^{mn}=\bra{w_{-k}^m}I\ket{w_k^n}$, and where $m,n$ run over occupied bands~\footnote{Througout the entire paper, indices in Wilson loops or sewing matrices run over occupied bands.}.

To determine the set $\{ \nu_i \}$ in the presence of inversion symmetry, it is sufficient to compute the irreducible representations (irreps) of inversion symmetry for the occupied bands at certain high-symmetry points (HSPs) in the Brillouin zone, rather than evaluating the Wilson loop. HSPs are the crystal momenta $\kv_I$ that are invariant under inversion, i.e., $\kv_I = -\kv_I$ modulo a reciprocal lattice vector. In one dimension, they are $k_I = \{0, \pi\}$; in two dimensions, for a square lattice, they are $\kv_I = \{ \Gamma = (0,0), X = (\pi,0), Y = (0,\pi), M = (\pi,\pi) \}$. At these points, Eq.~\eqref{eq:InversionSymmetryHermitian} implies that $[I, h(\kv_I)] = 0$, i.e., the Bloch Hamiltonian commutes with the inversion operator. As a result, the energy eigenstates at HSPs can be chosen to be simultaneous eigenstates of the inversion operator. Thus, for the sewing matrix at ${\kv}_I$, we have
\begin{align}
    0=&\bra{w_{\kv_I}^m}h(\kv_I) I-I h(\kv_I) \ket{w_{\kv_I}^n}\nonumber \\
    =&(\epsilon_{\kv_I}^m-\epsilon_{\kv_I}^n)\bra{w_{\kv_I}^m} I \ket{w_{\kv_I}^n}\nonumber \\
    =&(\epsilon_{\kv_I}^m-\epsilon_{\kv_I}^n)[B_{\kv_I}]^{mn},
    \label{eq:sewing matrix Hermitian inversion main text}
\end{align}
i.e., $B_{\kv_I}$  is diagonal over non-degenerate eigenstates and has eigenvalues $\pm 1$.

Ref.~\onlinecite{Alexandradinata} shows that the irreducible representations (irreps) of inversion symmetry for the occupied bands at all inversion-invariant momenta $\kv_I$—equivalently, the eigenvalues of $B_{\kv_I}$—can be used to determine the corresponding Wannier centers, which we now reproduce for completeness. The results are summarized in Tables ~\ref{tab:table0}.

\subsubsection{Protection of Wannier centers from symmetry representations at HSPs}

For a Wilson loop with base point at $k_i=-\pi$, we separate the sweep over the BZ into two Wilson lines
\begin{align}
    W_{\pi\leftarrow -\pi}&=W_{\pi\leftarrow 0}W_{0\leftarrow -\pi}\nonumber\\
   &=B_\pi W_{0\leftarrow -\pi}^{\dag} B_0 W_{0\leftarrow -\pi}.
   \label{eq:Hermitian Wilson loop Inversion}
\end{align}
Since Wilson lines are unitary for a Hermitian Hamiltonian, we assume $W_{0\leftarrow -\pi}$ has the form
\begin{equation}
    W_{0\leftarrow -\pi}=e^{\textrm{i}\alpha}\left(
\begin{array}{cc}
c & d \\
 -d^* & c^* \\
\end{array}
\right),
\label{eq:forms of half wilson loop}
\end{equation}
where $|c|^2+|d|^2=1$ to ensure unitarity. As mentioned below Eq~\eqref{eq:sewing matrix Hermitian inversion main text}, sewing matrices $B_0$ and $B_{\pi}$ are diagonal. Assuming that
\begin{equation}
    B_0=\left(
\begin{array}{cc}
\xi_0^1 & 0 \\
 0 & \xi_0^2 \\
\end{array}
\right), B_\pi=\left(
\begin{array}{cc}
\xi_\pi^1 & 0 \\
 0 & \xi_\pi^2 \\
\end{array}
\right),
\label{eq: forms of sewing matrix}
\end{equation}
where $\xi_0^{1,2}$ ($\xi_\pi^{1,2}$) are the irreps of IS at $k=0$($k=\pi$) for Hermitian crystals, which can only take values $\pm 1$.
Combining the the form of sewing matrices~\eqref{eq: forms of sewing matrix} with~\eqref{eq:Hermitian Wilson loop Inversion} and~\eqref{eq:forms of half wilson loop}, we can find the Wilson loop to be
\begin{align}
    \mathcal{W}_{\pi\leftarrow -\pi}=\left(
\begin{array}{cc}
\xi_\pi^1(|c|^2\xi_0^1+|d|^2\xi_0^2) & c^* d \xi_\pi^1(\xi_0^1-\xi_0^2) \\
 cd^*\xi_\pi^2(\xi_0^1-\xi_0^2) & \xi_\pi^2(|c|^2\xi_0^2+|d|^2\xi_0^1) \\
\end{array}
\right).
\end{align}
By going over all possible combinations of $\{\xi_{k_I}^{1,2}\}$, we arrive at Table \ref{tab:table0}. 

In what follows, we consider non-Hermitian (NH) Hamiltonians, for which inversion symmetry ramifies into inversion and pseudo-inversion.

\begin{table}[b]
\caption{
$I$ eigenvalues of occupied bands of Hermitian Bloch Hamiltonians at $k_I=\{0,\pi\}$ and corresponding Wannier centers $\nu_1,\nu_2$ for two occupied bands, $N_\mathrm{occ}=2$. This table is also valid upon the exchange $+ \leftrightarrow -$.
}
\begin{ruledtabular}
\begin{tabular}{cc}
\textrm{$I$ eigenvalues at $\{0,\pi\}$} & $\nu_1,\nu_2$\\
\colrule
$\{(++)(++)\}$ &$0,0$ \\
$\{(++)(+-)\}$ & $0,\pi$  \\
$\{(++)(--)\}$ & $\pi,\pi$  \\
$\{(+-)(+-)\}$ &  $\nu,-\nu$
\end{tabular}
\end{ruledtabular}
\label{tab:table0}
\end{table}

\subsection{Inversion symmetry for NH Hamiltonians}

The similarity-type branch of inversion symmetry for NH Hamiltonians is defined as
\begin{equation}
    h(k)=I_{s}h(-k)I_{s}^{-1},
\end{equation}
where $I_s^{\dagger}=I_{s}$ and $I_{s}^2=\mathbb{I}$. If $\ket{u^{R}_{k,n}}$ is an energy eigenstate with eigenvalue $E_{k,n}$, we have
\begin{equation}
    h(-k)I_s\ket{u^{R}_{k,n}} = I_sh(k)\ket{\psi^{R}_{k,n}} = E_{k,n}I_s\ket{u^{R}_{k,n}},
\end{equation}
so $I_s\ket{u^{R}_{k,n}}$ is a right eigenstate at $-k$ with the same eigenvalue $E_{k,n}$. Now, consider the projector onto the occupied subspace associated with momentum $-k$,
\begin{equation}
    P_{1,-k} = R_{-k}L_{-k}^{\dagger} = \sum_{n=1}^{N_{\mathrm{occ}}}\ket{u^{R}_{-k,n}}\bra{u^{L}_{-k,n}}. \label{ProjectorMinusk}
\end{equation}
Since $I_s\ket{u^{R}_{k,n}}$ lies in the subspace spanned by $\ket{u^{R}_{-k,n}}$, the object in \eqref{ProjectorMinusk} acts trivially on $I_sR_k$:
\begin{align}
    I_sR_k &= R_{-k}L_{-k}^{\dagger}I_sR_k \nonumber \\
    &= R_{-k}B^{s}_{k}, \label{Brelation}
\end{align}
where we have defined the \emph{sewing matrix}
\begin{equation}
    B^{s}_{k} \equiv L_{-k}^{\dagger}I_sR_k.
\end{equation}

If $\ket{u^{L}_{k,n}}$ is an left eigevector of $h(k)$ with eigenvalue $E_{k,n}$ we have
\begin{equation}
    \bra{u^{L}_{k,n}}I_sh(-k) = \bra{u^{L}_{k,n}}h(-k)I_s = E^{*}_{k,n}\bra{u^{L}_{k,n}}I_s,
\end{equation}
so $I_s\bra{u^{L}_{k,n}}$ lies in the subspace spanned by $\bra{u^{L}_{-k,n}}$. Therefore, we may also obtain a sewing relation for the left eigenstates:
\begin{align}
    L^{\dagger}_kI_s &= L^{\dagger}_kI_sR_{-k}L_{-k}^{\dagger} \nonumber \\ 
    &= B^{s}_{-k}L_{-k}^{\dagger}. \label{SewingLeft}
\end{align}

Now, acting with $I_s$ on \eqref{Brelation} and using $I_s^2=\mathbb{I}$, we get
\begin{align}
    R_{k}& = I_s^2R_k \nonumber \\
    &= I_sR_{-k}B^{s}_k.
\end{align}
But we know, also from \eqref{Brelation}, that $I_sR_{-k}=R_kB^{s}_{-k}$, so
\begin{align}
    R_{k} = R_{k}B^{s}_{-k}B^{s}_{k}.
\end{align}
Multiplying both sides by $L_{k}^{\dagger}$ and using the biorthonormality condition $L_{k}^{\dagger}R_k=\mathbb{I}$ yields
\begin{equation}
    B^{s}_{-k}B^{s}_{k} \Leftrightarrow (B^s_k)^{-1} = B^{s}_{-k}.
\end{equation}

Using $I_s^2=\mathbb{I}$ and the sewing relations \eqref{Brelation} and \eqref{SewingLeft} allow us to express the biorthogonal Wilson line element \eqref{WilsonLineElement} as
\begin{align}
    G_{k} &= (L_{k+\Delta}^{\dagger}I_s)(I_sR_k) \nonumber \\
    &= (B^{s}_{-k-\Delta}L^{\dagger}_{-k-\Delta})(R_{-k}B^{s}_k) \nonumber \\
    &= (B^{s}_{k+\Delta})^{-1}G^{-1}_{-k}B^{s}_{k}. \label{InversionLineElement}
\end{align}
Therefore, the Wilson loop becomes
\begin{align}
    \mathcal{W}_k &= G_{k}G_{k+\Delta}\cdots G_{k+2\pi-\Delta} \nonumber \\
    &= \prod_{j=0}^{N-1}
       \big[(B^{s}_{k_j+\Delta})^{-1} G^{-1}_{-k_j} B^{s}_{k_j}\big] \nonumber \\
    &= (B^{s}_{k+2\pi})^{-1}
       \big[G_{-k} G_{-(k+\Delta)} \cdots G_{-(k+2\pi-\Delta)}\big]^{-1}
       B^{s}_{k} \nonumber \\
    &= (B^{s}_k)^{-1}\,\mathcal{W}_{-k}^{-1}\,B^{s}_k
     = B^{s}_{-k}\,\mathcal{W}_{-k}^{-1}\,B^{s}_k,
\end{align}
that is, it is pseudo-involutory at HSPs.

Now, let us investigate how the possible symmetry representations at HPSs constrain the Wannier for two occupied bands.

\subsubsection{Protection of Wannier centers from inversion representations at HSPs}\label{Appendix:InversionIrreps}

We study the topological protection of IS at HSPs for NH crystals. Let $k_i=-\pi$ be the starting point of the Wilson loop. We split the Wilson loop into two Wilson lines, each spanning half BZ. Combining with \eqref{InversionLineElement}, we have
\begin{align}
    W_{\pi\leftarrow -\pi}&=W_{\pi\leftarrow 0}W_{0\leftarrow -\pi}\nonumber\\
    &=B_\pi^a W_{0\leftarrow -\pi}^{-1} B_0^a W_{0\leftarrow -\pi},
    \label{eq:Inv Sym HSP}
\end{align}
where $B_{0,\pi}$ are the sewing matrices. Note that the Wilson line $W_{0\leftarrow -\pi}$ is no longer unitary. We assume $W_{0\leftarrow -\pi}$ has the most general form
\begin{equation}
    W_{0\leftarrow -\pi}=\left(
\begin{array}{cc}
a & b \\
 c & d \\
\end{array}
\right)
\label{eq:forms of half wilson loop NH IS}
\end{equation}
where $a,b,c,d\in \mathbb{C}$. Again, we take the sewing matrices to be of the form
\begin{equation}
    B_0^a=\left(
\begin{array}{cc}
\xi_0^1 & 0 \\
 0 & \xi_0^2 \\
\end{array}
\right), B_\pi^a=\left(
\begin{array}{cc}
\xi_\pi^1 & 0 \\
 0 & \xi_\pi^2 \\
\end{array}
\right),
\label{eq: forms of sewing matrix NH IS}
\end{equation}
where $\xi_0^{1,2}$ ($\xi_\pi^{1,2}$) are the irreps of IS at $k=0$($k=\pi$) for NH crystals, which can only take values $\pm 1$. Combining the form of sewing matrices~\eqref{eq: forms of sewing matrix NH IS} with~\eqref{eq:Inv Sym HSP} and~\eqref{eq:forms of half wilson loop NH IS}, we can find the Wilson loop
\begin{align}
    \mathcal{W}_{\pi\leftarrow -\pi}=\frac{1}{ad-bc}\left(
\begin{array}{cc}
(ad\xi_0^1-bc\xi_0^2)\xi_\pi^1 & bd(\xi_0^1-\xi_0^2)\xi_\pi^1 \\
 ac(\xi_0^2-\xi_0^1)\xi_\pi^2 & (ad\xi_0^2-bc\xi_0^1)\xi_\pi^2 \\
\end{array}
\right).
\end{align}
By going over all possible combinations of $\{\xi_{k_I}^{1,2}\}$, we obtain Table \ref{tab:table1} in the main text.

\subsection{Pseudo-inversion symmetry for NH Hamiltonians}\label{Appendix:pInversion}

We define pseudo-inversion symmetry as
\begin{equation}
    h(k)^{\dagger}=I_{c}h(-k)I_{c}^{-1},
\end{equation}
where again $I_s^{\dagger}=I_{c}$ and $I_{c}^2=\mathbb{I}_N$. If $\ket{u^{R}_{k,n}}$ is a right eigenvector of $h(k)$, then
\begin{equation}
    h(-k)^{\dagger}I_c\ket{u^{R}_{k,n}} = I_c h(k)\ket{u^{R}_{k,n}}=E_{k,n}I_c
    \ket{u^{R}_{k,n}},
\end{equation}
so that $I_c\ket{u^{R}_{k,n}}$ is a right eigenstate of $h(-k)^{\dagger}$ with eigenvalue $E_{k,n}$, and therefore a left eigestate of $h(-k)$ with eigenvalue $E^{*}_{k,n}$. Therefore, $I_cR_{k}$ lies in the subspace spanned by $L_{-k}$. For that reason, the projector
\begin{equation}
    P_{2,-k}=L_{-k}R_{-k}^{\dagger} = \sum_{n=1}^{N_{\mathrm{occ}}}\ket{u^{L}_{-k,n}}\bra{u^{R}_{-k,n}}
\end{equation}
acts trivially on $I_cR_k$. Hence,
\begin{align}
    I_cR_k &= L_{-k}R_{-k}^{\dagger}I_cR_k \nonumber \\
    &= L_{-k}B^{c}_{k},
\end{align}
where we have defined the sewing matrix
\begin{equation}
    B^{c}_{k}=R_{-k}^{\dagger}I_cR_k \label{Bk}.
\end{equation}

Similarly, one can show that $I_cL_k$ lies in the subspace spanned by $R_{-k}$, so the projector
\begin{equation}
    P_{1,-k}=R_{-k}L_{-k}^{\dagger} = \sum_{n=1}^{N_{\mathrm{occ}}}\ket{u^{R}_{-k,n}}\bra{u^{L}_{-k,n}}
\end{equation}
acts trivially on $I_cL_k$. Hence,
\begin{align}
    I_cL_k &= R_{-k}L_{-k}^{\dagger}I_cL_k \nonumber \\
    &= R_{-k}\tilde{B}^{c}_{-k},
\end{align}
where
\begin{equation}
    \tilde{B}^{c}_{-k} = L_{-k}^{\dagger}I_cL_k. \label{Btildek}
\end{equation}
Using \eqref{Bk}, \eqref{Btildek} and $I_c^2 = \mathbb{I}_N$, we apply $I_c$ twice to $R_k$:
\begin{align}
    R_k &= I_c^2 R_k \nonumber \\
        &= I_c (I_c R_k) \nonumber \\
        &= I_c (L_{-k} B^c_k) \nonumber \\
        &= (I_c L_{-k}) B^c_k \nonumber \\
        &= (R_k \tilde B^c_k) B^c_k \nonumber \\
        &= R_k \tilde B^c_k B^c_k.
\end{align}
Left-multiplying by $L_k^\dagger$ and using biorthonormality $L_k^\dagger R_k = \mathbb{I}_{N_\mathrm{occ}}$, we obtain
\begin{equation}
    \mathbb{I}_{N_\mathrm{occ}}
    = L_k^\dagger R_k
    = L_k^\dagger R_k \tilde B^c_k B^c_k
    = \tilde B^c_k B^c_k.
\end{equation}
Thus $\tilde B^c_k$ is the (left) inverse of $B^c_k$ on the occupied subspace. Since these are finite-dimensional matrices, the left inverse coincides with the right inverse, and we conclude
\begin{equation}
    \tilde B^c_k = \big(B^c_k\big)^{-1}.
\end{equation}

Now, we can write the biorthogonal Wilson line element \eqref{WilsonLineElement} as
\begin{align}
    G_k &= (L^{\dagger}_{k+\Delta}I_c)(I_cR_k) \nonumber \\
    &= (B_{k+\Delta})^{-1}r_{-k-\Delta}L_{-k}B^{c}_k \nonumber \\
    &= (B_{k+\Delta})^{-1}G^{\dagger}_{-k-\Delta}B^{c}_k.
\end{align}
Therefore, the Wilson line from $k_i$ to $k_f$ as defined in \eqref{WilsonLine} becomes
\begin{align}
    W_{k_f \leftarrow k_i} &= (B^{c})_{k_f}^{-1}W^{\dagger}_{-k_f \leftarrow -k_i-\Delta}B^{c}_{k_i} \nonumber \\
    &\approx (B^{c})_{k_f}^{-1}W^{\dagger}_{-k_f \leftarrow -k_i}B^{c}_{k_i} \label{PseudoWilsonLine}
\end{align}
in the thermodynamic limit $\Delta \to 0$. Let the Wilson line sweep the entire BZ, and we obtain the following relation for the Wilson loop,
\begin{equation}
    \mathcal{W}_{k}=(B^{c}_{k})^{-1}W^{\dagger}_{-k}B^{c}_{k},
\end{equation}
so it is pseudo-Hermitian at HSPs.

\subsubsection{Protection of Wannier centers from pseudo-inversion representations at HSPs}\label{Appendix:PseudoInversionRepresentations}

By \eqref{PseudoWilsonLine}, we have
\begin{align}
    W_{\pi\leftarrow -\pi}&=W_{\pi\leftarrow 0}W_{0\leftarrow -\pi}\nonumber\\ 
    &=B_\pi^b W_{0\leftarrow -\pi}^{\dag} B_0^b W_{0\leftarrow -\pi}.
    \label{eq:pseudo Inv Sym HSP}
\end{align}
Again, we let Wilson line $W_{0\leftarrow -\pi}$ take the general form \eqref{eq:forms of half wilson loop NH IS}, and we assume that the sewing matrices have the form
\begin{equation}
    B_0^b=\left(
\begin{array}{cc}
\xi_0^1 & 0 \\
 0 & \xi_0^2 \\
\end{array}
\right), B_\pi^b=\left(
\begin{array}{cc}
\xi_\pi^1 & 0 \\
 0 & \xi_\pi^2 \\
\end{array}
\right),
\label{eq: forms of sewing matrix NH pIS}
\end{equation}
where $\xi_0^{1,2}$ ($\xi_\pi^{1,2}$) are the irreps of pIS at $k=0$($k=\pi$) for NH crystals, which can only take values $\pm 1$.
Then we have
\begin{align}
    \mathcal{W}_{\pi\leftarrow -\pi}=\left(
\begin{array}{cc}
|a|^2\xi_0^1 \xi_\pi^1 + |c|^2\xi_0^2\xi_\pi^1 & a^*b\xi_0^1\xi_\pi^1+c^*d\xi_0^2\xi_\pi^1 \\
 ab^*\xi_0^1\xi_\pi^2+cd^*\xi_0^2\xi_\pi^2 & |b|^2\xi_0^1\xi_\pi^2+|d|^2\xi_0^2\xi_\pi^2 \\
\end{array}
\right).
\end{align}
By going over all possible combinations of $\{\xi_{k_I}^{1,2}\}$, we have Table \ref{tab:table2} in the main text.

\end{document}